\documentclass[10pt,a4paper]{book}
\usepackage[centertags]{amsmath}
\usepackage{amsfonts}
\usepackage{amssymb}
\usepackage{amsthm}
\usepackage{graphicx}
\usepackage{fancyhdr}
\usepackage{titlesec}

\newcommand{\blankline}{\vskip .3cm}
\newcommand{\beq}{\begin{equation}}
\newcommand{\eeq}{\end{equation}}
\newcommand{\matrice}{\begin{pmatrix}}
\newcommand{\ematrice}{\end{pmatrix}}
\newcommand{\bea}{\begin{eqnarray}}
\newcommand{\eea}{\end{eqnarray}}

\let\a=\alpha \let\b=\beta  \let\g=\gamma  \let\d=\delta
     \let\th=\theta \let\k=\kappa \let\l=\lambda
\let\m=\mu    \let\n=\nu         \let\p=\pi    
\let\s=\sigma      
    
\let\G=\Gamma \let\D=\Delta   \let\L=\Lambda 
    \let\Si=\Sigma     
  \let\ee=\epsilon

\let\==\equiv
\let\io=\infty
\def\ie{{i.e. }}

\let\dpr=\partial

\def\TT{{\cal T}}\def\NN{{\cal N}}

\def\to{\rightarrow}
\def\la{\left\langle}
\def\ra{\right\rangle}

\def\Tr{{\rm Tr}\,}
\def\la{\langle}
\def\ra{\rangle}

\titleformat{\chapter}[display]{}{\bf\LARGE\chaptername\ \thechapter}{1pc}{\titlerule \vspace{1pc} \bf\Huge}

\titleformat{name=\chapter,numberless}[display]{}{}{1pc}{ \bf\Huge}[\vspace{1pc} \titlerule]

\begin{document}

\frontmatter

\thispagestyle{empty}

\vspace*{-3cm}

\vskip3.truecm
\centerline{\LARGE \textbf{Quantum Gravity from Simplices:}}\vskip.2truecm
\centerline{\LARGE \textbf{Analytical Investigations}}\vskip.2truecm
\centerline{\LARGE \textbf{of Causal Dynamical Triangulations}}
\vskip2.truecm

\centerline{\large Dario Benedetti}
\blankline \blankline \centerline{\it Spinoza Institute and Institute for Theoretical Physics,}
\centerline{\it Utrecht University,}
\centerline{\it Leuvenlaan 4, NL-3584 CE Utrecht, The Netherlands}
\centerline{\tt d.benedetti@phys.uu.nl}

\blankline\blankline \blankline

\vspace{1cm}

\centerline{ABSTRACT}
\vspace{.2cm}

\noindent{A potentially powerful approach to quantum gravity has been developed over the last few years
under the name of Causal Dynamical Triangulations. Numerical simulations have given very interesting
results in the cases of two, three and four spacetime dimension.
The aim of this thesis is to give an introduction to the subject (Chapter 1), and try to push the analytical
understanding of these models further. This is done by first studying (Chapter 2) the case of a (1+1)-dimensional
spacetime coupled to matter, in the form of an Ising model, by means of high- and low-temperature expansions.
And after (Chapter 3) by studying a specific model in (2+1) dimensions, whose solution and continuum limit are presented.
}

\clearpage
\thispagestyle{empty}


\cleardoublepage
\thispagestyle{empty}
\vskip2.truecm

\centerline{\LARGE \textbf{Quantum Gravity from Simplices:}}
\centerline{\LARGE \textbf{Analytical Investigations}}
\centerline{\LARGE \textbf{of Causal Dynamical Triangulations}}
\vskip3.truecm
\centerline{\Large \textbf{Quantum Gravitatie uit Simplices:}}
\centerline{\Large \textbf{Analytisch Onderzoek}}
\centerline{\Large \textbf{naar Causale Dynamische Triangulaties}}

\vskip5.truecm

\centerline{Proefschrift}

\vskip1.truecm

\centerline{ter verkrijging van de graad van doctor aan de Universiteit Utrecht op gezag van de rector}
\centerline{magnificus,
prof.dr.  W.H. Gispen, ingevolge het besluit van het college voor promoties in}
\centerline{ het openbaar te verdedigen op maandag
11 juni 2007 des middags te 2.30 uur}

\vskip1.truecm
\centerline{door}

\vskip2.truecm

\centerline{\large Dario Benedetti}

\vskip1.truecm

\centerline{geboren op 10 juli 1978}
\centerline{te Rome, Itali\"e}

\clearpage
\thispagestyle{empty}

\vskip2.truecm

Promotor:       Prof. Dr. R. Loll

\vskip20.truecm

\noindent Dit proefschrift werd mogelijk gemaakt met financiële steun van de Stichting voor\\
Fundamenteel Onderzoek der Materie (FOM).

%


\pagestyle{plain}
\pagenumbering{roman}
\mbox{}
\newpage
\setcounter{page}{1}

\tableofcontents




\pagestyle{empty}


\newpage
\mainmatter

\pagenumbering{arabic}
\setcounter{page}{1}

\pagestyle{fancy}

\addcontentsline{toc}{chapter}{Preface}

\chapter*{Preface}

Since ancient times man has always been fascinated by the quest for the ultimate
nature of things; the need for a deeper understanding has brought philosophers and
scientists to look where eyes cannot see, be it the very distant or the very small.

Meditating on the very small, Democritus, already twenty-four centuries ago formulated an
atomic hypothesis. Many centuries later, the study of atomic phenomena had a fundamental role
in the birth of quantum mechanics, which today is the basis of all microscopic
physics. In particular, the Standard Model, which gives a complete description of the strong,
the weak and the electromagnetic interactions, is built upon the rules of the quantum theory.

On the other hand it was by looking at large distances that Newton was able to formulate his
classical theory of the gravitational force, the fourth and the last of the known fundamental
interactions. Newton's theory of gravity is nowadays known to be valid only in the special
case of small velocities and masses, otherwise being replaced by Einstein's theory of general
relativity. The relativistic theory of gravity has changed our understanding of the universe,
and is a necessary ingredient in cosmology.

Given the very different scales at which the theories of quantum mechanics and general
relativity are typically relevant, it would not be a problem that they have different natures,
quantum and indeterministic in the first instance, and classical and deterministic in the second instance.
Since the ratio of the gravitational to electromagnetic force between two
protons is of the order of $10^{-38}$, gravity has never appeared in microscopic physics experiments.
However, quantum mechanics has taught us that in order to probe short distances we need high energy.
Additionally, general relativity has taught us that energy gravitates. It should then occur that looking
at shorter and shorter scales we would concentrate enough energy to make the gravitational field
come into play. A simple calculation shows how important this would be if we would try
to probe a distance of the order of the Planck length $L_P\simeq 1.6\times 10^{-33}cm$,
since doing that would concentrate enough energy to create a black hole.
If we do not want to run into the paradox of a classical and a quantum theory coexisting at the
same energy scale, close to the Planck scale gravity must be quantized too.

The need for a quantum theory of gravity was advocated already at the dawn of the quantum
revolution,
but has since then revealed itself as one of the most challenging tasks for theoretical physicists,
remaining unsettled up to now.

The reasons for this situation are manifold. Already Heisenberg foresaw troubles
because the coupling constant of gravity, Newton's constant $G_N$, has inverse-square energy
dimensions\footnote{Throughout this thesis I will use units $\hbar =c=1$, so that $G_N=[E^{-2}]$
and the Planck units are all defined just in terms of Newton's constant, $L_P=T_P=M_P^{-1}=\sqrt{G_N}$,.}.
Hence the effective coupling in a perturbative expansion is $G_N E^2$, and the perturbative
theory is expected to break down when $G_N E^2\sim 1$, that is, at the Planck energy.
In modern language we say that the theory is non-renormalizable.

Furthermore, and besides the many technical difficulties ultimately related to the highly non-linear
nature of general relativity, there are fundamental problems raised by the novel ideas that these two
theories have revealed and which appear in contrast with each other.
As already mentioned the classical nature of general relativity contrasts with the discovery that
the world is indeterministic.
On the other hand, in quantum physics the role of spacetime as an inert stage upon which everything
takes place is in contrast with the beautiful discovery that spacetime is itself a dynamical object.

Physicists have embraced many different attitudes toward these problems. The most popular is
to think that either one or the other of the two theories has to be modified. Most commonly it
is thought that general relativity is just a low-energy effective theory for some new and more fundamental
theory. It is also believed that the new theory should realize a
unification with the other three fundamental interactions: the strong, weak and electromagnetic.
The most popular candidate for such a theory seems to be string theory.
Alternatively there are also proposals where quantum physics is derived from some new physics
at the Planck scale.

Another attitude is to be more conservative and try new ways of looking at the same theory.
The point here is that maybe it is just the perturbation theory that is wrong and not the whole idea
of quantizing general relativity as it stands. In such case a non-perturbative treatment would be necessary.
Of course going non-perturbative is like opening Pandora's box, since most of the times
in physics the best we are able to do is a perturbative expansion around one of the few cases we can solve
exactly. Therefore the main problem in this approach is that of devising new tools for the non-perturbative quantization
of general relativity.

The models I describe and investigate in the present thesis take this second point of view as their starting point,
and are known as Causal Dynamical Triangulations.
In order to get a handle on non-perturbative quantization these models start from a basic
observation, namely, that computability in general requires a finite number of degrees of freedom.
This is something which is experienced in everyday life, for example whenever we look at the computer screen
where all images are represented as a finite collection of pixels.
In the case of quantum gravity the degrees of freedom are in the configurations of geometry
at each spacetime point, and are therefore infinite. The proposal is to sample these degrees of freedom by replacing
the continuous spacetime with a coarse-grained one built with identical triangles.
Such a triangulation has to be dynamical, since otherwise it would not encode any degree of freedom.
Finally causality is imposed, in a precise sense, as a physical ingredient.

These models have been studied intensely in the last decade and have shown to lead to very interesting results.
The triangulation of spacetime does what it is meant to do, in the sense that it produces
a model which can be studied on a computer. Of course at some point one would like to make these triangulations finer
and finer in order to recover the continuum theory. The computer simulations indicate that a continuum
limit exists and that it has at least some of the good properties we expect it to have.
The drawback to this approach is that we have basically no analytical method to investigate the properties
of such models in 4-dimensional spacetime. Computer simulations are our only tool, and they are limited
by the computer power. This is certainly an unsatisfactory situation.

The status quo is radically different in two spacetime dimensions, where the model can be solved exactly in
a variety of ways and the continuum limit taken analytically. This is maybe no surprise because in two dimensions
things are a lot easier and we have many techniques for studying discrete systems. Nevertheless, exactly solvable
models are not so abundant and it is a lucky case that causal dynamical triangulations are one of those.
A possible objection to such models is that in two dimensions there is no classical general relativity
to start with, and that the quantum theory is renormalizable because Newton's constant is dimensionless.
On the positive side, the two-dimensional model can be considered as an interesting playground for the development
of tools and ideas, with the hope to generalize them at a later stage to the higher-dimensional case.

Unfortunately any attempt to extend the analytical methods developed in two dimensions to higher dimensions
has up to now failed. Not even in three spacetime dimensions are we able to solve the full model nor to compute
any observables.

The aim of this thesis, besides the non-marginal one of making the author deserve the PhD degree,
is to try to take a small step beyond this impasse.

After reviewing in Chapter 1 in more detail what has been briefly mentioned in this preface, in particular giving an
introduction to the methods and the results of causal dynamical triangulations, we will proceed in the next
two chapters to illustrate in detail two original pieces of work which both represent
a small step away from the special case of pure gravity in two dimensions.

In Chapter 2 we tackle the problem of coupling matter to causal dynamical triangulations in two dimensions.
To fix a starting point we choose one of the simplest, and certainly the most notorious, matter fields
on a lattice, the Ising model.
We show how an old method of investigation for spin systems on a lattice, the high- and low-temperature
expansion, can be applied on the dynamical lattice represented by the triangulation. The application
is not trivial as it requires an ensemble average over the lattices for the graphs appearing in the expansion,
but we will develop a scheme that allows the computation of such averages.
The method is quite general and can in principle be used for other kinds of matter models.

The work presented in Chapter 3 goes into the realm of higher dimensions.
We introduce a particular model of three-dimensional causal dynamical triangulations and show that
it can be solved in an asymptotic limit and a continuum quantum Hamiltonian is found.
This is an important step in the understanding of these kinds of models, because this is the first case
in which, for a dimension larger than two, a continuum limit has been obtained analytically, although
many things remain to be understood and some approximations still removed.

Together, the works presented in Chapter 2 and 3 constitute an advance in the analytical understanding
of causal dynamical triangulations and have the appealing feature of indicating directions for further
development.


\clearpage 
\cleardoublepage


\chapter[Introduction to Causal Dynamical Triangulations]{Introduction to Causal Dynamical\\ Triangulations}

{\small
In this Chapter I will review the Causal Dynamical Triangulations (CDT) approach to quantum gravity.
First I will briefly review the problems we face in quantizing gravity, then give a short introduction
to the non-causal predecessor of CDT.
In Sec.~\ref{2d-cdt} I introduce the causal construction for the (1+1)-dimensional case, and proceed
to review what is known about the higher-dimensional cases in the next two sections.

Several excellent reviews exist on quantum gravity in general (see for example \cite{carlip-4d},
or \cite{rovelli-history} for an historical overview) and on CDT in particular (see \cite{loll-history}
and \cite{ajl-art}).
Here I will try to review all the necessary background in the attempt to make
this thesis as self-contained as possible. In particular the first section sets the background
of facts and ideas against which the research on non-perturbative quantum gravity takes place,
while the other sections recall the technical aspects needed for the two
final chapters.
}

\section{The non-perturbative quantum gravity program}

With the term quantum gravity I will refer
to the quantum theory, to be constructed in one way or another, based on the classical Einstein-Hilbert action
\beq \label{S_EH}
  S_{EH}[g_{\m\n}] = \frac{1}{16 \pi G} \int_M d^4x \sqrt{|g|}( 2 \L - R) \ ,
\eeq
where $G$ and $\L$ are respectively the Newton's and the cosmological constants, $M$ is the spacetime manifold,
$g_{\m\n}$ the spacetime metric, $g$ its determinant and $R$ the associated Ricci scalar curvature.

One of the most powerful ways to quantize a classical field theory is the path integral formalism.
For the case of gravity we would formally write something like
\beq\label{1}
  \int \mathcal D [g] e^{iS_{EH}[g]}\ ,
\eeq
where the integration should be over  the (diffeomorphism-equivalence classes $[g]$ of) metrics on spacetime.
Such a functional integration is rather formal and one still has to make sense of it in some way.

Usually in quantum field theory we would add a source term in the action, something like the $\int d^4 x j(x)\phi(x)$
for the scalar field, and use the path integral as the generating function of the Wightman functions,
$i.e.$ the vacuum expectation values of products of the field.
Knowledge of the generating function would then mean full knowledge of our theory.
In general one can only hope to be able to compute it for the free theory, and switch to the interacting
one by studying the perturbative expansion. In writing down the perturbative expansion for a Wightman
function the path integral formalism turns out to be extremely useful.
One can think of the path integral as a book-keeping device that tells us what to compute at every order.
We would then have to actually do these computations, and usually would get divergent quantities. If we are lucky the
theory is renormalizable and these divergences can be removed order by order by a suitable redefinition of a finite
number of parameters. If the theory is not renormalizable, at every order there will be new parameters
to readjust and the predictive power would be limited.
For this reason we consider non-renormalizable theories as non-fundamental, although they can still be
of use as effective low-energy theories (as, for example, the Fermi theory of weak interaction).

In applying the idea of perturbative quantization to gravity, one has to overcome a large number of
technical difficulties. Efforts towards the perturbative quantization of gravity have led to
important techniques, ranging from the introduction of the ``ghost" fields \cite{feynman} to the development
of the background field method \cite{dewitt}, which have subsequently found a fundamental role in the quantization
of Yang-Mills theories.
The greatest difficulty is certainly that the coupling constant has the dimension of inverse-square mass, which makes
the theory non-renormalizable. By a dimensional counting argument we expect to have worse and worse
divergences at higher orders of perturbation theory, and so new counterterms will be needed at each order.
One could still hope that by some miracle the divergences could be reabsorbed,
$i.e.$ physical quantities be made finite just by field redefinitions, without having to introduce
new parameters. This is exactly what happens at one-loop, as 't Hooft and Veltman \cite{hooft-veltman}
have shown. However the miracle did not last long, since Goroff and Sagnotti \cite{goroff-sagnotti}
showed that at two loops divergences appear that require a new counterterm, which
cannot be reabsorbed in any way.
Furthermore the dream of a finite perturbative quantum gravity was already broken in the presence of matter
even at one loop \cite{hooft-veltman}.

The problem of non-renormalizable divergences has produced two main lines of thinking: one is to think of some
other theory, which has general relativity only as a special limit and try to quantize the new theory
(supergravity, strings, etc.);
the other is to think that maybe the quantum theory of gravity still exists in a non-perturbative sense
(asymptotic safety, canonical loop quantum gravity, etc.).
The CDT model fits in this last perspective.

As argued by Weinberg \cite{weinberg}, there is a well-defined scenario under which a non-perturbative
theory of gravity could still make sense.
Namely, if the renormalization group flow of a perturbatively non-renormalizable theory would
admit a non-Gaussian ultraviolet fixed point, it could still be considered a meaningful theory,
in principle with the same predictive power as a renormalizable one. Such a scenario goes under the name
of ``asymptotic safety", referring to the fact that such a theory would be safe from divergences as
the cut-off is removed.

Not many examples of such theories are known.
The most relevant for our discussion is the case of gravity in $2+\ee$ dimensions, for small $\ee$ \cite{weinberg}.
The big question is whether gravity is asymptotically safe also at $\ee=2$. There is no answer to it yet, although
some positive indications have come out in recent years (see \cite{nieder-reuter} for a review),
but definitely this idea is at the heart of all non-perturbative approaches to quantum gravity.

Another great stimulus in the direction of non-perturbative quantization came from (2+1)-dimensional gravity.
At a first look the theory seems
to have all the problems of its higher-dimensional analogue, in particular since Newton's constant has dimension
of length.
Despite that, as Witten first showed in \cite{witten}, the non-perturbative theory not only makes sense but
is actually an exactly soluble system. Of course a more careful look shows that the two theories have a fundamental
difference, with the lower-dimensional one having no local degrees of freedom, once the constraints have been enforced.
Because of this the quantum theory can be made meaningful in several ways, for example, by working in reduced
phase space or imposing the constraints after quantization (see \cite{carlip-3d}
for a review). There are still several difficulties, for
example, with higher topologies or with the path integral over metrics, but for sure it is encouraging.

A lot of effort has been spent over the years to build a meaningful non-perturbative theory of quantum gravity
in four spacetime dimensions, but unfortunately without conclusive results yet.
One reason for this situation is that already in ``easier" theories a non-perturbative study of the path
integral is an extremely difficult task.

In general the path integral has no rigorous mathematical definition.
For the non-relativistic free quantum particle, the analogue of (\ref{1}) can be made meaningful by rotating
time to imaginary time. The quantum theory then becomes a stochastic theory and the integral has a measure-theoretic
definition in terms of Wiener measure, whose properties are known. For example, its support
is given by everywhere non-differentiable paths.
The Wiener measure is usually obtained by discretizing the
time line, which turns the path integral into a finite product of ordinary integrals, and then by taking
the continuum limit.

In quantum field theory the situation is mathematically less safe. For the Wick-rotated free scalar theory
it is known that the support of the theory is on distributions.
For the interacting theory, no rigorous result is known, but
one expects troubles because the fields will be at least distributions, whose products are ill defined.
This is connected to the problem of ultraviolet divergences in the perturbative approach,
which in turn is solved by renormalization if the theory is renormalizable.
If we want to address non-perturbative questions, like the confinement in QCD, or if - like for the case of
gravity - the theory is non-renormalizable and we want to explore the asymptotic safety hypothesis,
we have to find a way to define the path integral.
What is usually done is to discretize the spacetime on a lattice in analogy with the time variable of the
non-relativistic case.
The Wick-rotated theory on the lattice is a well-defined object which we can put on a computer
to make calculations. It also has the form of a partition function for a statistical mechanical system,
making it possible to borrow methods from statistical mechanics, like the renormalization group.
The quantum field theory should be defined as the continuum limit.
This limit, where the short-distance cut-off $a$ is sent to zero, should be such as to keep the physical mass finite.
In the statistical mechanical language this means we have to go to the regime of infinite
correlation length $\xi=(m_{phys} a)^{-1}$, $i.e.$ we have to tune the theory to the critical surface.

Clearly the task of constructing a non-perturbative quantization of gravity along these lines is extremely
challenging, and even at a quick look already presents a considerable number of difficulties.
I list some of the most relevant obstacles one encounters in trying to construct a non-perturbative theory of gravity:
\begin{itemize}
\item[(i)] no notion of preferred time is present in general relativity, so it is not possible to perform a
standard Wick rotation;
\item[(ii)] we  would like to have a regularization that preserves the diffeomorphism invariance of
general relativity; this may be not a must, as also ordinary lattice field theory breaks Lorentz symmetry,
but then one should recover it in the continuum;
\item[(iii)] as in gauge theory we expect divergences from the gauge volume (the infinite volume of diffeomorphism
group here);
a good gauge fixing would be needed and the associated Faddeev-Popov determinant should be computed;
alternatively, explicit field coordinates on the physical configuration space $\frac{\rm Metrics(M)}{\rm Diff(M)}$
of diffeomorphism-equivalence classes of metrics $[g_{\mu\nu}]$ (the so-called {\it geometries}) should be
found;
\item[(iv)] in case we manage to define a Wick rotation or in case we start with the Euclidean theory from the
beginning, we still have to face the problem that the action is unbounded from below, due to the
sign of the kinetic term for the conformal mode.
\end{itemize}

Although since the pioneering work of Leutwyler \cite{leutwyler}
on a lattice path integral formulation for gravity many things have been learned and understood,
these issues still remain an open problem.

An old proposal by Hawking to deal with (i) is to start with an Euclidean theory from the beginning
\cite{hawking-euclid}, from which physical information can directly be recovered, like for example on the
ground state of the theory. The hope was also that a solution to the problem of rotating back to Minkowskian
signature would be found once the Euclidean theory is solved.

The problem (iv) was also addressed by Hawking and others in the context of formal Euclidean path integrals
with the proposal to analytically continue the conformal factor to imaginary value \cite{hawking-conform}.
A more promising and less ad-hoc solution seems instead the idea that the cure would be naturally given
by a non-perturbative treatment of the path integral.
In this case, non-trivial measure contributions arising from Faddeev-Popov determinants can in principle
cancel the conformal divergence, as indicated by some formal calculations \cite{mazur-mottola,dasgupta-loll},
inspired by the generally covariant method used by Polyakov for the bosonic string \cite{polyakov}.

Several discretized models have been proposed over time, see \cite{loll-discrete} for an overview.
One of the most studied has been probably the quantum Regge calculus, because it is close to the
spirit of general relativity, coordinate independent, and can be used classically to approximate any geometry.
A path integral based on it presents several problems.
In order to avoid some of these the Dynamical Triangulation model was proposed in the '80s,
and eventually led to the CDT model at the end of the '90s.

The CDT model is a very concrete and promising way to face the above-mentioned problems, with a coordinate-independent
discrete path integral over geometries admitting a Wick rotation.
I will introduce it in this chapter, after having recalled the necessary ingredients borrowed from
its predecessors.

\section{The Dynamical Triangulations approach}

It is well known that Feynman's path
integral for quantum mechanics can be defined as a limit
for $N$ going to infinity of $N$ ordinary integrals for the position at time $t_k=k \frac{t}{N}$ (for $k=1,...,N$)
of a particle during its path from an initial to a final point in a total time $t$. At finite $N$ only a finite
number of points, those at time $t_k$, in the path are specified and the path at intermediate times
can be thought of as straight. This is then called a piecewise linear path. The limit as $N\to\infty$
is the continuum limit.

\begin{figure}[ht]
\centering
\vspace*{13pt}
\includegraphics[width=7cm]{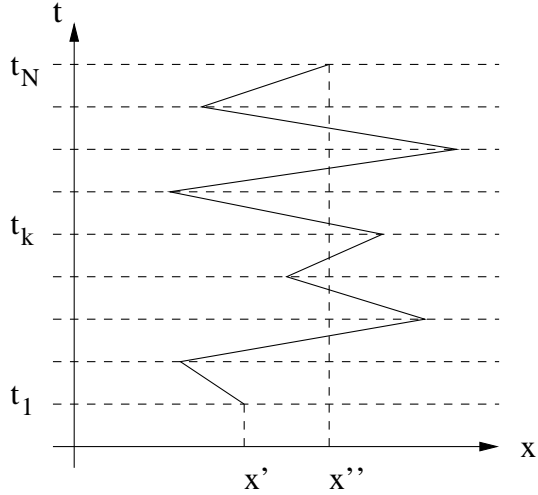} 
\vspace*{13pt}
\caption{\footnotesize A piecewise linear path appearing in the definition of the Feynman path integral.}
\label{path}
\end{figure}

The idea of dynamical triangulations for the quantization of gravity is simply to generalize
this procedure to higher dimensions.

The analogue for spacetime geometries of the piecewise linear paths of the Feynman path integral is given
by {\it piecewise linear manifolds}, which I am going to recall in the following subsection.

\subsection{Preliminaries: simplicial manifolds and Regge calculus}

A piecewise linear $n$-dimensional manifold is a collection of $n$-dimensional (flat)
polytopes (higher-dimensional analogue of polygons ($n=2$) and polyhedra ($n=3$)) glued together along
their $(n-1)$-dimensional faces in such a way as to preserve the topological dimension.
In general it will not be possible to embed the object obtained in this way isometrically in $\mathbb{R}^n$,
meaning that there will be curvature defects.

To be more specific and to simplify issues, piecewise linear manifolds are generally taken to be
simplicial manifolds, $i.e.$ the constituent polytopes are chosen to be simplices\footnote{
One advantage of simplices is that their geometry is completely determined by the specification of their
edge lengths.
Appealing to ``universality", this specific choice of building block should not affect the continuum results.
There is very good evidence of universality in two-dimensional DT models, but the property needs
to be verified for individual models.}.

A {\it simplex} is the higher-dimensional generalization of a triangle, a polytope with the minimal number
of faces in any given dimension. For example, in three dimensions it is a tetrahedron.
In general, a $n$-dimensional simplex $\s_n$ can be defined by its embedding in flat space as
the convex hull of $n+1$ affinely independent points (meaning that no more than $m+1$ of these points
are in the same $m$-dimensional plane) of $\mathbb{R}^d$, for $d\geq n$.
Denoting these points by $v_1,...,v_{n+1}$, the simplex is given by
\beq
\s_n = \left\{ \vec x \in  \mathbb{R}^d,\ d\geq n\ |\ \vec x=\sum_{i=1}^{n+1}\l_i v_i\ ;\ \l_i\geq 0,
 \ \sum_{i=1}^{n+1}\l_i=1\right\}.
\eeq

A {\it subsimplex} of $\s_n$ is a simplex $\s_k$ (for $k<n$) whose vertices are a subset of the {\it vertices}
$v_i$ of $\s_n$. Subsimplices of dimension $n-1$ are called {\it faces}, while those of dimension $n-2$
are called {\it bones} or {\it hinges}. Regardless of $n$ we will call $\s_1$ also {\it edges} or {\it links}.

A collection of simplices glued along their subsimplices is called a {\it simplicial complex}.
Finally, a {\it simplicial manifold} is a simplicial complex in which the neighbourhood of any vertex, $i.e.$
the set of simplices sharing that same vertex, is homeomorphic to $B_n$, an $n$-dimensional ball in $\mathbb{R}^n$.
We will construct simplicial manifolds by gluing together no more than two simplices along the same face
(faces belonging to only one simplex will belong to the boundary of the simplicial manifold).
Even with this restriction, gluing together simplices at random can give rise to singular
points, with a neighbourhood not homeomorphic to $B_n$. For example, in three dimensions such a neighbourhood
may have a higher-genus surface as its boundary. We will not be interested in this issue in what follows,
since our models by construction are given in terms of simplicial manifolds.

It is useful to introduce the so-called\footnote{The name derives from the fact that in mathematics its elements
are usually denoted $f_i$, rather than $N_i$ as we are used to in physics.} {\it f-vector} of the simplicial
manifold $M$,
\beq
f(M)=\{ N_0, N_1, ...,N_n\}\ ,
\eeq
where $N_k$ is the number of $k$-dimensional (sub-)simplices in $M$.
If the manifold has a boundary $\dpr M$ we also introduce the $f$-vector of the boundary,
\beq
f(\dpr M)=\{ N_0^{\dpr}, N_1^{\dpr}, ...,N_{n-1}^{\dpr}\}\ .
\eeq
The elements of the $f$-vectors are not all independent, but have to satisfy topological relations.
The most famous one is Euler's relation
\beq
\sum_{k=0}^n (-1)^k N_k = \chi (M)\ ,
\eeq
where $\chi(M)$ is the Euler characteristic of the manifold, which for $n=2$ is $\chi(M)=2-2g-b$, with
$g$ being the number of handles (the genus) and $b$ being the number of boundaries, and for $n$ odd is given by
 $\chi(M)=\chi(\dpr M)/2$.

Besides Euler's relation also the Dehn-Sommerville relations must be satisfied (see \cite{dehn-somm} for details),
namely,
\beq
N_k^{\dpr}=(1-(-1)^{n-k}) N_k + \sum_{j=k+1}^n (-1)^{n-j-1} \matrice j+1 \\ k+1 \ematrice N_j\ .
\eeq
Because Euler's relation is independent of the Dehn-Sommerville relations only in even dimension,
we find that among $f(M)$ and $f(\dpr M)$ only $n$ element are free if $n$ is even, and $n+1$ if $n$ is odd
(respectively $n/2$ and $(n+1)/2$ if there is no boundary, since then only $[(n+1)/2]$ of the Dehn-Sommerville
relations are linearly independent).

We can fix the metric properties of a simplicial manifold by fixing its edge lengths, and remembering that
the inside of the simplices is flat, it is trivial to see how this fixes distances between points.
Furthermore, the notion of curvature can be implemented too, as first shown by Regge in \cite{regge},
and later made more rigorous in \cite{cheeger}.

\begin{figure}[ht]
\centering
\vspace*{13pt}
\includegraphics[width=11cm]{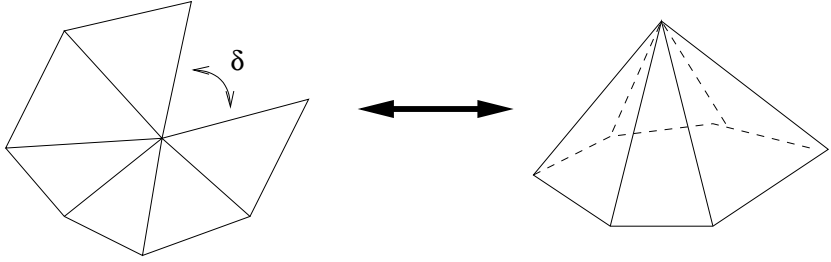}
\vspace*{13pt}
\caption{\footnotesize An example of a positive deficit angle.}
\label{deficit-angle}
\end{figure}

The idea is to replace the continuum notion of curvature in the following way:
\beq \label{regge-curv}
\int_M d^n x \sqrt{g}\ R \rightarrow 2 \sum_{\text{hinges }h} V_h \d_h\ ,
\eeq
where $\d_h$ is the deficit angle at $h$,
\beq
\d_h=2\pi-\sum_{\s_n\supset h}\th_{\s_n\rhd h}
\eeq
(the sum is over all $n$-dimensional simplices $\s_n$ sharing the hinge $h$, and $\th_{\s_n\rhd h}$ is the angle
between the two faces of $\s_n$ sharing $h$). Fig. \ref{deficit-angle} illustrates the two-dimensional case.

The formula (\ref{regge-curv}) can be understood by recalling that intrinsic curvature (as encoded in the Riemann
tensor) is associated with the variation a vector undergoes when parallel transported around an infinitesimal
closed loop spanned by two vectors.
One can then easily realize that the curvature is zero inside a simplex (since we have taken it to be flat
from the beginning) and that it is also zero at all points lying inside the faces (as Fig. \ref{deficit-angle}
illustrates at the faces there is only extrinsic curvature).

The curvature is therefore concentrated at the hinges, and one may think of (\ref{regge-curv}) as saying
that the Ricci scalar curvature is given by a linear combination of delta functions with support
on the hinges.
This fact should not constitute a worry, it is reminiscent of the distributional character
of fields in quantum field theory.

As the title in \cite{regge} suggests the original motivation behind the introduction of simplicial
manifolds in physics was to have a coordinate-independent approximation scheme for classical General Relativity.
With the notion of curvature at hand, one can introduce the discrete analogue of the Einstein-Hilbert
action (\ref{S_EH}), called the Regge action\footnote{In case the manifold has a boundary, boundary terms
must be added in the action. They are essential in the quantum theory in order that the transition amplitudes
satisfy the correct composition law and have the correct classical limit, as was pointed out in \cite{gibbons-hawk}.
The full continuum action is fixed by the requirements of additivity ($i.e.$ given two contiguous regions with
metrics $g$ and $g'$ the action for the union of the two regions must be $S[g+g']=S[g]+S[g']$) and
of equivalence of the classical equation of motions with the stationarity of the action under variation of the
metric at fixed boundary geometry. It is the standard Einstein-Hilbert action plus the so-called Gibbons-Hawking
term for the boundary,
\beq \label{boundary-action}
S[g]=\frac{1}{16\pi G}\int_M d^n x\sqrt{g}\left(2\L-R\right)-\frac{1}{8\pi G}\int_{\partial M} d^{n-1}x\sqrt{h} K\ ,
\eeq
where $h$ is the determinant of the induced metric on the boundary, and $K$ is the associated extrinsic curvature.
For the case of a simplicial manifold, the corresponding discrete action was given in \cite{hartle-sorkin}.
It amounts to adding the following term to the Regge action:
\beq
S_{\partial M}=\frac{1}{8\pi G} \sum_{\text{hinges }h} V_h \psi_h\ ,
\eeq
where $\psi_h=\pi -\sum_{\s_n\supset h}\th_{\s_n\rhd h}$.
},
\beq
  S_{R}=\frac{1}{8\pi G} \sum_{\text{hinges }h} V_h \delta_h - \frac{\L}{8 \pi G} \sum_{\s_n} V_{\s_n}\ ,
\eeq
vary it with respect to the edge lengths, and obtain a simplicial analogue of the Einstein equations
which is explicitly independent of coordinates.

Although this discrete formulation of gravity has had a number of classical applications,
it received a boost of interest in the eighties with the proposal of using it to define a path
integral quantization of gravity \cite{rocek-williams,hamber-williams}.

The prescription for defining the path integral (\ref{1}) from Regge calculus is to take for $M$ a piecewise
linear manifold with fixed connectivity, and to integrate over the edge lengths, with the exponential of the
Regge action as weight.

Quantum Regge calculus has usually been constructed and studied for Euclidean signature, even if a Lorentzian
version has also been attempted \cite{williams-lorentzian}.
It was shown to reproduce the standard continuum theory in the weak field limit but understanding its
non-perturbative properties remains an open problem.

Many choices are possible for the measure, a common choice being of the type
\beq
\mathcal D [g] \to \prod_{\text{edges} \ e} \frac{dl_e^2}{l_e^{\b}}\ ,
\eeq
and there is no agreement on the correct one.
Furthermore it has also been strongly argued that a gauge-fixing and related Faddeev-Popov determinant
are needed \cite{menotti-peirano}, but in this case (at least in two dimensions) the measure becomes non-local,
making even computer simulations difficult.

A weakness of quantum Regge calculus is that no analytic treatment is possible, not even in
two dimensions.

\subsection{Dynamical Triangulations}

One of the potential problems of quantum Regge calculus
is the use of a fixed simplicial manifold. This may prevent the path integration from
exploring entire regions of the space of metrics where the metric is very different from those realizable
for the chosen connectivity of the triangulated manifold.
On the other hand, other metrics are overcounted and would need a gauge fixing (think for example
of the flat two-dimensional triangulation  and how many ways we have
to (over-)count it by varying the edge lengths in such a way that the vertices
remain in the flat plane).

A possible solution to the first problem is to include also a sum over simplicial manifolds.
The idea of having variable connectivity {\it and} variable edge lengths was revived recently under the name
of random Regge calculus \cite{carf-dapp-marz}, but apart from not solving the other problems of Regge calculus,
it is extremely difficult to handle.

To simultaneously solve the various problems and introduce a simplification, it was proposed in the mid-eighties
to fix completely the edge lengths to be all equal and only keep the sum over simplicial manifolds
\cite{DT}.
This is the approach we call Dynamical Triangulations (DT), and it is thoroughly reviewed in the book
\cite{DT-book}.

Contrary to Regge calculus, this formalism is not well suited for classical approximations, but
is specifically designed as a functional integral over geometries. It is expected that the ensemble
of dynamical triangulations is more evenly distributed on the space of all geometries than
that of the quantum Regge calculus on a given simplicial manifold \cite{rom-zahr}.

Another advantage of dynamical triangulations is that it introduces a natural UV cut-off $a$,
the edge length common to all simplices.
This provides a regularization of the path integral and allows for a clear definition of the continuum limit
procedure, which I will explain after having given some more details.

\blankline
In this approach the integral over geometries is turned into a discrete sum over simplicial manifolds,
\beq
  \int \mathcal D [g] \rightarrow \sum_{\TT}\ .
\eeq

Furthermore, due to the restriction to identical building blocks, the action simplifies a lot,
basically reducing to a counting of global variables. The volumes and internal angles of the simplices
can only assume certain discrete values, and it is an easy exercise to check that the Regge action
for Euclidean gravity on a closed manifold becomes
\beq
  S_R=\k_n N_n - \k_{n-2} N_{n-2}\ ,
\eeq
where we have introduced the dimensionless couplings
\beq
\begin{split}
 & \k_n= \L_0 V_{\s_n} + \frac{1}{2} n (n+1) \frac{\arccos\frac{1}{n}}{16\pi G_0} V_{\s_{n-2}}\\
 & \k_{n-2}=\frac{V_{\s_{n-2}}}{8 G_0}\ ,
\end{split}
\eeq
where $\L_0$ and $G_0$ are the {\it bare} cosmological and gravitational constants,
and where we have used the fact that each simplex has $n(n+1)/2$ internal dihedral angles equal to
$\arccos\frac{1}{n}$.
The path integral (\ref{1}) is reduced to a combinatorial object, which takes the form of a generating function
for the number of simplicial manifolds with given numbers of simplices and hinges:
\beq \label{Z-DT}
Z(\k_{n-2},\k_n)=\sum_{\TT}\frac{1}{C(\TT)}e^{-S_R(\TT)}=\sum_{\TT}\frac{1}{C(\TT)}e^{-\k_n N_n + \k_{n-2} N_{n-2}}\ ,
\eeq
where we have also taken into account $C(\TT)$, the order of the automorphism group of the triangulation $\TT$,
to avoid overcounting of manifolds with special symmetries\footnote{This factor can be
understood in the following way. If we would label all the subsimplices of the triangulation
and do the sum over labelled triangulations, we would have to factor out the number of ways
a triangulation can be labelled. In general this number is $N_0!N_1!...N_n!$, but for triangulations
with some symmetry it is less, precisely $N_0!N_1!...N_n!/C(\TT)$. The factor $1/C(\TT)$ thus survives in the
case where we sum over unlabelled triangulations, and therefore can be thought of as what remains
when factoring out the volume of the diffeomorphism group.}.

Due to the existing relations among the $N_i$, the action is now bounded above and below
for fixed volume $N_n$ (to put a rough bound, note that since there are $n(n+1)/2$  hinges per simplex, it follows
that $N_{n-2}\leq \frac{n(n+1)}{2} N_n$). If we rewrite (\ref{Z-DT}) as
\beq \label{Z-DT-2}
Z(\k_{n-2},\k_n)=\sum_{N_n}e^{-\k_n N_n}\sum_{\TT_{|N_n}}\frac{1}{C(\TT)}e^{\k_{n-2} N_{n-2}}
 \equiv\sum_{N_n}e^{-\k_n N_n} Z(\k_{n-2},N_n)\ ,
\eeq
we see that $Z(\k_{n-2},N_n)$ is only a finite sum with respect to $N_{n-2}$, and so is well defined.
In order to check whether $Z(\k_{n-2},\k_n)$ is well defined too, we need to check the convergence with
respect to the sum over $N_n$.
Clearly (\ref{Z-DT-2}) will converge for some values of the coupling constants if and only if $Z(\k_{n-2},N_n)$
grows at most exponentially, like $Z(\k_{n-2},N_n)\sim e^{c N_n}$, for come constant $c$.
A necessary condition for this to be true is that the topology must be fixed, otherwise $Z(\k_{n-2},\k_n)$
would grow factorially\footnote{In two dimensions, a sum over topologies can still be defined after having
first performed the sum for any fixed genus, in a kind of resummation of the series, but only in a double-scaling
limit where the weight of higher-genus sums goes to zero like a specific power of the edge length $a$.
Since this procedure is neither unique nor seems to be generalizable to higher dimensions (where topologies
are not even classified), topology is usually fixed in models of dynamical triangulations.
We will also stick to a fixed topology in this thesis.
}. That this condition is also sufficient is proven only for $n=2$, but
numerical results support it also in higher dimension \cite{exp-bound}. The existence of an exponential bound is
now generally believed to be true.

At this point it is worth noting once more that the regularized Euclidean version of the path integral,
defined in (\ref{Z-DT}),
is a partition function for a model of statistical mechanics. I will from now on often refer to
it and its model-dependent variations as {\it partition functions}.
The relation between $Z(\k_{n-2},N_n)$ and $Z(\k_{n-2},\k_n)$ is seen to be that between a canonical
and a grand canonical formulation. Because in general relativity the volume of spacetime is not fixed we
are usually interested in the grand canonical partition function.

Going back to the issue of convergence, we will now see that it is closely related to the way in which we can
formulate the continuum limit.
For fixed topology, if we assume for the canonical partition function a large-volume behaviour of the form
\beq
Z(\k_{n-2},N_n)\sim N_n^{\g(\k_{n-2})-3}e^{\k_n^c(\k_{n-2}) N_n}\times (1+O(1/N_n))\ ,
\eeq
the partition function is convergent for $\k_n>\k_n^c(\k_{n-2})$, and near the critical line $\k_n^c(\k_{n-2})$
we get\footnote{This expression also defines the {\it susceptibility exponent} $\g(\k_{n-2})$.
The name comes from the fact that it is the exponent of the non-analytic part of the second derivative
of the partition function, which is usually called susceptibility.}
\beq \label{Z-sing}
Z(\k_{n-2},\k_n)\sim (\k_n-\k_n^c(\k_{n-2}))^{2-\g(\k_{n-2})}\ .
\eeq
Since $\k_n$ is conjugate to the number of $n$-simplices, we can extract the average lattice volume in the ensemble
of triangulations from the partition function as
\beq
\langle N_n \rangle = -\frac{\partial \ln Z(\k_{n-2},\k_n)}{\partial \k_n}\ .
\eeq
From (\ref{Z-sing}) it is then clear that the average number of $n$-simplices diverges at the critical line like
\beq
\langle N_n \rangle \sim \frac{\g(\k_{n-2})-2}{\k_n-\k_n^c(\k_{n-2})}\ ,
\eeq
and it is exactly there that we can send the cut-off $a$ to zero, while keeping the physical volume
of the universe finite. More precisely, we have to approach the critical line according to a proper scaling, like
\beq
\k_n\sim \k_n^c(\k_{n-2})+\L a^n\ ,
\eeq
so that
\beq
\langle V \rangle = a^n \langle N_n \rangle \sim \frac{\g(\k_{n-2})-2}{\L}\ ,
\eeq
where $\L$ is the renormalized cosmological constant, conjugate to the volume in the continuum.

Tuning $\k_n$ to obtain a finite continuum volume may not be enough to arrive at an interesting continuum limit.
We may still need to tune also $\k_{n-2}$ to some specific value, for example, in order to get
a divergent (lattice) correlation length between local variables\footnote{A classical example in this respect
is the Ising model, where the limit of infinite lattice volume and zero lattice spacing is not sufficient
to obtain a continuum field theory with local propagating degrees of freedom. This is only achieved
by also fine-tuning the temperature to a critical value corresponding to a second-order phase transition.}.
In two dimensions, taking the limit of infinite lattice volume turns out to be sufficient, leading automatically to
a continuum theory of two-dimensional quantum gravity. Of course $n=2$ is a special case since the
gravitational constant couples to the Euler characteristic and thus plays no role when the topology is fixed.
The situation in higher dimension is not completely clear, numerical simulations for $n=4$ indicate that
different phases are separated by first-order transitions \cite{bialas,catterall} so that either a non-trivial
continuum limit (one with propagating degrees of freedom) is automatically reached with the infinite-volume limit
or otherwise there is no chance to get one.

In the infinite-volume limit one finds for $n=4$ two phases, a crumpled one (at $\k_2<\k_2^c$) with a very large
{\it Hausdorff dimension}\footnote{The Hausdorff dimension is defined as the leading power in $R$ with which the
volume of a ball $B(R)$ of radius $R$ scales with the radius. More precisely, $B(R)$ is defined as the set of points $x$ at
geodesic distance $d(x,x_0)\leq R$ from a given point $x_0$; taking first the manifold average of the volume of
$B(R)$ with respect to the center $x_0$, and then the ensemble average over the triangulations we
define the Hausdorff dimension $d_H$ as
\beq
\langle V(R)\rangle \sim R^{d_H}
\eeq
in the limit of large $R$. If $\langle V(R)\rangle$ grows faster than any power of $R$, we set $d_H=\infty$.}
 $d_H$ ($\approx\infty$) and an elongated phase (at $\k_2>\k_2^c$) with the characteristics of a branched polymer
and $d_H\approx 2$. Since as just mentioned the transition between the two is only first-order nowhere
a 4-dimensional spacetime can be found, and the model does not seem to have good chances of describing our world.

\begin{figure}[ht]
\centering
\vspace*{13pt}
\includegraphics[width=8cm]{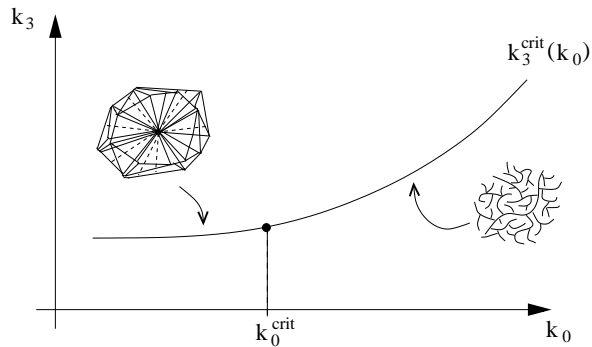} 
\vspace*{13pt}
\caption{\footnotesize The phase diagram of 3- and 4-dimensional Euclidean dynamical triangulations.
The region of convergence, from which the critical line is approached, is above the critical line.}
\label{phaseeu}
\end{figure}

\subsection{Dual complex and matrix models}

Before moving to the causal version of dynamical triangulations, which has proven to be better behaved
and which is the main topic of this thesis, it is useful to recall some of the techniques used in the
investigations on dynamical triangulations.

Something that we will use very often in this thesis is the {\it dual} complex of a triangulation,
which encodes all the combinatorial information we need.
The dual complex is associated with a $n$-dimensional simplicial manifold via a dual mapping,
in the sense that the dual complex of the dual complex is again the original simplicial manifold.
The dual mapping associates to any $k$-dimensional subsimplex $\s_k$ a $(n-k)$-dimensional dual polytope
$\s_{n-k}^*$ whose sub-polytopes are dual to higher-dimensional subsimplices.
In particular, the dual to a $n$-simplex is a vertex, the dual to a face $\s_{n-1}$ is an edge,
and the dual to a hinge $\s_{n-2}$ is a polygon.

\begin{figure}[ht]
\centering
\vspace*{13pt}
\includegraphics[width=7cm]{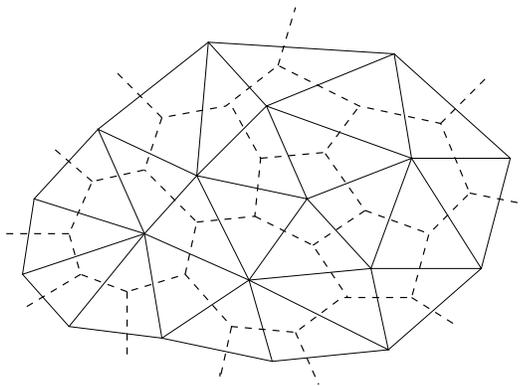} 
\vspace*{13pt}
\caption{\footnotesize A portion of a triangulation and the corresponding dual graph.}
\label{dual}
\end{figure}

If in a triangulation all the building blocks are of the same kind, but its vertices have all possible
coordination numbers, the situation in the dual complex is reversed, namely, all kind of polytopes
appear but all the vertices have coordination $n+1$.
This turns out to be a useful property in many situations.
In two dimensions, it allows us to regard the dual of a triangulation as the Feynman diagram of
a $\phi^3$-theory. This is precisely the link between two-dimensional quantum gravity and matrix
models, on which a lot of literature exists (see, for example, \cite{2dQG-review}).

Matrix models are a powerful method, which enables us to solve exactly the two-dimensional
DT model for the pure gravity case \cite{brezin}, and for gravity coupled to matter in the form of the Ising
model \cite{kazakov}, Potts model \cite{kazakov-potts} or hard particles \cite{difra-hardobj2}.
The partition function for the DT model is identified with the generating functional for the connected
diagrams of the corresponding matrix model (in the large-$N$ limit),
\beq
Z(\l,...)=\ln \int d^{N^2}A\ d^{N^2}B ... e^{-N \Tr V(A,B,...)}\ ,
\eeq
where for example in the pure case the potential is given by
\beq
V(A)=\frac{1}{2}A^2-\frac{g}{3}A^3\ ,
\eeq
with $g=e^{-\l}$.

A generalization of matrix models for higher-dimensional DT has been constructed, but it has not led
to any advance because of major problems like the fact that there is no equivalent of the large-$N$ limit,
it is not possible to restrict to manifolds, and there is no reduction to eigenvalues.

In two dimensions, other methods are possible but it is very difficult to generalize any of them to higher
dimensions, where the most valuable method remains that of Monte Carlo simulations.

\section{Enforcing Causality: the (1+1)-dimensional case} \label{2d-cdt}

\subsection{The model}

As I have argued in the previous section, the naive idea of gluing together simplicial building blocks
without any restriction in general does not work. To begin with one may not get a manifold at all.
So we have to\footnote{
The necessity of this point may be argued. In the Group Field Theory approach to quantum gravity
the appearance of conical singularities does not seem to be considered a major problem \cite{oriti}.}
restrict to gluings that have the properties of a simplicial manifold.
Also this is not enough, and if we want to meaningfully define a partition function we have
to fix the topology.
Even if this makes the model well defined, the statistical ensemble is dominated by very pathological geometries
with an effective dimension which is too large or too small.
One possible reason for this is that the class of geometries being summed over is still
too broad.

\begin{figure}[h]
\begin{center}
$\begin{array}{c@{\hspace{.4cm}}c}
\multicolumn{1}{l}{} &
    \multicolumn{1}{l}{} \\ [-0.53cm]
\includegraphics[width=6cm]{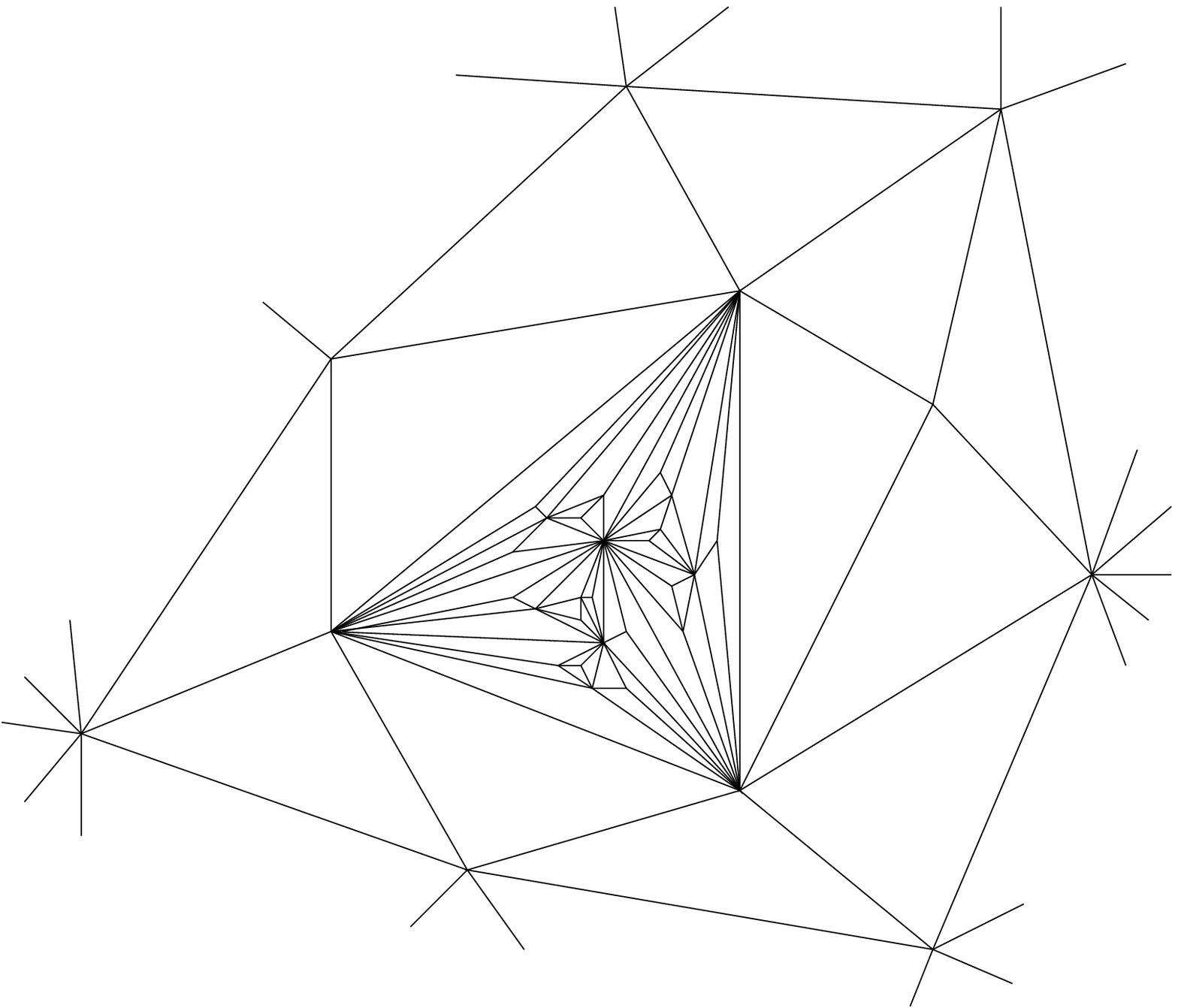} &
\includegraphics[width=8cm]{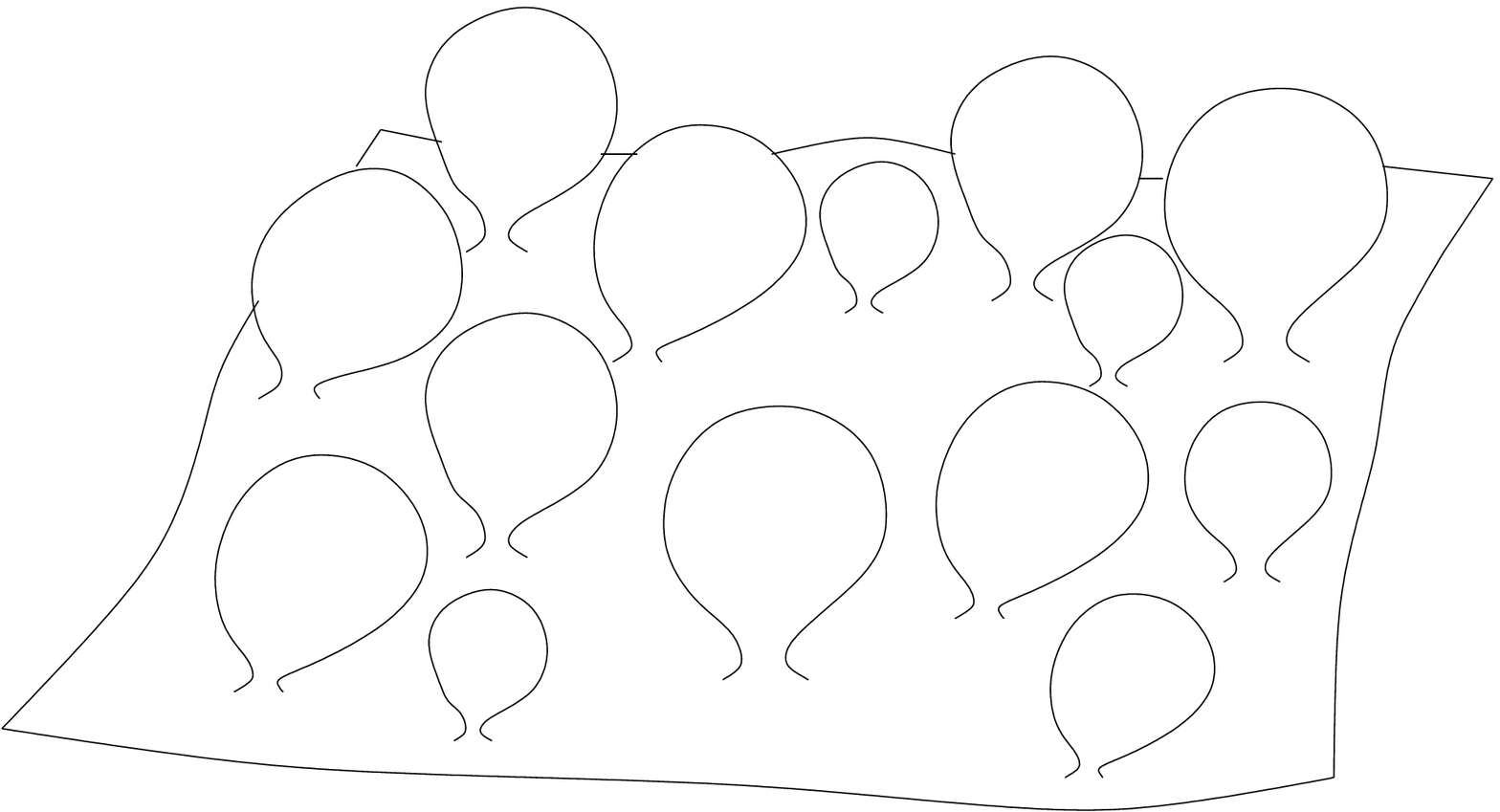} \\ [0.4cm]
\mbox{\bf (a)} &  \mbox{\bf (b)}
\end{array}$
\end{center}
\caption{\footnotesize \mbox{\bf (a):} A portion of a triangulation with a baby universe
appearing inside a loop made of three links. It should be understood that, according to the DT construction,
all the triangles should be of equal size. When trying to draw a baby universe on a plane like here we are forced
to draw smaller and smaller triangles, generating some kind of fractal picture. \mbox{\bf (b):} A pictorial
representation of baby universes attached everywhere on the mother universe.}
\label{babies}
\end{figure}

A clear example of this point is the $n=2$ DT model. There the effective dimension $d_H=4$ can be explained
in terms of baby universes. A {\it baby universe} is a triangulated manifold with a loop of minimal length
(of the order of three edges) as its boundary, along which it is glued to a {\it mother} universe (see
Fig.~\ref{babies}a).
When gluing triangles without any restriction apart from the fixed topology, it turns out that these baby universes
dominate in the partition function. In the continuum limit we roughly speaking have a small two-dimensional universe
attached to each point of the two-dimensional mother universe, giving the effective dimension 4.
An artist's impression of this situation is given in Fig.~\ref{babies}b.

These baby universes cannot be removed, for example, by imposing that no more than one triangle can be contained
inside a three-edge loop of the mother universe, because then baby universes will simply appear on four-edge loops
and so on. It turns out that they can instead be removed, at least in two dimensions, by introducing a new principle
in the construction of the geometries included in the sum.

In \cite{2d-ldt} Ambj\o rn and Loll introduced such a principle and thereby defined the
model of Causal Dynamical Triangulations (CDT).
The idea is to restrict the path integral to geometries with a well-defined causal structure.

Although at the classical level causality is considered a necessary condition for a theory to be physical
(for example, a spacetime with pathologies in the causal structure, like the G\"odel spacetime, is usually
considered unphysical), this condition has often been put aside in the construction of a quantum theory
of gravity, especially in its Euclidean formulation. Nevertheless the importance of causality
has been stressed by many, in particular by Teitelboim \cite{teitelboim} in the path integral approach,
and has led to the formulation of new approaches like that of Causal Sets \cite{causalset}.

The first step in implementing causality into dynamical triangulations is to work with Lorentzian spacetime
geometries from the outset, because otherwise it would make no sense to talk of causality.
To do that in simplicial geometry amounts to using flat simplices with Lorentzian signature.
I illustrate this first in two dimensions, postponing the cases $n=3,4$ to the following sections.
A triangle with Lorentzian signature is roughly speaking a triangle cut out of (1+1)-dimensional Minkowski space.
Regge calculus can be defined on such pseudo-Riemannian structures, as has been known for a long time \cite{sorkin}.
In a dynamical triangulations approach we want to take these triangles to be all equal to each other. A convenient
choice is therefore to have two edges of time-like type and a space-like one, say, of squared length
$l_t^2=-\a\hspace{.05cm} a^2$ (with $\a>0$ introduced for convenience) and $l_s^2=a^2$ respectively, and where
gluing is only allowed between edges of the same type.
However, this signature implementation is not enough to implement causality,
since, for example, time-like lines forming closed loops are still present.
Neither is it in general enough to change the properties of the model with respect to the Euclidean formulation.
An example of such a model can be formulated \cite{beirl-johnst} as a matrix
model with a matrix $A$ associated to space-like edges and a matrix $B$ associated to time-like edges, and a potential
\beq
U=\Tr \left[\frac{1}{2}A^2+\frac{1}{2}B^2-g A B^2  \right]\ .
\eeq
The integration over $A$ can be performed explicitly, because it is just a Gaussian and we obtain
\beq \label{2d-squares}
U=\Tr \left[\frac{1}{2}B^2-\frac{g^2}{2} B^4  \right]\ ,
\eeq
which describes a DT model with squares instead of triangles (the squares are given by pairs of triangles
glued along the space-like edge), and which is known to lie in the same universality class as the original Euclidean
one with triangles \cite{brezin}.

\begin{figure}[ht]
\centering
\vspace*{13pt}
\includegraphics[width=11cm]{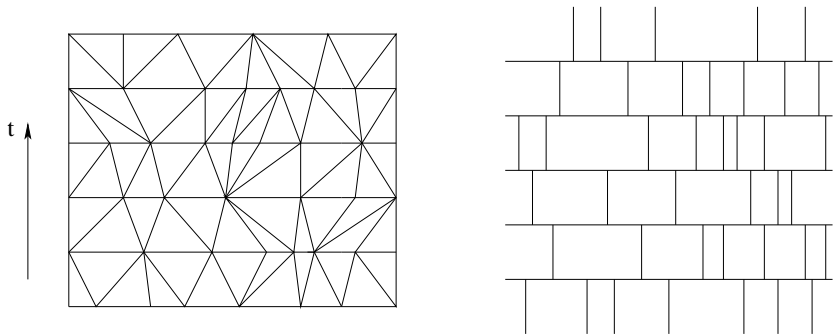}
\vspace*{13pt}
\caption{\footnotesize Example of a triangulated piece of spacetime in the CDT model
(left), and the corresponding dual graph (right).
The picture captures the way triangles are glued together, but does not represent
faithfully intrinsic distances, and therefore the curvature properties of the simplicial
geometry. If we were to respect this property, we would not be able to draw the picture
in a plane, because the model's triangles are all equal.}
\label{cdt}
\end{figure}

To enforce causality and to select geometries in a different way, the following construction, for the two-dimensional
case, was introduced in \cite{2d-ldt}.
A triangulation is made of $t$ ``strips", a strip being a triangulation of topology $\Si\times [0,1]$ (with
$\Si$ chosen to be either $S^1$ or $[0,1]$ once and for all) where the space-like edges belong either
to the initial boundary $\Si\times \{ 0\}$ or to the final boundary  $\Si\times \{ 1\}$. Different strips are
glued along these boundaries (with the constraint that the lengths of the glued boundaries must match),
and the topology of the full triangulation is again $\Si\times [0,1]$ (or $\Si\times S^1$ if
we impose periodicity in time).
An example of such a construction is shown in Fig.~\ref{cdt}, together with the dual graph, which is obtained
with the same rules as explained in the previous section.

It turns out that this new principle for the selection of geometries is relevant for the critical properties
of the model. Thanks to it the baby universes are eliminated, and the effective dimension of the geometries
is $d_H=2$ \cite{2d-ldt}.
As I will review in the following section, also in higher dimensions it is found that $d_H=n$ for n-dimensional
CDT ($n=3,4$).

\begin{figure}[ht]
\centering
\vspace*{13pt}
\includegraphics[width=10cm]{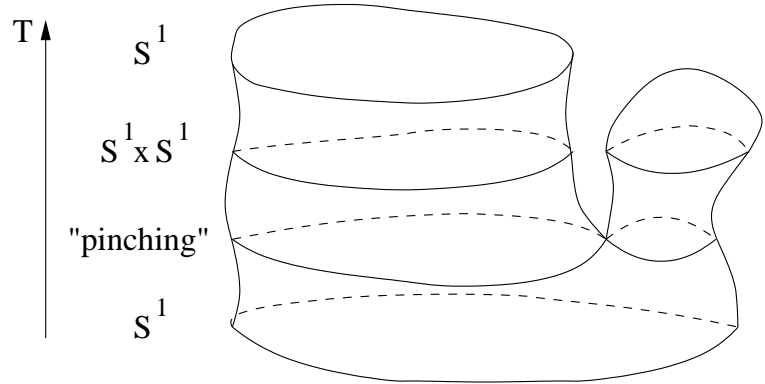}
\vspace*{13pt}
\caption{\footnotesize The branching of a universe and the appearance of a baby universe as seen from
a time-slicing point of view.}
\label{branching}
\end{figure}

Two features are extremely important in this construction: there is a global time $t$, with respect to which there
is a foliation into space plus time, and there is no topology change of the slices with time.
It is thanks to these features that the model belongs to a different universality class than the Euclidean one.
As was shown in \cite{2d-ldt},
if one lifts these constraints by allowing the slices to branch from $S^1$ into several $S^1$'s and forcing
all of them but one to eventually contract to a point and disappear without rejoining with any of the others
(in this way no handles are formed and the topology of the total manifold is unchanged, see Fig.~\ref{branching}),
these branchings dominate and the model ends up again in the same universality class of the Euclidean DT.

Having formulated the model with Lorentzian signature, one may wonder whether we end up having problems
in defining the path integral, because of the oscillatory behaviour of the integrand.
It turns out that instead in our discrete framework we can easily go back and forth from Lorentzian
to Euclidean by simultaneously changing the sign of all squared time-like lengths.
In order to do that, we simply have to change the sign of the parameter $\a$, which amounts to an analytical
continuation of it in the complex plane.
A detailed analysis of the dependence of the geometric variables, and thus of the action, on
the parameter $\a$ goes beyond the scope of this thesis, where we are mainly concerned with applying analytical
methods after having made the continuation to Euclidean signature.
It suffices to say that under the analytical continuation to negative $\a$,
the weights appearing in the sum over geometries transform according to
\beq
e^{i S_R^{Lor}}\to e^{-S_R^{Euclid}}\ .
\eeq
For details see \cite{ajl-def}.

It should be emphasized that this ``Wick rotation" is well defined because we are in a discrete setting and because
we have introduced a distinction between time- and space-like distances. The issue of how to perform the inverse
Wick rotation comes back once the continuum limit is performed. In the two-dimensional case, it is possible to
perform a standard analytic continuation in time $T\to -iT$ and obtain a unitary theory \cite{ajl-nato}.

In the rest of this thesis we will only use Euclidean signature.

\subsection{The solution} \label{2d-solution}

In two spacetime dimensions, the total curvature is a topological invariant proportional to the Euler characteristic,
and thus we can omit it from the path integral when working at fixed topology.
The partition function then takes the simple form
\beq \label{pure}
Z_t(\l)=\sum_{N_2} e^{-\lambda N_2} \sum_{\TT_{N_2,t}} \frac{1}{C_{\TT_{N_2}}}\ ,
\eeq
where now the triangulations $\TT_{N_2,t}$ (with fixed number of triangles $N_2$ and of given extension
$t$ in time) have to be constructed according to the causal rules just explained.

The model was first solved in \cite{2d-ldt}, but it can be solved in various other ways too,
including generalizations, like the inclusion of additional weights or of different building blocks
(see \cite{difra-integr,difra-hardobj,difra-calogero}).
This indicates a rich integrability structure and supports the universality hypothesis for these models.
I will not review them all, but rather will recall the method and results we will need in
the developments of Chapter 2 and 3.

In \cite{difra-hardobj} a formula was proven expressing the partition function of the (1+1)-dimensional CDT model,
in its dual formulation and for open boundary conditions in the space direction, as the inverse
of that of a hard-dimer model in one dimension:
\beq \label{fund2}
  Z_t(u=e^{-2\l})= \frac{1}{Z_t^{hd}(-u)}
\eeq
where
\beq\label{hd}
  Z_t^{hd}(u)= \sum_{{\rm hard}\ {\rm dimer}\ {\rm config.}\ D} u^{|D|},
\eeq
and where $e^{-\l}$ is the weight assigned to each triangle and $t$ is the number of time steps.
I rederive the proof of the inversion formula in appendix \ref{App-inversion} for the
three-dimensional case, which we will study in Chapter 3.

This result is not accidental, it is quite general and indeed could already be found in
\cite{viennot}, in a more general mathematical context, where the notion of {\it heap of pieces}
was introduced. Roughly speaking a heap of pieces is a partially ordered set whose elements can be seen as occupying
columns out of a finite set of columns and where direct order relations only exist between
elements in the same column or in neighbouring ones.
It is not difficult to realize that the structure of the (1+1)-dimensional
CDT model is (in its dual picture) that of a heap of dimers\footnote{Note that a dimer is the dual
of two triangles glued along their space-like edge, that is, it is the analogue of the 4-vertex
in (\ref{2d-squares}). Observe that in order to describe the CDT model in terms of a matrix model,
one would have to impose on it the requirement that in the generated graphs the vertices
form a heap of pieces. It is not yet known how to implement such a constraint in a matrix model.}.
Stated differently this simply means that in the dual picture of Fig.~\ref{cdt} we are free to slide
the vertical segments horizontally, with the constraint that segments in the same slice or in neighbouring
ones can neither cross nor touch each other.

One can also give to the partition function (\ref{hd}) a transfer matrix formulation.
This is done by introducing a two-dimensional vector space associated to each site,
with base given so that $(1,0)$ corresponds to the empty state and $(0,1)$ to
a dimer state. If we then associate a weight 1 to the transition empty-empty, a weight
$\sqrt{u}$ to the transition dimer-empty or empty-dimer, and a weight 0 to the transition dimer-dimer,
we can write the transfer matrix $M$ around a site. Since the dimer partition function
is given, for the case of periodic boundary conditions, by $Tr M^t$ we can rewrite
(\ref{fund2}) as:
\beq\label{Z-2d-1}
  Z_t(u)=\frac{1}{Tr \matrice 1 & {\rm i} \sqrt{u} \\
{\rm i} \sqrt{u} & 0\ematrice^t}\ .
\eeq

Since $Tr(M^t)=\lambda_+^t+\lambda_-^t$, where
\beq \label{lambda-pm}
\lambda_{\pm}=(1\pm\sqrt{1-4u})/2
\eeq
are the eigenvalues of $M$ (with $\sqrt{u}$ replaced by ${\rm i}\sqrt{u}$ as in (\ref{Z-2d-1})),
we have for large $t$
\beq \label{Z-2d-2}
  Z_t(u)\sim \frac{1}{\lambda_+(u)^t}
\eeq
which is not analytical at $u=1/4$, the critical point.

The critical point can already be found by analyzing the one-step propagator.
We know that in (1+1) dimensions\footnote{Following standard notation, we reserve the letter $Z$ for
the partition function, and use the letter $G$ for the propagator.}
\beq\label{}
  G(l_1,l_2 ;t=1| \l)=\sum_{strip-\TT_{l_1,l_2}} \frac{1}{C_{\TT}} e^{-S_{EH}}
\eeq
reduces to
\beq\label{}
  G(l_1,l_2 ;t=1| \l)=e^{-\lambda (l_1+l_2)}\sum_{strip-\TT_{l_1,l_2}} 1\ ,
\eeq
and this can be obtained by the generating function
\beq\label{gen2}
  G(x,y ;t=1| \l)=\sum_{l_1,l_2\geq 0} x^{l_1} y^{l_2} e^{-\l (l_1+l_2)}\sum_{strip-\TT_{l_1,l_2}} 1\ ,
\eeq
which in turn can now be written with the use of the inversion formula (and
considering $x$ and $y$ as the square roots of the weights of the
boundary dimers) as\footnote{We could of course also obtain the final result by simply counting
the number of triangulations of the strip with open boundary conditions and obtain
\beq
  G(x,y ;t=1| \l)=\sum_{l_1,l_2\geq 0} x^{l_1} y^{l_2} e^{-\l (l_1+l_2)}\matrice l_1+l_2\\l_2\ematrice
  =\frac{1}{1-e^{-\l}(x+y)}\ ,
\eeq
which coincides with (\ref{gen2bis}). The same is true for the other calculations presented in this
section, since most of the results in (1+1) dimensions can be derived in several different ways.
We concentrate on the inversion method because it is the one that we will be able to generalize
to a particular model of (2+1)-dimensional CDT in Chapter 3.}
\beq\label{gen2bis}
  G(x,y ;t=1| u=e^{-2\lambda})=\frac{1}{\matrice 1 & {\rm i} y\ematrice
  \matrice 1 & {\rm i} \sqrt{u} \\
  {\rm i} \sqrt{u} & 0\ematrice \matrice 1 \cr {\rm i} x \ematrice}
  =\frac{1}{1-\sqrt{u}(x+y)}\ .
\eeq

We see that if we put $x=y=1$ ($i.e.$ considering the partition
function for $\Delta t=1$ with free boundary conditions on the slices) we get a
singularity in $\sqrt{u_c}=\frac{1}{2}$. This is exactly the critical point,
and the continuum limit has to be taken by fine-tuning $\lambda$ to
its critical value $\lambda_c=\ln{2}$ as explained in the following subsection.

\subsection{The continuum limit}

As I have already argued, in order to remove regularization artifacts and obtain a continuum theory
we must tune the coupling constants to a critical value as $a$ goes to zero.
In two dimensions we only have one coupling constant, the cosmological constant, which is conjugate to
the volume. Tuning it to its critical value achieves the infinite-volume limit and as I will show also
leads to a divergent correlation length for the dynamical degrees of freedom.

From (\ref{Z-2d-2}) it is easy to see that
in the limit $\l\rightarrow \l_c=\ln 2$ (equivalent to $u\rightarrow u_c=1/4$) the volume diverges like
\beq \label{Vol-2d-discr}
\langle N_2\rangle=-\frac{\partial \ln Z_t}{\partial\lambda}\sim
 \frac{ 4 t e^{-2\l}}{(1+\sqrt{1-4e^{-2\l}})\sqrt{1-4e^{-2\l}}}
 \simeq \frac{t}{\sqrt{2(\l-\l_c)}}\left(1+O(\l-\l_c)\right)\ .
\eeq
If we scale the cosmological constant like $\l\simeq\l_c+\frac{1}{2}\L a^2$, with $\L$ the
renormalized cosmological constant, and the number of time steps like $t=T/a$, keeping $T$ finite
(a continuous time interval), we get a finite continuous volume\footnote{This expression
is valid for large $T$. To be precise, we should take into account also the contribution from
$\l_-(u)$ in (\ref{Z-2d-1}), leading to
\beq
\langle V\rangle=\frac{T}{\sqrt{\L}}\left(1 -\frac{e^{-2T\sqrt{\L}}}{1+e^{-2T\sqrt{\L}}} \right) \ .
\eeq
Note however that for any value of $T$ the correction gives a multiplicative factor between 1 and 1/2
which does not change the qualitative result (\ref{Vol-2d-cont}). However, the correction given
by $\l_-(u)$ can become fundamental when looking at other quantities, as is shown by the computation
of the correlation function following in the text.}:
\beq \label{Vol-2d-cont}
\langle V\rangle=\lim_{a\rightarrow 0}a^2\langle N_2\rangle\sim \frac{T}{\sqrt{\L}}\ .
\eeq
This expression should be compared with the Euclidean one $\langle V\rangle_E\sim T/\L^{3/4}$,
where the geodesic time has to scale anomalously like $T=a^{1/2} t$, signaling a fractal dimension
(see \cite {DT-book} for details).

We can see that this infinite-volume limit is sufficient to obtain a continuum theory by evaluating
the average and correlation for the length of the spatial slices (the only degree of freedom in two dimensions).
These can be derived with the same inversion method by assigning some weight $z_1$ to the dimer
at time $t_1$, that is, by introducing the generating function for triangulations with a slice of length $l(t_1)$
at time $t_1$, and which we denote by $ F_{t,t_1}(u,z_1)$.

By the inversion formula we have
\beq
\begin{split}
  F_{t,t_1}(u,z_1)&\equiv \sum_{N_2} e^{-\lambda N_2} \sum_{\TT_{N_2,t}} \frac{1}{C_{\TT_{N_2}}}z_1^{l(t_1)}\\
  &=\frac{1}{Tr\left[\matrice 1 & {\rm i} \sqrt{u} \\{\rm i} \sqrt{u} & 0\ematrice^{t_1}
  \matrice 1 & 0 \\ 0 & z_1 \ematrice
   \matrice 1 & {\rm i} \sqrt{u} \\{\rm i} \sqrt{u} & 0\ematrice^{t-t_1}\right]}\ ,
\end{split}
\eeq
from which we find
\beq \label{mean-L}
\langle L(T_1)\rangle=\lim_{a\to 0} a\frac{\partial \ln F_{t,t_1}(u,z)}{\partial z_1}_{|z_1=1}
 \sim \frac{1}{\sqrt{\L}}\ ,
\eeq
which does not depend on $T_1$ because of the periodic boundary conditions.

Similarly, from
\beq
\begin{split}
  F_{t,t_1,t_2}(u,z_1,z_2)&\equiv \sum_{N_2} e^{-\lambda N_2}
   \sum_{\TT_{N_2,t}} \frac{1}{C_{\TT_{N_2}}}z_1^{l(t_1)}z_2^{l(t_2)}\\
  &=\frac{1}{Tr\left[ \matrice 1 & 0 \\ 0 & z_1 \ematrice
  \matrice 1 & {\rm i} \sqrt{u} \\{\rm i} \sqrt{u} & 0\ematrice^{t_2-t_1}
  \matrice 1 & 0 \\ 0 & z_2 \ematrice
   \matrice 1 & {\rm i} \sqrt{u} \\{\rm i} \sqrt{u} & 0\ematrice^{t-(t_2-t_1)}\right]}\ ,
\end{split}
\eeq
where we have used the cyclic property of the trace to move the element $M^{t_1}$ to the end of the trace
to make the dependence on $t_2-t_1$ explicit, we find
\beq
\langle L(T_1) L(T_2)\rangle=\lim_{a\to 0} a^2\frac{1}{F_{t,t_1,t_2}(u,z_1,z_2)}
\frac{\partial^2 F_{t,t_1,t_2}(u,z_1,z_2)}{\partial z_1\partial z_2}_{|z_1=z_2=1}
 \sim \frac{1}{\L}\left( 1+e^{-2(T_2-T_1)\sqrt{\L}}\right)\ .
\eeq

Putting things together, we find for the correlation function of two lengths
\beq \label{correlation}
\langle L(T_1) L(T_2)\rangle-\langle L(T_1)\rangle\langle L(T_2)\rangle
 \sim \frac{e^{-2(T_2-T_1)\sqrt{\L}}}{\L}\ ,
\eeq
which is the continuum limit of the corresponding correlator,
\beq
\langle l(t_1) l(t_2)\rangle-\langle l(t_1)\rangle\langle l(t_2)\rangle
 \sim \frac{\l_+\l_- e^{-(t_2-t_1)\ln\frac{\l_+}{\l_-}}}{(\l_+-\l_-)^2}\ ,
\eeq
from where we identify the lattice correlation length as
\beq
\xi = (\ln\frac{\l_+}{\l_-})^{-1}\ .
\eeq
This diverges at $u_c=1/4$, where $\l_+(u_c)=\l_-(u_c)$, demonstrating the existence of a good continuum limit,
and leading to a finite continuum correlation length equal to $\bar\xi=\lim_{a\to 0}a\xi=(2\sqrt{\L})^{-1}$.

\begin{figure}[ht]
\centering
\vspace*{13pt}
\includegraphics[width=1.5cm]{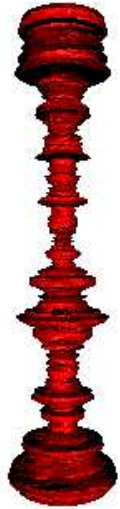} 
\vspace*{13pt}
\caption{\footnotesize A typical (1+1)-dimensional spacetime as obtained from numerical simulations of
the CDT model, at volume $N_2=18816$ and total time $t=168$ \cite{loll-history}.}
\label{2d-tube}
\end{figure}

The existence of correlations between successive slices suggests the presence of a Hamiltonian
operator for the length of the slices.
At the same time, from the result (\ref{correlation}) we see that the variance
of the length at $T_2=T_1$, given by
\beq
 \s_{L(T_1)}\equiv \sqrt{\langle L(T_1)^2\rangle-\langle L(T_1)\rangle^2}
 \sim\frac{1}{\sqrt{\L}}\ ,
\eeq
is of the same order as the expectation value (\ref{mean-L}) of the length itself, which implies that the
quantum fluctuations will dominate and mask any ``semiclassical" behaviour (this is in accordance with the fact
that we do not expect to see any classical behaviour since in two dimensions there is no classical theory of gravity).
This phenomenon is nicely illustrated by Fig.~\ref{2d-tube}.

\subsection{Transfer matrix, propagator and Hamiltonian in (d+1)-dimensions}

Because of the global proper time we are dealing with a standard evolution framework.
In any dimension, the discrete CDT model  can be formulated in terms of a transfer matrix (see the following
section \ref{d+1-dim}).
This is obtained constructing a vector space $W$ out of the geometrical information characterizing the constant-time
slices, $i.e.$ we define a 1-to-1 correspondence between geometries of the slices and state-vectors $|g\rangle$
which we use as basis of $W$\footnote{Remember
that at the discrete level, and for finite volume, the number of different geometries is finite
and therefore we get a finite-dimensional vector space $W_V$. If we include the sum over volume, we just get
$W=\bigoplus_V W_V$, which is (countably) infinite.}.
We can then define the {\it transfer matrix} as that matrix whose elements in such a basis are given
by the one-step propagator with fixed boundary data\footnote{Note that for lack of letters or fantasy I
use the same letter for time, triangulation and transfer matrix, but with some variation: $t$ is the discrete time,
$T$ the continuous time, $\TT$ indicates a triangulation, $\hat T$ the transfer matrix and $T_{ij}$ its
elements in a given basis. I hope this does not lead to any confusion.},
\beq \label{T-matrix}
T_{i j}\equiv\langle g_i|\hat T |g_j\rangle =\sum_{\TT:g_j\to g_i,\ \D t=1}\frac{1}{C(\TT)}e^{-S_R(\TT)}\ .
\eeq
The transfer matrix can be iterated to obtain the $t$-step propagator
\beq \label{propagator}
G(g_i,g_j;t|\l)\equiv\langle g_i|\hat T^t |g_j\rangle =\sum_{\TT:g_j\to g_i,\ \D t=t}\frac{1}{C(\TT)}e^{-S_R(\TT)}\ ,
\eeq
eventually with periodic boundary conditions in time, giving the partition function
\beq
Z_t(\l)=\Tr \hat T^t\ .
\eeq
If $\hat T$ is a bounded, symmetric and positive operator, we can define a lattice Hamiltonian by
\beq \label{H-lat}
\hat H_{lat}=-\frac{1}{a}\ln \hat T\ .
\eeq
That the definition (\ref{T-matrix}) does actually lead to a bounded and symmetric operator is easy
to prove \cite{ajl-def}, while in general (for any dimension) the positivity has only been shown
for the case where the one-step propagator is replaced by a two-step propagator \cite{ajl-def},
$i.e.$ it has only been shown explicitly that the square of the transfer matrix is positive.
However, it is then sufficient to modify the definition (\ref{H-lat}) into
\beq
\hat H'_{lat}=-\frac{1}{2 a}\ln \hat T^2\ .
\eeq
In general $\hat H_{lat}$ (or $\hat H'_{lat}$) is a very complicated operator which depends on the details
of the discretization,
but luckily we are not really interested in most of them. Rather, we are interested in its continuum limit
(for which it also should not matter whether we started from $\hat H_{lat}$ or from $\hat H'_{lat}$)
\beq \label{H-quant}
\hat H = \lim_{a\to 0} \hat H_{lat}\ ,
\eeq
to which we refer as the {\it quantum Hamiltonian} of the system (see also \cite{montvay}).

\subsection{Back to (1+1)-dimensions}

In the continuum formulation, pure quantum gravity in (1+1)-dimensions can be thought of as a string theory
with a zero-dimensional target space. In general, coupling to $d$ scalar fields corresponds to having a $d$-dimensional
target space. In this case, many results are available from the continuum, starting with Polyakov's
seminal work \cite{polyakov}.
Whenever a comparison is possible, it shows that two-dimensional Euclidean DT in the continuum limit agrees
with the continuum results
of non-critical ($d\neq 26$) string theory in conformal gauge $g_{\m\n}=e^{\phi}\d_{\m\n}$ (which, as shown by Polyakov,
is a quantum Liouville theory for the Liouville field $\phi$).

What about the continuum limit of (1+1)-dimensional CDT? It turns out that also CDT reproduces the
continuum results, although in a non-standard gauge.
The presence of a time slicing, with fixed distance between the slices, suggests
comparison with results obtained in the proper-time gauge
\beq
g_{00}=1\ , \ \ \ g_{01}=g_{10}=0\ , \ \ \ g_{11}=\g(x^0,x^1)\ .
\eeq
Two-dimensional quantum gravity in this gauge was studied in \cite{nakayama} and the results agree
with the continuum results from CDT (see \cite{2d-ldt} and \cite{difra-integr})\footnote{There are some interesting
subtleties, which are probably relevant for a better understanding of the relation between Euclidean
and Causal DT in two dimensions, but which I will not discuss since it is not the topic of this thesis.}.
The main lesson I want to take from \cite{nakayama} is that two-dimensional quantum gravity
in proper-time gauge amounts to a problem of quantum mechanics rather than quantum field theory.
The only degree of freedom surviving such a reduction is the length of the slices at constant proper-time,
$i.e.$
\beq
L(x^0)=\int_0^{\pi}dx^1 \sqrt{\g(x^0,x^1)}\ ,
\eeq
and the Hamiltonian depends only on this variable (and its conjugate momentum).

Clearly, CDT fit well into this picture, since having all edges of the same length corresponds to having
no $x^1$-dependence in $g_{11}$.

It also brings about a simplification, which allows us to handle the combinatorial problem present in the discrete
approach.
Having to keep track only of a finite number of (discrete) degrees of freedom (just one in the specific case),
we can rely on the method of generating functions.
It is well known in mathematics (and physics) that if we want to study a sequence $\{ a_n\}$, it is usually easier
to investigate its generating function $f(x)=\sum_n a_n x^n$ instead.

The quantities we want to study in quantum gravity are the propagators
\beq
G(g_{\m\n}(1),g_{\m\n}(2),T)=\int_{g_{\m\n}(2)\to g_{\m\n}(1)}\mathcal D [g] e^{iS_{EH}[g]}\ ,
\eeq
which in our discretization are given by the $t$-step propagators (\ref{propagator}).
In (1+1) dimensions, the states $|g_i\rangle$ are just labelled by the length, which in lattice units is
just a natural number and which we will denote by $l$ as before.
We thus have
\beq
G(l_1,l_2;t|\l)=\langle l_1|\hat T^t|l_2\rangle
\eeq
for the propagator, and
\beq \label{discr-Lapl}
G(x,y;t|\l)=\sum_{l_1,l_2\geq 0} x^{l_1} y^{l_2}G(l_1,l_2;t)
\eeq
for its generating function.
Once the generating function is known, we can recover the propagator by
\beq \label{discr-invLapl}
\begin{split}
G(l_1,l_2;t|\l)&=\frac{1}{l_1!}\frac{\partial^{l_1}}{\partial x^{l_1}}
   \frac{1}{l_2!}\frac{\partial^{l_2}}{\partial y^{l_2}}  G(x,y;t)_{|x=y=1}=\\
   &=\oint_{\g}\frac{dx}{2\pi \rm i}\oint_{\g}\frac{dy}{2\pi \rm i}\frac{G(x,y;t)}{x^{l_1+1}y^{l_2+1}}\ ,
\end{split}
\eeq
where in the last step we have used Cauchy's formula, and where the contour of integration $\g$ goes around
the origin without encircling any singularity of $G(x,y;t)$ (which is analytical in a sufficiently small
neighbourhood of the origin).

Propagator and generating function are translated into continuum functions by use of the
canonical scaling relations
\beq \label{2d-scaling}
l = \frac{L}{a}\ , \ \ \ t=\frac{T}{a}\ , \ \ \ \l=\l_c+ \frac{1}{2}\L a^2
\eeq
for the geometric variables and the (dimensionless) cosmological constant,
by introducing the boundary cosmological constants through\footnote{Note that in (1+1) dimensions the
boundary cosmological constants do not need to be renormalized since the boundaries have no entropy.}
\beq
x=e^{-a X}\ , \ \ \ y=e^{-a Y}\ ,
\eeq
and by employing a multiplicative renormalization for the propagator and generating function\footnote{
The power of $a$ is fixed by the requirement of obtaining (\ref{cont-Lapl}) and  (\ref{cont-invLapl})
from  (\ref{discr-Lapl}) and  (\ref{discr-invLapl}), given (\ref{2d-scaling}).
If this would not lead to a finite propagator, it would mean that (\ref{2d-scaling}) has to be changed,
signaling an anomalous scaling.} according to
\beq
G(X,Y;T|\L)=\lim_{a\to 0} a G(x,y;t|\l)\ , \ \ \ G(L_1,L_2;T|\L)=\lim_{a\to 0} \frac{1}{a} G(l_1,l_2;t|\l)\ .
\eeq
The relations (\ref{discr-Lapl}) and (\ref{discr-invLapl}) become in this limit that of a Laplace and inverse
Laplace transform respectively:
\beq \label{cont-Lapl}
G(X,Y;T|\L)=\int_0^{\infty}dL_1\int_0^{\infty}dL_2 e^{-X L_1} e^{-Y L_2}  G(L_1,L_2;T|\L)
\eeq
\beq \label{cont-invLapl}
G(L_1,L_2;T|\L)=\int_{-{\rm i}\infty}^{+{\rm i}\infty}\frac{dX}{2\pi \rm i}
 \int_{-{\rm i}\infty}^{+{\rm i}\infty}\frac{dY}{2\pi \rm i} e^{X L_1} e^{Y L_2}  G(X,Y;T|\L)\ .
\eeq

We have already seen how to compute the generating function with the inversion formula, namely,
\beq \label{}
\begin{split}
  G(x,y;t | u &=e^{-2\lambda})=\frac{1}{\matrice 1 & {\rm i} y\ematrice
  \matrice 1 & {\rm i} \sqrt{u} \\
  {\rm i} \sqrt{u} & 0\ematrice^t \matrice 1 \cr {\rm i} x \ematrice}=\\
  &=\frac{\l_+-\l_-}{\D_{t+1}-\sqrt{u}\D_t(x+y)-(\D_{t+1}-\D_t)xy}\ ,
\end{split}
\eeq
where $\l_{\pm}$ are as in (\ref{lambda-pm}) and $\D_t\equiv\l_+^t-\l_-^t$.
From this we get\footnote{The factor $2^{-t}$ is necessary because in (\ref{gen2d-cont})
we have to divide by two in order to get the identity.}
\beq
\begin{split}
G(X,Y;& T|\L)=\lim_{a\to 0} \frac{a}{2^t} G(x,y;t|\l)=\\
&=\frac{\sqrt{\L}}{\L\sinh (\sqrt{\L}T) +\sqrt{\L}\cosh (\sqrt{\L}T) (X+Y) + \sinh (\sqrt{\L}T) X Y}\ ,
\end{split}
\eeq
and from its inverse Laplace transform
\beq \label{propagator-T}
G(L_1,L_2;T|\L)=\sqrt{\L}\ \frac{e^{-\sqrt{\L} (L_1+L_2)\coth (\sqrt{\L}T)}}{\sinh (\sqrt{\L}T)}
 I_0\left(\frac{2 \sqrt{\L L_1 L_2}}{\sinh (\sqrt{\L} T)}\right)\ ,
\eeq
where $I_0(z)$ is a modified Bessel function of the first kind.
This expression agrees perfectly with the continuum calculation (equation (44) of \cite{nakayama}
with $m=-1/2$, where $m+1$ is some kind of winding number that in our case is one-half, because we have
considered open rather than cylindrical topology -- see \cite{difra-integr} for an interpretation
of such winding number in CDT).

With the same method we can also compute the Hamiltonian. According to the definition (\ref{H-quant}),
all we need is the one-step propagator, whose generating function was given in (\ref{gen2bis}).
From $\hat T=e^{-a \hat H_{lat}}$ we see that in the continuum limit we have
\beq \label{T-matrix-transf}
\begin{split}
G(X,Y; & T=a |\L)=\int_0^{\infty}dL_1\int_0^{\infty}dL_2 e^{-X L_1} e^{-Y L_2}
 \langle L_1 | \hat T |L_2 \rangle =\\
 &=\int_0^{\infty}dL_1\int_0^{\infty}dL_2 e^{-X L_1} e^{-Y L_2}
 \langle L_1 | \left( \hat 1 - a \hat H +O(a^2) \right) |L_2 \rangle\ ,
\end{split}
\eeq
and hence from the generating function of the transfer matrix can get in the continuum limit the
Laplace transform of the Hamiltonian (in the ``L-representation").
We find from (\ref{gen2bis})
\beq \label{gen2d-cont}
G(X,Y; T=a |\L)=\lim_{a\to 0} \frac{a}{2} G(x,y;t=1|\l)=
 \frac{1}{X+Y}-a \left( \frac{2\L-X^2-Y^2}{2(X+Y)^2}  \right) + O(a^2)\ ,
\eeq
which upon inverse Laplace transformation yields
\beq
\langle L_1 | \hat T |L_2 \rangle\ = \d(L_1-L_2)
 - a \left( -L_2\frac{\partial^2}{\partial L_2^2}-\frac{\partial}{\partial L_2} + \L L_2 \right) \d(L_1-L_2)
 +O(a^2)\ .
\eeq
We recognize the Hamiltonian
\beq
\hat H=-L_2\frac{\partial^2}{\partial L_2^2}-\frac{\partial}{\partial L_2} + \L L_2\ ,
\eeq
which is self-adjoint (with respect to the measure $dL$) and bounded from below, as it should be.
Its eigenfunctions and eigenvalues can be found without too much pain, and are respectively
\beq
\psi_n(L)=\sqrt[4]{4\L}\ e^{-\sqrt{\L}L}L_n(2\sqrt{\L}L)\ ,
\eeq
\beq
E_n=\sqrt{\L}(2n+1)\ ,
\eeq
where $n$ is a non-negative integer, $L_n(x)$ is the $n$'th Laguerre polynomial and the eigenfunctions
$\psi_n(L)$ are orthonormal.
They can be used for a consistency check, showing that
\beq
G(L_1,L_2;T|\L)=\langle L_1 | e^{-T \hat H} |L_2 \rangle = \sum_{n=0}^{\infty} e^{-T E_n} \psi_n^*(L_1)\psi_n(L_2)
\eeq
indeed reproduces the result (\ref{propagator-T}).

The model is thus completely solved. From the combinatorics of the discrete triangulation we have
recovered in the continuum limit the quantum-mechanical proper-time formulation of two-dimensional gravity.
The Hamiltonian has been found and diagonalized.

\subsection{Coupling CDT to matter}

Although highly non-trivial, pure quantum gravity may be considered a rather academic topic,
in the sense that it is a theory of empty space.
In order to make contact with the real world, it is important to investigate the properties of
the coupled system of gravity and matter.
In addition, in a regime where matter and geometry interact strongly, the presence of matter can
in principle alter the results of the pure gravity case.
For example, in the context of the perturbative approach, we know that the one-loop finiteness result
for the pure gravity case is spoiled by the coupling to matter \cite{hooft-veltman}.
In a non-perturbative context, like that of CDT, the addition
of matter degrees of freedom on the triangulations can in principle affect the short-distance behaviour.

Besides understanding how matter affects the geometrical properties of spacetime, it is also
important to understand how in turn the behaviour of matter is affected by the fluctuations of spacetime.
Firstly, this will provide additional cross-checks for a correct classical limit of the gravitational dynamics,
and secondly, it will be relevant for potential observable effects in nature.

Matter coupling in quantum Regge calculus and DT has been investigated mostly by numerical simulations
(exceptions are the already mentioned matrix model methods in two-dimensional DT).
I will not review here the variety of results that has been found, and will move straight to CDT;
the interested reader is suggested to start from the review \cite{loll-discrete}.

Although a two-dimensional spacetime would not be made more physical by a coupling to matter,
it is a natural starting point and testing ground for analytical and numerical methods,
as usual with the hope of extending them to higher dimensions at a later stage.

Matter coupling in two-dimensional CDT models has been introduced and investigated in \cite{aal1} and \cite{aal2}.
As a prime example of a simple matter model, the Ising model was chosen,
probably the most thoroughly investigated matter system on fixed lattices.
Both on a regular lattice and on the DT ensemble this model has been solved exactly.
Unfortunately no exact solution has yet been found for the Ising model on CDT.

In \cite{aal1} the Ising model on two-dimensional CDT was investigated via Monte-Carlo simulations
and a high-temperature expansion. These methods provided strong evidence that the model belongs to the same
universality class as the Ising model on a fixed square lattice, and that the Hausdorff dimension of the geometries
is $d_H=2$, and therefore unaffected by the presence of matter.
In \cite{aal2} these investigations were extended by coupling eight copies of Ising models to CDT.
It was observed in Monte-Carlo simulations that even though the matter exponents, and hence the universality class,
remain unaltered, the geometry has undergone a phase transition, to a phase in which $d_H=3$.

In Chapter 2, I will present some new findings that strengthen the results of \cite{aal1}.
The method used will be that of high- and low-temperature expansions, whose application to the CDT model
will be developed in detail. In comparison with the preliminary results of \cite{aal1}, the method will also be
made more algorithmic.

\section{(d+1)-dimensional CDT} \label{d+1-dim}

In higher dimensions, CDT are constructed according to the same principle as in two dimensions,
only that now the spatial manifold $\Si$ is higher-dimensional and the triangles are replaced by
tetrahedra, four-simplices and so on \cite{ajl-def}.

\begin{figure}[ht]
\centering
\vspace*{13pt}
\includegraphics[width=11cm]{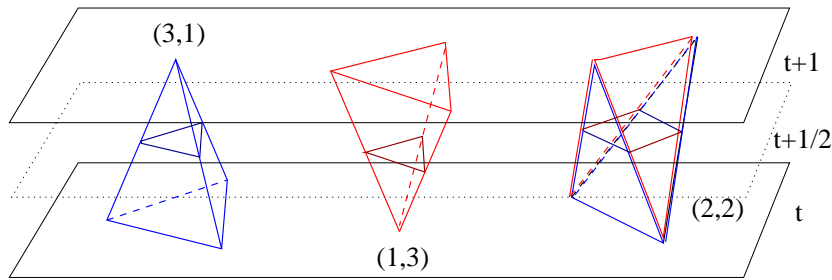} 
\vspace*{13pt}
\caption{\footnotesize The three types of tetrahedral building blocks and their
intersections with the $t+1/2$ plane.}
\label{blocks}
\end{figure}

At integer times we have different triangulations $\TT_t(\Si)$ of $\Si$, constructed of $d$-dimensional completely
space-like (Euclidean) simplices $\s_d$.
The spacetime between two such triangulations at times $t$ and $t+1$ is interpolated
by $(d+1)$-dimensional simplices $\s_{d+1}(d-k,k)$ having a $\s_{d-k}$ and a $\s_k$ as subsimplices
at time $t$ and $t+1$, for $k=0,1,...,d$. We distinguish $d+1$ types of $(d+1)$-dimensional simplices
in CDT, to which we will also refer as ``of type $(d-k,k)$". The three types appearing in (2+1) dimensions are
shown in Fig.~\ref{blocks}.

\subsection{The combinatorics of the (2+1)-dimensional transfer matrix}

We have seen that in two dimensions it is possible to solve the combinatorial problem
for the $t$-step propagator directly. In three dimensions the same task does not seem feasible at the moment.
We have mentioned that in the Euclidean case this has been tried by a generalization of matrix models,
but up to now without success. In the CDT case we know that such a matrix model-like description
is unlikely to work because of the anisotropy of the triangulations. Furthermore, none of the methods used
to solve (1+1)-dimensional CDT seem to be generalizable in an obvious way.

On the other hand, we have also seen how in CDT - once we have obtained a continuum Hamiltonian from
the transfer matrix - it is possible to construct the finite-time propagator directly in the continuum.
This approach seems to have more chances of being soluble because the transfer matrix is encoded in the triangulation
of the one-step propagator.
With a single layer of tetrahedra confined between two
adjacent slices of constant time, the combinatorial problem is simplified, but still by no means simple!

The nice thing about the one-step propagator of (2+1)-dimensional CDT is that it can be formulated as
a two-dimensional problem ``with colours".
In order to do this, we introduce a space-like slice at constant half-integer time $t+1/2$, $i.e.$ a slice that
cuts the $\s_3(3-k,k)$ tetrahedra in the middle as shown in Fig.~\ref{blocks}.
This slice is by construction a piecewise linear manifold too, but in general not a simplicial one.
It will consist of squares (coming from the intersection of the time-like faces of tetrahedra of type (2,2)
with the plane at $t + \frac{1}{2}$) and of triangles (coming from tetrahedra of type (3,1) and (1,3)).
We can give a colour-code to the links to distinguish whether they come from
triangles of type (2,1) (blue links) or of type (1,2) (red links), as in Fig.~\ref{blocks}.

\begin{figure}[ht]
\centering
\vspace*{13pt}
\includegraphics[width=9cm]{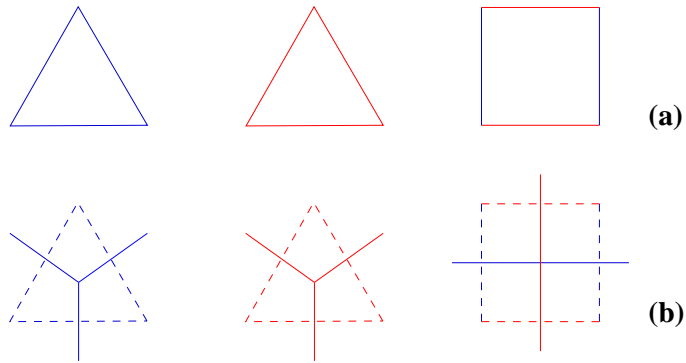} 
\vspace*{13pt}
\caption{\footnotesize The $t+1/2$ figures and their dual.}
\label{vertices}
\end{figure}

Again it is useful to switch to the dual picture.
This is done by drawing dual links which connect the centre of a triangle or
square with the centre of its first neighbours and colouring them
with the same colour as the link they intersect (see Fig.\ref{vertices}).
The dual of a blue (red) triangle is a blue (red) 3-vertex, and
the dual of a square is a bi-coloured 4-vertex.

The dual of the intersection pattern is a graph with blue and red 3-vertices, bi-coloured 4-vertices
and no bi-coloured 2-vertices. This can also be thought of as the superposition of two three-valent graphs
of different colours.
The blue and red graphs are respectively the dual of the triangulation at time $t$ and at time $t+1$.

The next question is which part of the information contained in these two graphs
we should keep track of and which part we need to sum over in order to obtain the transfer matrix.
In principle the in- and out-states are labelled by the whole combinatorial information of the blue
and red graphs, $i.e.$ by the full geometric details of the in- and out-triangulation.
If it was really the case that we needed to keep track of this complete boundary information, we probably
would not have gained much with our construction of coloured graphs.
We would have to store the information about the two graphs, for example, in the form of
adjacency matrices, and count in how many inequivalent ways they can be superimposed.
It would certainly be a challenging task to relate this complex information to analytic data
and a quantum Hamiltonian.


In this situation, the special properties of three-dimensional continuum
gravity come to our help.
We know from the canonical analysis of (2+1)-dimensional general
relativity that the constraints
reduce the number of degrees of freedom from infinite to finite (see, for
example, \cite{carlip-3d}).
The remaining degrees of freedom correspond to the Teichm\"uller
parameters of the spatial slices.
How this is reflected in the quantum theory is less clear.
The classical canonical constraint analysis can be implemented formally in
a
phase space path integral, by integrating over the lapse $N$
and the shift $N_i$ variables of the ADM decomposition.
However, the starting point of the non-perturbative CDT path integral is
quite different.
Firstly, it is the non-perturbative implementation of a {\it configuration
space} path integral,
and secondly, it contains by construction {\it no} gauge degrees of
freedom relating
to coordinate transformations, and works entirely in terms of geometries,
that is, on the
quotient space of metrics modulo diffeomorphisms.

How the degree-of-freedom counting of classical, three-dimensional gravity
is
implemented in a
configuration space path integral (never mind whether discrete or
continuous) is not
immediately obvious. For example, in a non-perturbative geometric
formulation like
CDT, it would be interesting to see explicitly how most of the degrees of
freedom
labelling the states of the theory (roughly speaking,
spatial geometries at a given time) are not propagating, and we are
therefore
dealing with a theory of quantum mechanics rather than quantum field
theory. In a formal continuum treatment of the path integral, it is clear
that
this property must be encoded in the non-trivial path integral measure
obtained after gauge-fixing. A continuum treatment in proper-time gauge
($N=1$, $N_i=0$),
which presumably comes closest to the coordinate-free CDT formulation,
does indeed
strongly suggest that the kinetic term
for the conformal factor of the spatial geometries (the would-be
propagating field degree of freedom)
is cancelled in the path integral by a Faddeev-Popov determinant in the
measure, leaving over only
a finite number of degrees of freedom \cite{dasgupta-loll}. Also the
numerical simulations of CDT in
three \cite{ajl-3d-lat} and four \cite{ajl-semi} dimensions show that in the phase where an
extended geometry is
found, there is no dynamical trace of the divergent conformal mode.


On the basis of these arguments we go back to the transfer matrix of CDT and integrate over all but
a finite number of variables for the in- and out-state. In particular if we choose a spherical topology
there are no Teichm\"uller parameters, and we will keep track of only one piece of information, the area of the slices.
We define a new set of states
\beq
|A\rangle = \frac{1}{\sqrt{\NN(A)}}\sum_{\TT_{|A}}|\TT_{|A}\rangle\ ,
\eeq
where $\NN(A)$ is the number of triangulations of given area $A$ (equal to the number of triangles $N_2$ on the
slice) and $\TT_{|A}$ any triangulation with a given total number of triangles $A$,
with the aim of computing the transfer matrix in this basis,
\beq
T_{A A'}=\langle A|\hat T |A'\rangle.
\eeq
In a general theory, this would be just a small part of the full transfer matrix. Only for the special case of
three-dimensional gravity, on the basis of continuum arguments, we may conjecture that
\beq \label{conjecture1}
\langle A|\hat T|\TT_{|A'}\rangle-\langle A|\hat T|\TT'_{|A'}\rangle\rightarrow 0 \hskip29pt \text{for}
\hskip15pt A,A'\rightarrow\infty\ ,
\eeq
for any two triangulations $\TT_{|A'}$ and $\TT'_{|A'}$ with the same area $A'$.
It follows that the completeness relation
\beq
\int dA |A\rangle\langle A|=\mathbb{I}
\eeq
would hold in the large-area limit and we may use $\langle A|\hat T|A'\rangle$ as our transfer matrix.

In that case we can again define a generating function for the transfer matrix
\beq
G(x,y;t=1|\l,\k)=\sum_{A_1,A_2} x^{A_1}y^{A_2}\langle A_1|\hat T|A_2\rangle\ ,
\eeq
where $\k$ is the bare dimensionless Newton constant,
having the interpretation of a generating function for the number of bi-coloured graphs with a given number
of 3-vertices and arbitrary number of 4-vertices.
Like in two-dimensional DT these graphs can be thought of as Feynman diagrams of a matrix model.
We can write
\beq \label{ABAB-model}
G(x,y;t=1 |\l,\k)=\lim_{N\to\infty} \frac{1}{N^2}\ln \int dA dB
 e^{-N \Tr(A^2+B^2-\frac{u}{3}A^3-\frac{u}{3}B^3-\frac{w}{2}ABAB)}\ ,
\eeq
where $u=x e^{c_1\k - b_1\lambda}$, $v=y e^{c_1\k -b_1\lambda}$ and $w= e^{-c_2\k -b_2\lambda}$
are the weights associated to $N_{31}$, $N_{13}$ and $N_{22}$, the number of $(3,1)$-, $(1,3)$- and
$(2,2)$-blocks respectively, which come from an evaluation of the action for the one-step propagator
similar to the one reported in appendix \ref{App-action} for the case studied in Chapter~3.

In a series of papers, Ambj\o rn et al. \cite{ABAB1,ABAB2,ABAB3} studied a modified version of CDT with pyramids
replacing the $(3,1)$- and $(1,3)$-tetrahedra, corresponding to a modified, exactly solvable version
of the matrix model (\ref{ABAB-model})
with quartic rather than cubic interactions \cite{zinn-just-1,zinn-just-2},
and analyzed its consequences.
The matrix model generates also many configurations which do not have an interpretation as regular slices of
a three-dimensional simplicial manifold, but instead correspond to configurations with ``wormholes".
According to the solution of the matrix model
there is a critical line (at $u=v$, otherwise it is a critical surface with a richer phase structure)
on which two different phases are realized, which in \cite{ABAB1} were interpreted in terms of a dominance
or otherwise of these wormhole geometries. Taking the continuum limit of the CDT model would then correspond to
approaching that part of the critical line where wormhole geometries are rare.

One main aim of these investigations was to extract, via the analogue of (\ref{T-matrix-transf}),
the Hamiltonian of the system.
Unfortunately it turns out to be highly non-trivial to even identify the leading-order behaviour giving rise to
the identity part of the transfer matrix \cite{ABAB3}, and up to now a derivation of the Hamiltonian has
not been possible.

In Chapter~3, I will present a detailed analysis of a different CDT model for three-dimensional gravity
for which some more progress can be made in this direction.

\subsection{Numerical results}

A valuable tool in CDT is represented by Monte Carlo simulations.
Since exactly solvable models are rare, numerical simulations play a vital role in physics.
Being able to study quantum gravity via simulations presents a great advance, which goes back to Regge calculus and DT.
Numerical results for CDT for the first time produced tangible evidence for the emergence of extended
and ``semiclassical" geometries in higher-dimensional non-perturbative gravity.

\begin{figure}[ht]
\centering
\vspace*{13pt}
\includegraphics[width=9cm]{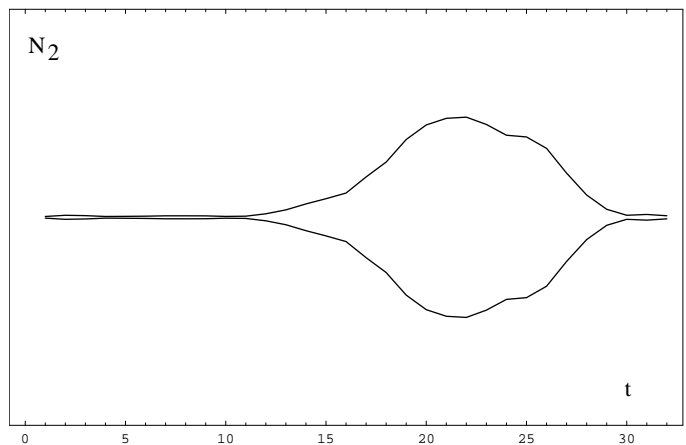} 
\vspace*{13pt}
\caption{\footnotesize The ``lump" of 3d geometry as produced by Monte-Carlo simulations of three-dimensional CDT
\cite{ajl-3d-lat}.
$N_2$ (vertical axis) is the number of triangles in the slices of constant time $t$,
$i.e.$ the area of the slices. The plot is reflected about the central $t$-axis to give a pictorial impression of the
emerging spacetime.}
\label{lump-3d}
\end{figure}

Monte Carlo simulations of (2+1)-dimensional CDT with topology $S^2\times [0,1]$ \cite{ajl-3d,ajl-3d-lat} showed
the existence of a single phase, where the typical geometry represents a ``lump" of spacetime with a three-dimensional
scaling behaviour. In Fig.~\ref{lump-3d} a plot of the area of the slices as a function of time is reproduced,
to illustrate the presence of such an extended lump.

From an analysis of the fluctuations of the area of successive spatial slices, evidence was gathered
for an effective description by an action of the form
\beq
S^{(2+1)d}_{eff}(A)=\int dt \left( \frac{1}{G}\frac{\dot A^2(t)}{A(t)} + \L A(t) \right)\ ,
\eeq
which is exactly the kind of action we would find following, for example, \cite{hosoya}.
Numerical results support the existence of a well-defined dynamics in terms of a reduced
set of degrees of freedom (just a single one for the sphere). Whether this has to be thought of as an effective
mini-superspace-like reduction or as a reduction coming from the constraints of the full theory
should be determined in some other way, for example, by checking the property (\ref{conjecture1}).

A similar result was obtained more recently for (3+1)-dimensional CDT \cite{ajl-rec},
where we expect to have all the field degrees of freedom of general relativity.
There seem to be three different phases, separated by lines of first-order transition.
Evidence for the emergence of an extended four-dimensional universe has been found in one of these phases,
the others being again dominated by degenerate geometries.

By an analysis similar to that in (2+1) dimensions, an effective action for the volume of spatial slices
(within the extended phase) was found to be given by \cite{ajl-semi}
\beq
S^{(3+1)d}_{eff}(V_3)=\int dt\left( \frac{c_1}{G}\frac{\dot V_3^2(t)}{V_3(t)}+\frac{c_2}{G}V_3^{1/3}+\L V_3(t)\right)\ .
\eeq
Introducing a scale factor $a(t)$ by $V(t)=a^3(t)$, one finds a simple mini-superspace action often used in
quantum cosmology, apart from an overall sign. The fact that such sign differs in the effective
action obtained in CDT (leading to an action bounded from below) from that obtained by simply substituting
a mini-superspace ansatz  in the action (resulting in an unbounded action)
supports the idea that a correct inclusion of the contribution of the measure in the path-integral
would also cure the conformal factor problem in a continuum path integral treatment.

Finally, I would like to mention another very interesting result found in (3+1) dimensions from the numerical
simulations.
As already mentioned, various notions of effective dimension can be measured within this approach.
By a global scaling of the volume-volume correlator it was found that in the extended phase the value of the
Hausdorff dimension is compatible\footnote{By numerical simulations it is of course impossible to prove that a
certain measured number is an integer.} with $d_H=4$ \cite{ajl-prl}. Another possible notion of effective dimension
is the so-called spectral dimension $d_S$, which I will not describe in detail here. While also for this
dimension a value compatible with $d_S=4$ on large scales was found,
the value of the spectral dimension turns out to be scale-dependent and decreases monotonically
from $d_S=4$ on large scales to $d_S\approx 2$ on short scales \cite{ajl-spectral}.
A similar result has subsequently been found within a non-perturbative renormalization group approach
\cite{reuter-spectral}.
Such a dynamical dimensional reduction is very appealing in that it indicates a way in which
gravity may cure its own ultraviolet divergences by virtue of its asymptotic safety. It is conceivable, at least
in principle, to construct a non-perturbatively improved perturbative expansion where due to the
dimensional reduction at short distances there would be no such divergences.

\section{Summary}

In this chapter I have given an introduction to the CDT approach to quantum gravity, together with its motivations,
background, and results.
Particular attention has been given to what is known analytically, since the purpose of the work presented
in the following chapters is to elaborate on it and to try to push the analytical understanding of
the CDT model further.

We have seen that starting from the idea of constructing a non-perturbative quantum theory of gravity in analogy
with the lattice formulation of quantum field theory, Regge calculus provides us with a natural discretization
of general relativity.
We reviewed the arguments that led from quantum Regge calculus, where the triangulation is fixed and the
edge-lengths are the dynamical degrees of freedom in the path integral, to dynamical triangulations,
where the edge-lengths are fixed and the combinatorial structure of the triangulation is being summed over.

The original Euclidean dynamical triangulation approach has a number of advantages over the quantum Regge calculus,
but ultimately leads only to macroscopically pathological geometries, the main reason for which it has mostly been
abandoned.
As I have tried to highlight, the causal version of DT offers a physically motivated cure to its troubles.
By contrast, CDT models have succeeded in recovering the large-scale effective dimension $d=n$ when starting from
$n$ dimensional building blocks, and have provided strong evidence of the emergence of a semi-classical spacetime from
the sum over geometries.

One of the great merits of dynamical triangulations is that they give a definite prescription for studying
quantum gravity by computer simulations, and by that obtain definite (albeit numerical) results.
On the other hand, the discrete nature of the model makes it difficult to obtain analytical
results, and the only CDT model so far whose continuum limit has been obtained by purely analytical methods is the
one in (1+1) dimensions without matter coupling.

I will introduce in Chapter~2 a ``semi-analytical" technique that can be used in two
dimensions to obtain information about the critical behaviour (and hence the continuum
limit) in the case of gravity-matter coupled systems.

I have reviewed the solution of the (1+1)-dimensional pure-gravity CDT model obtained by use of the inversion
formula for heaps of pieces. The same technique will be of great value in Chapter~3, where it will be applied to
the evaluation of the one-step propagator for a special version of (2+1)-dimensional CDT, in the same spirit
as the work on the ABAB matrix model mentioned in the last section, demonstrating that a continuum
limit can be performed analytically also in that case.

%


\chapter{Adding matter in two dimensions}

{\small
In this chapter I will present work on spin systems coupled to  Causal Dynamical Triangulations (CDT),
more specifically, on the Ising model on a (1+1)-dimensional CDT.
It is shown how a method invented to analyze the critical behaviour of spin systems on flat lattices
can be adapted to the fluctuating ensemble of curved spacetimes underlying the CDT approach to quantum gravity.
A systematic counting of embedded graphs is developed to evaluate the thermodynamic functions of the
gravity-matter models in a high- and low-temperature expansion.
This chapter is based on publications \cite{mio-ising1} and \cite{mio-ising2}.
}

\section{Coupling quantum gravity to matter}

If a quantum theory of gravity is to describe properties of the real world,
it must tell us if and how matter and spacetime interact at extremely high energies.
Current, incomplete models of quantum gravity usually approach the problem by
first trying to construct a quantum theory of spacetime's geometric degrees of
freedom alone, and then adding matter degrees of freedom in some way. One is
then interested in whether and how the ``pure" gravity theory gets modified and
how the dynamic aspects of quantum geometry may influence the matter
behaviour. A priori, different scenarios at very short distance scales are
thinkable: the behaviour of quantum geometry may be completely dominant,
geometric and matter degrees of freedom may become indistinguishable, or the
matter-coupled theory may have no resemblance with the pure theory at all.

In the previous chapter we have seen how difficult it is to study the quantum behaviour
of pure gravity. As one may expect, coupling matter to geometry does not improve the situation,
and understanding how the known behaviour of matter models on flat spacetime is affected
by the fluctuations of geometry is a challenging problem.
Most of the difficulties stem from the fact that in the case of quantum gravity, conventional
calculational methods have to be adapted to a situation where there is no fixed
background spacetime, and instead ``geometry" is among the dynamical degrees of
freedom.

In a context where we try to make sense of quantum gravity by a regularization of the path integral
in terms of discretized geometries, like is the case of CDT, matter fields are replaced
by lattice models.
Typical lattice models are spin systems on a lattice, which have been extensively
studied in the case of regular (flat) lattices, and for which many techniques exist.
In this case, the problem of studying matter on a fluctuating geometry is transformed into the problem of studying
spin systems on an ensemble of random lattices, in particular, the causal triangulations.
It is then not clear which, if any, of the techniques developed for the case of a fixed lattice
can be generalized to the case of a fluctuating lattice.

In this chapter we will look in detail at an example of such a generalized method,
for a particular class of quantum systems of gravity and matter.
As it should be clear, the gravity part will be played by the CDT model, and, as announced
in the title of the chapter, we will work in the simplified context of two spacetime dimensions,
where we have full control on the pure-gravity model, and where many exact
results for spin systems on fixed lattices are available.

We will see how a time-honoured method of estimating the critical behaviour of a
lattice spin system, namely, by considering finitely many terms of a weak-coupling
(i.e. a high-temperature) expansion of
its thermodynamic functions can be adapted to the case of ``quantum-gravitating"
lattices. The method consists in classifying local spin configurations (which can be
represented by diagrams consisting of edges or dual edges of a triangulated
spacetime), and determining their weight, i.e. the probability to find such a configuration
in the ensemble of dynamical triangulations. Diagrammatic techniques have also
been employed recently in another coupled model of geometry and matter in three
dimensions, in the context of a so-called group field theory, an attempt to
generalize matrix model methods to describe ensembles of geometries in
dimension larger than two (see \cite{oriti} for a recent review).

The model we will be studying in detail is that of an Ising model whose spins
live on either the vertices or triangles of a two-dimensional simplicial lattice
and interact with their nearest neighbours.
Summing over all triangulated lattices in the usual CDT ensemble, each with
an Ising model on it, gives rise to the matter-coupled system whose properties
we are trying to explore.
An exact solution to this Lorentzian model has not yet been found. This has to do with
the fact that in terms of the randomness of the underlying geometry the model
lies in between that on a fixed regular lattice (solved by Onsager long time ago \cite{onsager})
and that on purely Euclidean triangulations (solved by Kazakov \cite{kazakov}),
but is sufficiently different from either to make the exact solution methods known
for these cases inapplicable.

As argued in Chapter~1, it is difficult to introduce additional restrictions concerning the local nature of
geometry into a matrix model without destroying its simplicity and, more importantly, its solubility.
As seen there, this does not prevent the pure-gravity model to be exactly solvable by other methods,
but unfortunately the same cannot be said up to now for the Ising-gravity coupled case.
Of course there is no obstacle to studying such a system by numerical simulations. This was
done in \cite{aal1,aal2}, where
strong evidence was found that the spin model behaves in the presence of gravity (at least as far as
its critical exponents are concerned) just like it does on a fixed, regular lattice.
This is supported by a ``semi-analytical" high-temperature
expansion, whose initial results were merely quoted in the original work of \cite{aal1}.
In the work I am reporting in this chapter, the details of how to perform the relevant graph
counting on CDT are developed, both for the standard Ising model on CDT and on its dual.
The algorithms found have allowed us to compute the diagrammatic contributions to the Ising susceptibility to
order 6 and order 12 for CDT lattices and their duals, respectively.

The pure counting results were first announced in \cite{mio-ising1}, where
we put forward the conjecture that the best way to extract information about
the universal properties of a matter system on {\it flat space} is by coupling
it to an ensemble of quantum-fluctuating causal triangulations!
The strongest evidence for this conjecture so far comes from the fact that an evaluation by
straightforward ratio method of a rather small number of susceptibility coefficients
already leads to results in surprisingly close agreement with the exactly known
value for the susceptibility. As will become clear below, the
evidence for this conjecture is still too limited to reach any definite
conclusion, and the issue is being studied further.

Generally speaking, one can look at our method and results also from a
purely condensed matter point of view, as an example of how the
universal properties of spin and matter systems are affected by introducing
random elements into its definition.
For example, the inclusion of random impurities in a lattice model leads to
Fisher renormalization \cite{fisher} of the critical exponents in the so-called
annealed case, where the disorder forms part of the dynamics.
By contrast, in the quenched case the Harris-Luck criterion
is conjectured to hold \cite{harrisluck}.
To what extent these ideas extend to the case of random connectivity of the
lattice -- which is the one relevant to non-perturbative quantum gravity models --
is still not very clear, despite a number of results in this direction,
involving spin models with (quenched) geometric disorder
coming from Euclidean dynamical triangulations and
Voronoi-Delaunay lattices (see, for example, \cite{jj,luck}).

One of the reference points for such investigations is that of the
two-dimensional Ising model on
fixed, regular lattices, which in absence of an external magnetic field can
be solved exactly in a variety of ways (see \cite{mccoywu,baxter}). In this
case, the critical exponents, which characterize the system's behaviour
near its critical temperature, are given by the so-called
Onsager values $\alpha=0$, $\beta=0.125$ and $\gamma=1.75$ for the
specific heat, spontaneous magnetization and susceptibility.
In sharp contrast,
in the case of the Ising model on {\it Euclidean} dynamical triangulations
mentioned earlier the disordering effect is so strong that the same
critical exponents are altered to $\alpha=-1$, $\beta=0.5$ and $\gamma=2$
\cite{kazakov}. On the other hand, despite the
annealed geometric randomness present, the corresponding Ising model
coupled to {\it causal} dynamical triangulations seems to share
the flat-space exponents, indicating a perhaps surprising
robustness of the Onsager universality class.\footnote{The study of fermions
coupled to two-dimensional causal dynamical triangulations was initiated
in \cite{Burda}.}

In the piece of work presented here, we will concentrate for simplicity and
definiteness on the CDT-Ising system, although our method should also apply
straightforwardly to other spin models
and may inspire similar diagrammatic
expansions in other, discrete models of quantum gravity.
Sec.~\ref{Ising-highT} introduces
the high-temperature expansion on regular and dynamical lattices.
In Sec.~\ref{graph-theory}, after recalling some general definitions and results of graph theory,
a set of rules is presented which in principle allow us to compute the weight
of any susceptibility graph on the fluctuating CDT lattice and its dual, and explain
the method with the help of illustrative examples. The finite number of terms
of the series expansions we obtain in this way are analyzed in Sec.~\ref{series} with
the help of the ratio, Dlog Pad\'e, and differential approximants methods.
Sec.~\ref{Ising-lowT} contains some remarks on the extension of our method to
low-temperature expansions.

\section{The Ising model and its high-$T$ expansion} \label{Ising-highT}

Given any lattice $G$ of volume $N$ with $v$ vertices and $l$ links (edges),
the Ising model on $G$ in the presence of
an external magnetic field $h$ is defined by the partition function
\beq
Z_N(K,H)=\sum_{\{\sigma_i=\pm 1\}_{i\in G}}e^{-\beta \mathcal H[\{\sigma\}]}
=\sum_{\{\sigma_i=\pm 1\}_{i\in G}}e^{K \sum_{\langle i j \rangle}\sigma_i\sigma_j +H\sum_i\sigma_i},
\eeq
where $ \mathcal H[\{\sigma\}]$ is the Hamiltonian of the spin system, $\beta:=1/(k_B T)$
the inverse temperature,
$\sigma_i$ the spin variable at vertex $i$, taking values $\sigma_i=\pm 1$, and
${\langle i j \rangle}$ denotes nearest neighbours. In standard notation,
we will use $K=\beta J$, where $J>0$ is the ferromagnetic spin coupling, and
$H=\beta mh$, where $m$ is the spin magnetic moment.
The fact that $\sigma_i\sigma_j=\pm1$ allows us to write
\beq
\begin{split}
e^{K\sigma_i\sigma_j} &=(1+u\; \sigma_i\sigma_j) \cosh (K),\\
e^{H\sigma_i} &=(1+\tau\; \sigma_i) \cosh (H),
\end{split}
\eeq
in terms of newly defined variables $u=\tanh (K)$ and $\tau=\tanh(H)$,
and rewrite the partition function as
\beq\label{high-exp}
\begin{split}
Z_N(K,H)
   &=\sum_{\{\sigma_i\}}\prod_{\langle i j \rangle}e^{K\sigma_i\sigma_j}\prod_i e^{H\sigma_i}\\
   &=\cosh^l(K) \cosh^v(H)\sum_{\{\sigma_i\}}\prod_{\langle i j \rangle}(1+u\; \sigma_i\sigma_j)
     \prod_i (1+\tau\; \sigma_i)\\
   &=\cosh^l(K) \cosh^v(H)\sum_{\{\sigma_i\}}\big[ (1+u\sum_{\langle i j \rangle}\sigma_i\sigma_j+
     u^2\sum_{\langle i j \rangle,\langle k l \rangle,j\neq l}\sigma_i\sigma_j\sigma_k\sigma_l+...)\times\\
   & \hspace{1.5cm}\times (1+\tau \sum_i\sigma_i+\tau^2\sum_{i\neq j}\sigma_i\sigma_j+...) \big]\\
   &=2^v \cosh^l(K) \cosh^v(H)\; F^{(N)}(u,\tau).
\end{split}
\eeq
Because of the sum over spin values $\pm 1$ it is clear that every term containing a spin at a
given vertex to some odd power will give a vanishing contribution. Thus we need to keep only
terms with even powers of $\sigma_i$'s, which are terms where each $\sigma_i$ belongs
either to an even number of nearest neighbour couples $\sigma_i\sigma_j$ (from the
$u$-terms) or to an odd number of them and to one of the spins coming from the
$\tau$-terms. It is not difficult to convince oneself that each such term
is in one-to-one correspondence with a graph drawn on the lattice whose length is
given by the power of $u$
and whose number of vertices with odd valence is given by the power of $\tau$.
The function $F^{(N)}(u,\tau)$ introduced in eq.\ (\ref{high-exp}) is then the generating
function for the number of such graphs that can be drawn on the lattice, and the factor
$2^v$ comes from the sum over the spin configurations.
The representation we have obtained in this manner is a high-temperature
expansion, since for infinite temperature $T$ we have $u=\tau=0$.
We will return to this graphic interpretation in the next section, after having recalled
some basic definitions from graph theory. In what
follows, we will only be interested in the case of vanishing external magnetic field.

All we have said up to now does not require any specific properties of the lattice
$G$, but works for any lattice, regular or not. This makes it
straightforward in the framework of CDT to couple the Ising model to gravity.
We simply associate a complete Ising model with each
triangulation of the ensemble, viewed as a lattice, by putting the spins at
the vertices of the triangles,
and then perform the sum over such triangulations, leading to
\beq
\tilde{Z}(K,H,N)=\sum_{T_N}\;\sum_{\{\sigma_i\}_{i\in T_N}}e^{-\beta \mathcal H[\{\sigma\}]}.
\label{sumtri}
\eeq
The situation where the volume is allowed to fluctuate is also of interest in
a quantum gravity context and described by the grand canonical partition function
\beq
\tilde{\mathcal Z}(K,H)=\sum_N e^{-\lambda N}\tilde{Z}(K,H,N),
\eeq
where the role of chemical potential is played by the cosmological constant $\lambda$.
As usual in dynamically triangulated models, the asymptotic behaviour as function of
$N$ of the canonical partition function in the infinite-volume limit $N\rightarrow\infty$
will determine the critical value of $\lambda$ at which the continuum limit can be performed.

In trying to understand the behaviour of the Ising model coupled to quantum
gravity in the form of causal dynamical triangulation, we will make crucial use
of the known probability distribution $P(q)$ of the coordination number $q$ of
vertices (the number $q$ of links meeting at a vertex) for the pure-gravity case.
This is relevant because we will be expanding about the point $\beta =0$, at
which the spin partition function (\ref{high-exp}) reduces to a trivial term $2^v$ and
the geometry is therefore that of the pure 2d Lorentzian gravity model.
The probability distribution in the thermodynamic limit was derived in \cite{aal1},
resulting in
\beq
P(q)=\frac{q-3}{2^{q-2}}.
\label{vertprob}
\eeq
\begin{figure}[ht]
\centering
\vspace*{13pt}
\includegraphics[width=11cm]{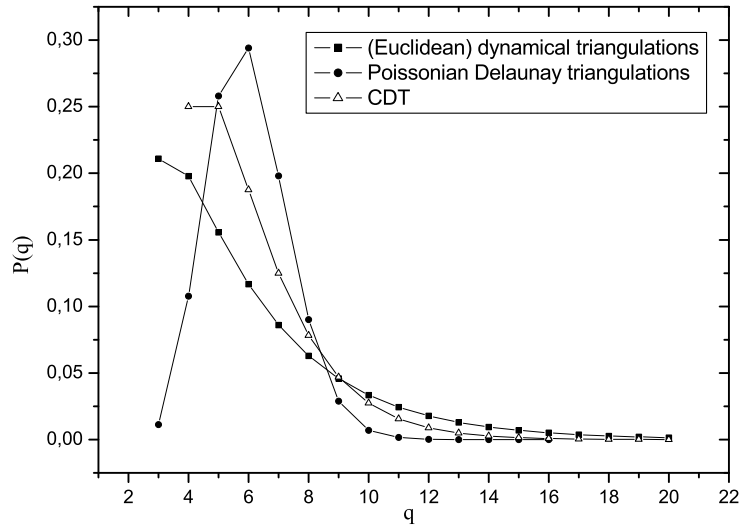}
\vspace*{13pt}
\caption{\footnotesize Comparing the probability distributions $P(q)$ of the
vertex coordination numbers $q$ for three different types of random triangulations:
Euclidean dynamical triangulations \cite{randomedt}, Poissonian Delaunay triangulations
\cite{drouffe} and causal dynamical triangulations \cite{aal1}.}
\label{coord-distr}
\end{figure}
The main idea is to use this distribution, together with some information about correlations
between distributions at different vertices, to re-express the sum over triangulations
in the high-temperature expansion of the partition function, (\ref{sumtri}), by an
average over coordination numbers. In terms of geometric randomness, the pure
CDT model lies in between the Voronoi-Delaunay triangulations
based on Poissonian random distributions of vertices and the planar
triangulations underlying the approach of
Euclidean dynamical triangulations. A comparative plot of the probability distributions
for the coordination numbers in the three different types of geometry
is shown in Fig.\ \ref{coord-distr}.

Let us point out a further subtlety which arises from the fact that the underlying
lattices are not fixed, but fluctuating, and therefore do not have a fixed ``shape".
In this case it can happen that in the graph counting for the high-temperature
expansion of some thermodynamic quantity topologically non-trivial graphs must
be taken into account. By this we mean closed non-contractible graphs which wind
around the
spacetime. This does not invalidate the method in
principle, but requires a more detailed knowledge of the global properties of the
spacetime. In two-dimensional CDT, where the spacetime topology is usually
chosen to be that of a cylinder (a spatial circle moving in time), this would be
closed loop graphs which wind one or more times around the spatial $S^1$, and
might be as short as a single link. Such pinching configurations certainly exist,
but their contribution to the graph counting has been shown to be irrelevant in
CDT, because it is subleading in $N$ in the thermodynamic limit \cite{aal1}.

\section{Graph embeddings} \label{graph-theory}

\subsection{Terminology of graph theory and embeddings}

To prepare the ground for our counting prescription, we will first
review some of the relevant terminology of graph theory and embeddings,
along the lines of reference \cite{domb1}.

A linear graph is a collection of $v$ vertices and $l$ lines (or edges)
connecting pairs of vertices.
A {\it simple graph} is a graph in which two vertices are connected by at most one line
and in which single lines are not allowed to form closed loops.
A graph is said to be {\it connected} if there is at least one path of lines between any two
vertices, and {\it disconnected} if for some vertex pair there is no such path.

The {\it cyclomatic number} $c$ of a connected graph $g$ is defined as
\beq
c(g)=l-v+1,
\eeq
and represents the number of independent cycles in the graph.

The {\it degree} (or {\it valence} or {\it coordination number}) of a vertex is the number
of edges incident on that vertex.

Two graphs are said to be {\it isomorphic} if they can be put into one-to-one
correspondence in such a way that their vertices and edges correspond. They are
called {\it homeomorphic} if they are isomorphic after insertion or suppression of any
number of vertices of degree 2 (this operation being defined on an edge in a trivial way).

\begin{figure}[ht]
\centering
\vspace*{13pt}
\includegraphics[width=12cm]{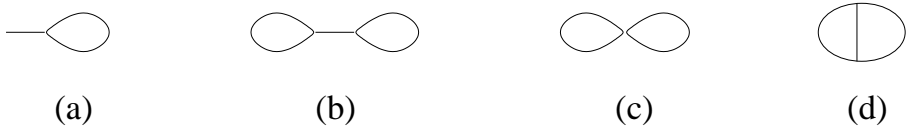}
\vspace*{13pt}
\caption{\footnotesize Four kinds of homeomorphs: (a) tadpole, (b) dumbbell, (c) figure eight,
and (d) $\theta$-graph.}
\label{topol}
\end{figure}

Homeomorphs are thus graphs with the same topology, in particular, with the same
cyclomatic number.
It is then possible to classify graphs in terms of {\it irreducible} graphs. An irreducible
graph is one in which all vertices of degree 2 have been suppressed.
(Note that in general it will no longer be a simple
graph since it will contain single-line loops.)
Typical irreducible graphs whose homeomorphs we will encounter are
{\it tadpoles, dumbbells, figure eights}
and {\it $\theta$-graphs} (Fig.\ \ref{topol}).

In the high-temperature expansion of an Ising model it is useful to think of the lattice
with its $v$ vertices and $l$ edges as a graph $G$. The graphs appearing in the
expansion of the thermodynamic function of interest are then graphs embedded in $G$.
A general definition of embeddings of graphs is the following.
Let $G$ be a general graph. A graph $H$ is a {\it subgraph} of $G$ when all its vertices
and edges are vertices and edges of $G$. An {\it embedding}\footnote{To be precise,
this defines a {\it weak} embedding. By contrast,
a {\it strong} embedding of $g$ in $G$ is defined as any section
graph $G^+$ of $G$ which is isomorphic to $g$, where a section
graph is a subgraph of $G$ consisting of a subset $A$ of vertices
and all the edges which connect pairs of (nearest-neighbour) vertices
of $A$. Since all our calculations will concern weak embeddings and the
corresponding weak lattice constants,
we will from now on drop this specification.}
of a graph $g$ in $G$ is a subgraph $G'$ of $G$ which
is isomorphic to $g$.

The {\it lattice constant} $(g;G)^{(N)}$ of a graph $g$ on $G$ is defined as the number
of different embeddings of $g$ in $G$. In practice it is often useful to work with
the lattice
constant {\it per site}, defined as $(g;G)=\frac{2}{N}(g;G)^{(N)}$.

With these definitions in hand we can now rewrite the function $F^{(N)}(u,\tau)$
introduced in (\ref{high-exp}) as
\beq
F^{(N)}(u,\tau)\equiv 1+\sum_{s=0}^{\infty}\sum_{l>0} D_{l,2s}^{(N)}u^l\tau^{2s},
\eeq
where $D_{l,2s}^{(N)}$ are the lattice constants of graphs of length $l$ and with $2s$ odd
vertices, for which $F^{(N)}(u,\tau)$ serves as a generating function.
It follows from the extensive nature of the free energy of the system that
in the thermodynamic limit
\beq
F^{(N)}(u,\tau)\stackrel{N\rightarrow\infty}{ \longrightarrow} {\rm e}^{N\Theta (u,\tau)}=
1+ N\Theta(u,\tau) +O(N^2),
\label{non}
\eeq
where the function $\Theta (u,\tau)$ does not depend on $N$.
From the usual combinatorial theory of non-embedded graphs it is well known that the
logarithm of the generating function is the generating function for connected graphs.
Because of the embedded nature of the graphs in the case at hand,
the disconnected graphs will still contribute to $\log F^{(N)}(u,\tau)$ with a
``repulsion" term of sign $(-1)^{(\# connected\ components)-1}$.

In our analysis of the Ising model on CDT we will be looking at a quantum-gravitational
average of the magnetic susceptibility at zero external field. For the
usual Ising model the susceptibility $\chi$ per unit volume is given by
\beq\label{chi}
   \chi(u)=\frac{1}{N}\frac{\partial^2\ln Z_N}{\partial H^2}_{\big|H=0}=
   \frac{1}{2}+\frac{2}{N}\frac{\sum_l D_{l,2}^{(N)}u^l}{1+\sum_l D_{l,0}^{(N)}u^l},
\eeq
where in the second step we have substituted in the high-temperature expansion
of the partition function. By virtue of (\ref{non}) it is clear that the susceptibility is
independent of $N$ in the infinite-volume limit (thus justifying our notation
$\chi(u)$), and that moreover the denominator in the last term in (\ref{chi})
does not contribute at lowest order in $N$, which is the one relevant
to the computation. To calculate the susceptibility, it is therefore sufficient
to compute the term of order $N$ in $\sum_l D_{l,2}^{(N)}u^l$, as is well known.

The computation in the gravity-coupled case proceeds completely analogously.
Performing the sum over all triangulations at fixed $N$, as in eq.\ (\ref{sumtri}),
and then letting $N\rightarrow\infty$, one obtains
\beq\label{chigrav}
   \chi_{\rm CDT}(u)\approx
   \frac{2}{N}\, \frac{\sum_l\left( \sum_{T_N}D_{l,2}^{(N)}(T_N)\right) u^l}
   {\sum_{T_N}1+\sum_l \left( \sum_{T_N}D_{l,0}^{(N)}(T_N)\right) u^l}=
    \frac{2}{N}\, \frac{\sum_l\langle D_{l,2}^{(N)}\rangle u^l}
   {1+\sum_l \langle D_{l,0}^{(N)}\rangle u^l}
\eeq
for the susceptibility in
presence of the gravitational CDT ensemble, where we have dropped the irrelevant
constant term and introduced an obvious notation for the ensemble average in
the second step. Whereas it was fairly straightforward to verify the cancellations
of terms of higher order in $N$ between numerator and denominator in the pure
Ising case in eq.\ (\ref{chi}) by an explicit calculation, the analogous computation
does not seem feasible in the gravity-coupled case, although we know from
the same general arguments that it must be realized here too. Again the terms
in the denominator of the expressions on the right-hand side of (\ref{chigrav}) do
not contribute at lowest order, which -- at least for this particular observable --
implies that its annealed and quenched gravitational averages coincide.

\subsection{Useful results on lattice constants}

We now want to identify which graph topologies appear in the
high-temperature expansion of the magnetic susceptibility.
For convenience we can divide our set of graphs into two subsets,
graphs with zero cyclomatic number (also called Cayley trees), and graphs
with $c>0$. Since the graphs we encounter in the susceptibility
series must have two and only two odd vertices, it follows that the first subset will
only contain connected non-selfintersecting open chains (in other words, self-avoiding
walks), while the second will contain everything else.
The connected graphs in the second subset are graphs with loops and one or no open ends,
while the disconnected ones take the form of a union of one of the previous two kinds
with any number of even graphs (graphs in which every vertex has even degree).

The problem we meet in the high-temperature expansion is the evaluation of the
lattice constants of such graphs. It turns out that there are two kinds of theorems
which can be applied profitably in the context of our random triangulated lattices.
The first is very general and reduces the calculation of the lattice constants for
disconnected graphs into that for connected graphs. This makes it possible to
compute the lattice constants for the second subset ($c>0$) in a unified way.
The second theorem is less general but powerful, since it gives us a recurrence
relation for the expansion coefficients of the susceptibility.

\begin{figure}[ht]
\centering
\vspace*{13pt}
\includegraphics[width=13cm]{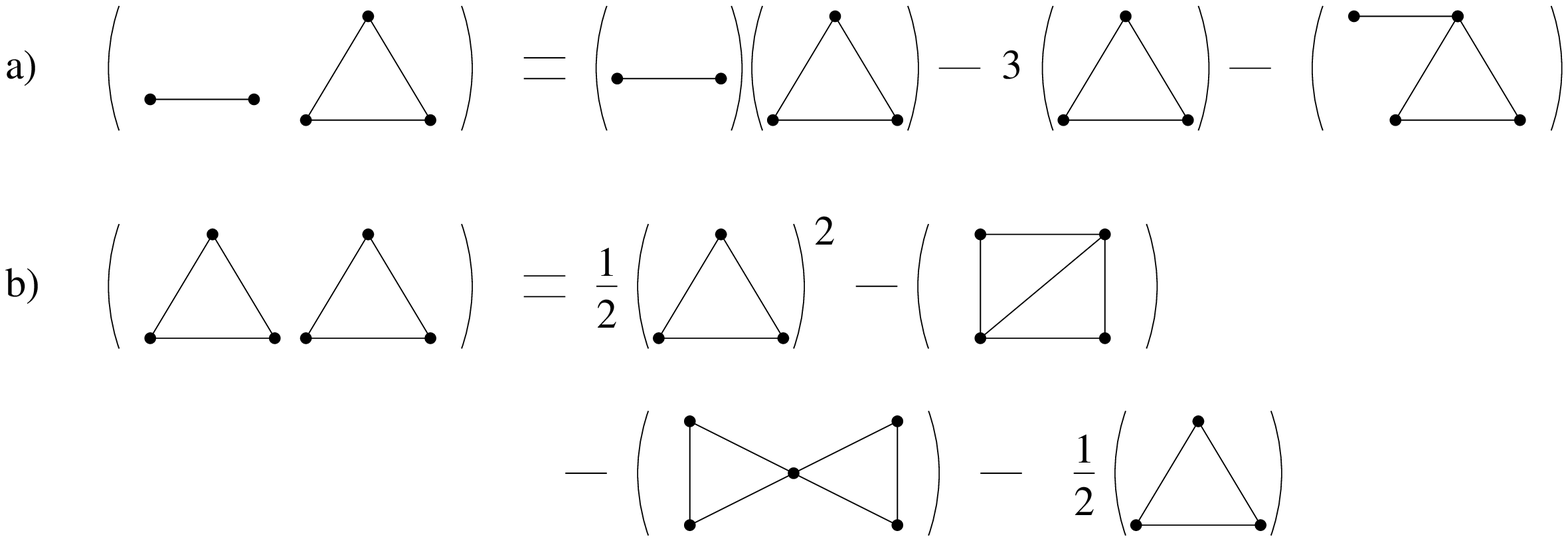}
\vspace*{13pt}
\caption{\footnotesize Two examples of the reduction theorem. The graph $G$
has been dropped in the notation.}
\label{reduct}
\end{figure}

The {\it reduction theorem} for disconnected graphs states that if $g_i$ and $g_j$
are two graphs $g_i\neq g_j$ and $G$ is any graph, then
\beq \label{reduction}
(g_i\cup g_j;G)=(g_i;G)(g_j;G)-\sum_k\{ g_i+g_j=g_k\}(g_k;G),
\eeq
where $g_i\cup g_j$ stands for the disjoint union of the two graphs,
$\{ g_i+g_j=g_k\}$ is the number of possible choices
of embeddings of $g_i$ and $g_j$ in $g_k$ having $g_k$ as their sum graph, and the
summation is over all graphs $g_k$ obtainable in this way.
If $g_i= g_j$ we can compute the right-hand side of (\ref{reduction}) as if the two graphs
had different colours
and thus obtain $2 (g_i\cup g_i;G)$. If $g_i$ or $g_j$ are themselves
non-connected graphs we can iterate
the theorem. In this way the lattice constant of the disconnected union of $n$ connected
graphs can be expressed
as a polynomial of order $n$ in the lattice constants of connected graphs (see Fig.\ \ref{reduct}
for some simple examples).

The {\it counting theorem} for the susceptibility series was first given by Sykes \cite{sykes} in 1961,
but only later proved by Nagle and Temperley \cite{nagletemp} as a special case of a more
general theorem of graph combinatorics. Closer examination reveals that the proof
makes no reference to the structure of the lattice, but only to its coordination number $q$,
which will turn out to be useful when considering the lattice dual to the triangulation.
The theorem states that the high-temperature series for the susceptibility can be written as
\beq
\chi(u)=1+(1-\sigma u)^{-2}\left[ q u (1-\sigma u)-2 (1-u^2) S_1 + 8(1+u)^2(S_2+S_3)\right],
\label{expa}
\eeq
where $q=\sigma +1$ is the coordination number, and the functions $S_i$ are defined as
\beq
\begin{split}
S_1 &= \sum_{l>0} l d_l u^l,\\
S_2 &= \sum_{l>0}\left[\sum_{w>0} w d_l(w)\right] u^l ,\\
S_3 &=\sum_{l>0} \left[\sum_{r\leq r'} rr'm_l(r,r')\right] u^l.
\end{split}
\eeq
The coefficient $d_l$ is the sum of lattice constants of even graphs with $l$ lines,
while $d_l(v)$
is restricted to even graphs with $l$ lines and vertices of degree $2q_1, 2q_2,...$,
characterized by the
weight $w=\sum_i q_i(q_i-1)/2$.
The function $m_l(r,r')$ is the sum of lattice constants of graphs with exactly two odd
vertices of degrees $2r+1$
and $2r'+1$.
Substituting the left-hand side of formula (\ref{expa}) by the expansion
$\chi (u)= 1+\sum_{n>0} a_n u^n$, one finds
a recursive relation for the coefficients $a_n$, namely,
\beq \label{recurr}
\begin{split}
a_l-2\sigma a_{l-1} &+\sigma^2 a_{l-2} = 2(l-2)d_{l-2}-2ld_l+ \\
&+8\left[\sum_{w>0} w d_l(w)+2\sum_{w>0} w d_{l-1}(w)
+\sum_{w>0} w d_{l-2}(w)\right]+\\
&+8\left[\sum_{r\leq r'} rr'm_l(r,r')+2\sum_{r\leq r'} rr'm_{l-1}(r,r')+
\sum_{r\leq r'} rr'm_{l-2}(r,r')\right].
\end{split}
\eeq
As a result, at every new order $l$ in our calculation the only new lattice constants we
have to evaluate are those
for graphs with no vertex of degree 1, reducing the computational effort considerably.

This formula can in principle be applied to any non-regular lattice whose vertices are
all of the same degree. For a fixed non-regular lattice this observation would usually
not be of much help, because one would still need to compute at each order the new
lattice constants depending on the complete, detailed information of the lattice geometry.
In our case this difficulty is not present, since we are summing over triangulations
and can simply substitute in (\ref{recurr}) the
lattice constants {\it averaged} over the triangulations, without the need of keeping
track of the lattice geometry for individual lattices.

\subsection{Counting graphs on the CDT triangular lattice}

Next, we will analyze the evaluation procedure for lattice constants in the
quantum-gravita\-tio\-nal CDT model.
Computations here are made difficult by the randomness of the coordination
number.
At the outset it is not even obvious that any of the known recurrence relations can be used.
Our task is to list all possible diagrams, and for each count the number of ways
it can be embedded in the lattice. On a regular lattice such an operation is tedious
but straightforward: starting from a fixed vertex, we trace out all possible sequences
of links of a given length, say, which do not self-intersect.
The discrete symmetries of the regular lattice simplify this task greatly.
In the simplest case, the counting has to be done only for a single initial vertex, with
all other graphs obtained subsequently by translational symmetry.
The analogous counting on the dynamical CDT lattices
is complicated by the fact that the lattice
neighbourhoods of different vertices will in general look different. We will deal
with this difficulty by setting up an algorithm to count the embedding of a given
diagram on the {\it ensemble} of CDT lattices, making use of the known
probability distribution of the vertex coordination numbers.

\begin{figure}[ht]
\centering
\vspace*{13pt}
\includegraphics[width=8cm]{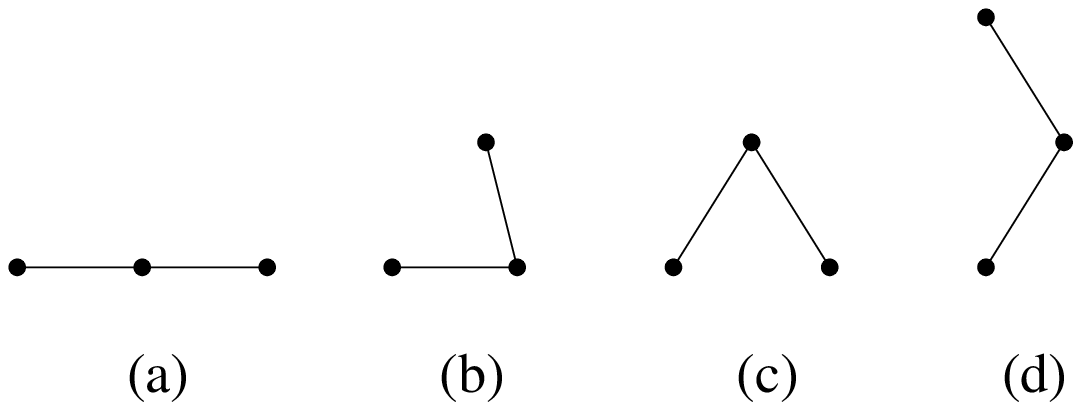}
\vspace*{13pt}
\caption{\footnotesize The possible typologies of embedding, up to reflection symmetry,
of a length-2 open chain.
The angle which the time-like links form with the vertical has no relevance for
the graph counting.}
\label{embed2}
\end{figure}

\begin{figure}[ht]
\centering
\vspace*{13pt}
\includegraphics[width=8cm]{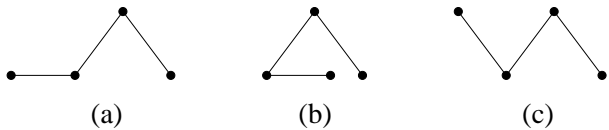}
\vspace*{13pt}
\caption{\footnotesize Three of the different typologies of possible embeddings of
a length-3 open chain.}
\label{embed3}
\end{figure}

Before presenting this algorithm we need some more notation.
CDT lattices have two kinds of links, time-like and space-like\footnote{
Even after performing the Wick rotation to Euclidean signature, these two
link types remain combinatorially distinguishable because of the special way
in which the original Lorentzian simplicial geometries were constructed.}.
In keeping with the usual representation of two-dimensional CDT lattices (c.f.
Fig.\ \ref{cdt}, left-hand side), we will draw time- and space-like links in embeddings
of graphs as (diagonally) upward-pointing and horizontal lines, respectively
(see Fig.\ \ref{embed2}). Furthermore we need to keep track of the relative
up-down or right-left orientation of consecutive links, to distinguish between
graphs like (c) and (d) in Fig.\ \ref{embed2} (which give
different contributions to the lattice constant).
In addition, the counting for a graph like (b) will be identical to that for its
mirror images under left-right and up-down reflections. These mirrored graphs will
be counted by multiplying the embedding constant of the graph by an
appropriate {\it symmetry factor}.
A symmetry factor 1 is assigned to graphs symmetric under both up-down
and left-right reflections (for example, graphs $(a)$ and $(d)$), a factor 2 to graphs
with only one of the two symmetries (like graph $(c)$) or symmetric with respect
to a composition of the two
(like graph $(c)$ of Fig.\ \ref{embed3}), and a factor 4 to graphs without
reflection symmetry (like graph $(b)$
of Fig.\ \ref{embed2} or $(a)$ and $(b)$ of Fig.\ \ref{embed3}).

The lattice constant $(g;T)$ of a graph $g$ embedded in a graph corresponding
to a triangulation $T$ is computed as
\beq\label{lat-cnst}
(g;T)=\sum_i s_i (g;T)_i,
\eeq
where $(g;T)_i$ is the number of ways a particular typology of embedding $(g)_i$
can be realized on $T$,
$s_i$ is the relevant symmetry factor, and the sum extends over all possible
typologies of embeddings of the graph $g$.
In the context of quantum gravity, we are interested in the average of
(\ref{lat-cnst}) over all triangulations, that is,
\beq
\langle g\rangle = \frac{\sum_T(g;T)}{\sum_T 1},
\eeq
which will take into account the
probability distribution of the vertex coordination.

\begin{figure}[ht]
\centering
\vspace*{13pt}
\includegraphics[width=8cm]{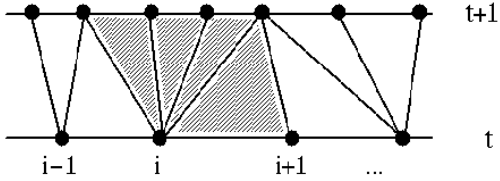}
\vspace*{13pt}
\caption{\footnotesize A strip in a triangulation of the CDT ensemble, and, in shadow, the triangles
contributing to the weight at the vertx $i$.}
\label{slice}
\end{figure}

As mentioned earlier, all calculations will be performed in the thermodynamic limit
where the cosmological constant is tuned to its critical value $\lambda_c=\ln 2$.
In this limit, the probability of having $k$ incoming
time-like links at a particular lattice vertex is given by
\beq \label{prob}
p(k)=\frac{1}{2^k},
\eeq
with an identical probability for having $k$ {\it outgoing} time-like links
at a vertex. This probability distribution can be derived easily in the following way.
At a vertex there is always at least one outgoing link, which can be identified with the right
vertical edge of the upward-pointing triangle to the left of the vertex (see Fig.~\ref{slice}).
To the right of such edge
there will be always another triangle, which in the critical limit will be with probability $1/2$
either upward- or downward-pointing. If it is upward-pointing, the edge in common to the two triangles
is the only vertical link coming out of the vertex in question, such a situation having probability $1/2$.
If it is a downward-pointing one, we can continue to its right and repeat the observation until a
upward-pointing triangle is found at the $k$-th step, giving a vertex with $k$ vertical outgoing links,
such a situation having probability $1/2^k$.

Moreover, {\it at one and the same vertex} the two probabilities (incoming and outgoing) are independent of each
other. In fact, not just for a single vertex are these two probabilities independent,
but the same is true for all vertices lying in the same space-like, horizontal slice.
By contrast, the outgoing probability at a vertex will in general condition the incoming
probability at another vertex on the subsequent horizontal slice. By construction,
the probability of having a space-like link to the left and to the right of a given
vertex is equal to 1.

Armed with this information we are now ready to compute any of the $(g;G)_i$ in
(\ref{lat-cnst}). We do not have a general counting formula in closed form, but
will formulate a number of rules and tools which will enable us to
do the counting recursively.
We will start with some illustrative examples. Examples 3 and 4 will serve as
our elementary building blocks in more complicated constructions.\\

\noindent\textbf{Example 1.} The simplest example is that of a length-1 graph $c_1$,
consisting of
a single horizontal or vertical link. The number of horizontal embeddings
is the number of horizontal links on a CDT lattice, which by virtue of the
Euler relation is given by half the number of triangles, i.e. $N/2$.
If we give the lattice constant per vertex, as is customary in the literature,
it is therefore $\langle c_1\rangle_h=1$.
By a similar topological argument we can immediately deduce the number
2 for the lattice constant of vertical embeddings, but it is instructive to compute it
from the probability distribution (\ref{prob}) instead. If at a vertex we have $k$
outgoing vertical links, there are precisely
$k$ ways to embed the one-link graph such that it emanates in upward direction from
this vertex. Since every vertical embedding is accounted for in this way, by
assigning the relevant probability and summing over
$k$ we obtain
\beq
\langle c_1\rangle_v=\sum_{k=1}^{\infty}\frac{k}{2^k}=2,
\eeq
in agreement with the earlier argument. The total contribution to the
lattice constant at order 1 is therefore
$\langle c_1\rangle_h+\langle c_1\rangle_v=3$.\\

\noindent\textbf{Example 2.} Next, let $c_2$ be a length-2 open chain, and consider the embedding
typology $c_{2,b}$ shown in Fig.\ \ref{embed2}$(b)$.
The result follows straightforwardly from example 1: there is one horizontal link per vertex,
and there are in the ensemble on average two ways of attaching an outgoing vertical
line to its right vertex, yielding $\langle c_2\rangle_b=2$.
For the embedding of Fig.\ \ref{embed2}$(a)$ there is nothing to compute; we have
$\langle c_2\rangle_a=1$. The embedding
of Fig.\ \ref{embed2}$(d)$ is the first configuration we encounter which extends over
two triangulated strips. Because of the independence
of the probabilities (\ref{prob}) in successive strips we obtain
\beq
\langle c_2\rangle_d= (\sum_{k=1}^{\infty}\frac{k}{2^k})^2=4.
\eeq
\\
\noindent\textbf{Example 3.} Now consider the embedding $c_{2,c}$ depicted in
Fig.\ \ref{embed2}$(c)$. If the vertex on top
has $k$ incoming links, there are exactly $k(k-1)/2$ ways to realize the embedding,
leading to
\beq
\langle c_2\rangle_c=\sum_{k=1}^{\infty}\frac{k (k-1)}{2^{k+1}}=2.
\eeq
Note that the same ``inverted-V" embedding on a flat triangular lattice
$T_{\rm reg}$ would give $(c_2;T_{\rm reg})_c=1$ instead.
Putting everything together and multiplying by the appropriate symmetry
factors we get
$\langle c_2\rangle_a+4\langle c_2\rangle_b+2\langle c_2\rangle_c+\langle c_2\rangle_d=17$
as a total contribution at order 2.

\begin{figure}[ht]
\centering
\vspace*{13pt}
\includegraphics[width=10cm]{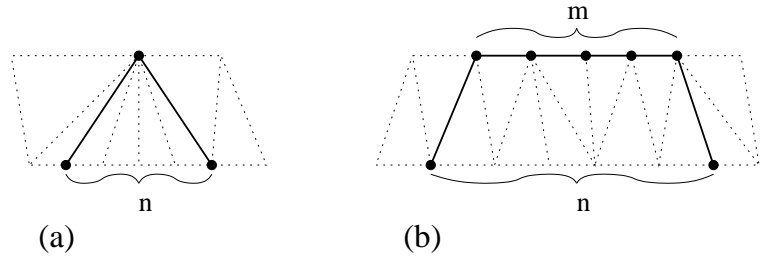}
\vspace*{13pt}
\caption{\footnotesize Two configurations contributing to the calculations of
examples 3 and 4. The labels m and n count the numbers of space-like links as
indicated.}
\label{kernel}
\end{figure}

In view of more complicated graphs, an alternative and more convenient way
of doing the calculation in example 3 is by focussing
on the probability distributions of the vertices lying on the {\it lower} space-like slice
of the triangulated strip (see also Fig.\ \ref{kernel}$(a)$).
The probability of having $n$ space-like links in between the two vertices
at the ends of the inverted V is given by
the probability that there are $n-1$ vertices in between with precisely one outgoing
link each, which is $1/2^{n-1}$.
Thus we obtain again
\beq
(c_2;G)_c=2\sum_{n=1}^{\infty}\frac{1}{2^{n-1}}=4.
\eeq
\\
\noindent\textbf{Example 4.} Consider an open chain of length greater than
2 embedded in such a way
that only the first and last edges lie along time-like links, with a horizontal chain of $m$ space-like
links in between. Both time-like links are supposed to lie in the same strip of the triangulation,
as illustrated in Fig.\ \ref{kernel}$(b)$. We are going to prove that the lattice constant of
such a configuration is given by
\beq\label{c_mn}
\langle c_{m+2}\rangle_b=\sum_{n=1}^{\infty}\frac{1}{2^{m+n-1}}\matrice m+n \\ m
\ematrice = 4- 2^{1-m}.
\eeq
The result $4-2^{1-m}$ could be obtained easily by (i) considering
the number of ways in which each of the two vertices at the top left and right can have an
incoming link (giving a total of $2\times 2=4$ possibilities) and (ii) subtracting from that
the probability for the two time-like links touching each other in their lower vertex,
which is $2^{1-m}$ (the probability that the $m-1$ intervening vertices have only one
incoming link each). However, we rather want to do the
counting in a way that keeps track of the number of links separating the two vertices
at the bottom, for reasons that will become clear soon.
In other words, we want to determine the lattice constant for a closed polygon
consisting of two
vertical edges, $m$ edges along the top and $n$ along the bottom.
From a combinatorial point of view this can be rephrased as the problem of counting
the number of different ways in which $n$ triangles and $m$ upside-down triangles can
be arranged to form a strip, with the well-known
binomial result $\big({m+n\atop m}\big)$.
To get the lattice constant for our random lattice we still have to include a probabilistic
factor $\frac{1}{2}$ for each vertical link in the polygon interior, leading to the factors
$2^{1-m-n}$ in formula (\ref{c_mn}).\\

\begin{figure}[ht]
\centering
\vspace*{13pt}
\includegraphics[width=10cm]{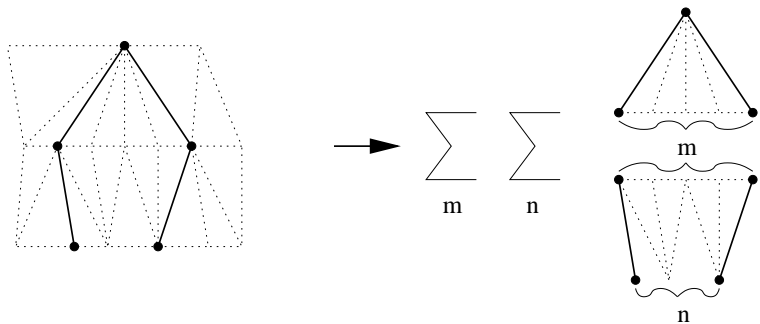}
\vspace*{13pt}
\caption{\footnotesize Example of a strip decomposition. The embedding on the left is split
into two strips, after which the strip contributions for given $m$ are multiplied pairwise
and summed over.}
\label{convol}
\end{figure}

\begin{figure}[ht]
\centering
\vspace*{13pt}
\includegraphics[width=10cm]{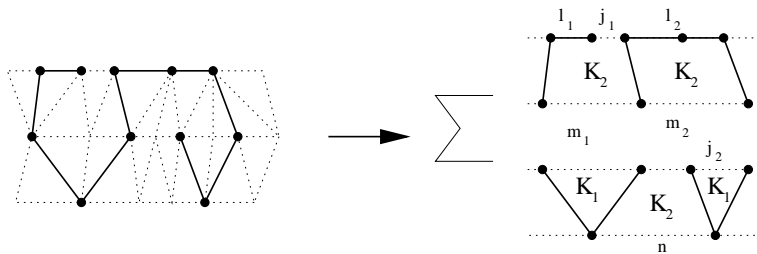}
\vspace*{13pt}
\caption{\footnotesize Example of a more complicated strip decomposition where each strip
consists of a sequence of elements of the kind illustrated in Fig.\ \ref{kernel}.
The sum refers to the total sum in the expression (\ref{ex-stripdec})
which has to be performed after the proper weights have been assigned
to each element of the graph.}
\label{convol2}
\end{figure}

\noindent\textbf{The strips decomposition.}
The embedding typology $(g)_i$ of any connected graph $g$ extends over a
well-defined number of strips in the CDT triangulation, and each part of $(g)_i$
belongs to a definite strip or horizontal slice.
Our first step in the counting procedure is to decompose the embedded graphs
into strip contributions, thus breaking the graph into pieces.
The example of Fig.\ \ref{convol} will help us explain the procedure.
It shows a particular embedding of $c_4$, how it is decomposed into two
pieces, and how its lattice constant can be computed accordingly.
All we have to do is multiply the probability of $c_{2,c}$ (example 2)
with $m$ links at the bottom with the probability of $c_{m+2,b}$ (example 3),
and then sum over $m$, resulting in
\beq
\langle c_4\rangle_x = \sum_{m=1}^{\infty}\frac{1}{2^{m-1}} (4- 2^{1-m}) = \frac{20}{3}.
\eeq
We can think of this operation as a product of an (infinite-dimensional)
vector and matrix,
\beq\label{k1k2}
\begin{split}
K_1(m) &=\frac{1}{2^{m-1}},\\
K_2(m,n) &=\frac{1}{2^{m+n-1}}\matrice m+n \\ m \ematrice,
\end{split}
\eeq
followed by taking the trace in order to impose the open boundary condition,
that is,
\beq
\langle c_4\rangle_x = {\rm Tr} ( K_1\cdot K_2).
\eeq
Fig.\ \ref{convol} is a simple case with only two vertical lines in each strip. In general
there will be more lines, like in Fig.\ \ref{convol2}.
However, there are only two possibilities for pairs of neighbouring vertical links:
they can be either disjoint or have exactly one vertex in common,
so that the probability associated with them in a particular embedding in a
particular triangulation will be given by either $K_2(m,n)$ or $K_1(n)$.
We can then associate a probability to a given pattern of vertical lines in a strip with
fixed distance between them (in terms of the horizontal links in the in- and out-slice),
which is the product of the corresponding $K_1$- or
$K_2$-probabilities, times a factor $\frac{1}{2}$ for each
shared vertical line between two neighbouring $K_i$-patterns in the same
strip (see Fig.\ \ref{convol2}).
In the end we have to glue the strips back together again,
with conditions such as to get the desired graph, and
sum over the allowed values of horizontal lengths.
As an illustrative example, we obtain for Fig.\ \ref{convol2} the multiple sum
\beq \label{ex-stripdec}
\sum_{j_1=1}^{\infty}\sum_{m_1=1}^{\infty}\sum_{m_2=2}^{\infty}
\sum_{j_2=1}^{m_2-1}\sum_{n=1}^{\infty}
[K_2(l_1+j_1,m_1) \frac{1}{2} K_2(l_2,m_2)] [K_1(m_1) \frac{1}{2} K_2(m_2-j_2,n) \frac{1}{2} K_1(j_2)],
\eeq
where we have grouped the factors corresponding to the same strip in square brackets.
As should be clear from this example, the gluing and the range of the summations
for the various graph embeddings
will have to be taken care of on a case-by-case basis.

Using these techniques we have repeated and independently
confirmed the calculation of the first five orders given in \cite{aal1}
and then extended it to order 6. Table~\ref{tab-coeff} gives the lattice constants
(times 2) per number of vertices (which we will
call the {\it susceptibility coefficients})
of the open chains, the other graphs, and of their grand total.
The effort involved in this extension is by no means minor.
There are 387 order-6 open chains that have to be counted individually,
with the identifications introduced in this section,
and for many of them the computation requires a careful use of the strip decomposition and
a subsequent evaluation of the sums involved.

\begin{table}[hbtp]
\begin{center}
\begin{tabular}{|r||r|r|r|}
\hline
{\rule[-3mm]{0mm}{8mm} $n$} & \hspace{1cm} open & closed + disconnected &
\hspace{.6cm} total ($a_n$)\\
\hline\hline
{\rule[-3mm]{0mm}{8mm} 1} & 6 & 0 & 6 \\
 \hline
{\rule[-3mm]{0mm}{8mm} 2} & 34 & 0 & 34 \\
\hline
{\rule[-3mm]{0mm}{8mm} 3} & 174 & 0 & 174 \\
 \hline
{\rule[-3mm]{0mm}{8mm} 4} & $859 \frac{1}{3}$ & $-12$ & $847 \frac{1}{3}$\\
 \hline
{\rule[-3mm]{0mm}{8mm} 5} & $4152 \frac{2}{3}$ & $-162 \frac{2}{3}$ & 3990\\
 \hline
{\rule[-3mm]{0mm}{8mm} 6} & $19800 \frac{19}{27}$ & $-1416 \frac{1}{9}$ & $18384 \frac{16}{27}$\\
  \hline
\end{tabular}
\caption{\footnotesize Susceptibility coefficients for the CDT lattice up to order 6.
The terminology `open', `closed' and `disconnected' refers to the topology
of the graphs. }\label{tab-coeff}
\end{center}
\end{table}

\subsection{Counting graphs on the dual CDT lattice}

On the dual lattice (an example of which is depicted in Fig.\ \ref{cdt}, right-hand side)
we can apply formula (\ref{recurr}), which for $q=3$ reduces to
\beq
\begin{split}
a_l-4\sigma a_{l-1} +4 a_{l-2} &= 2(l-2)d_{l-2}-2ld_l+ \\
&+8\left[m_l(1,1)+2m_{l-1}(1,1)+m_{l-2}(1,1)\right].
\end{split}
\eeq
Because of the low coordination number the only even graphs that can appear are
connected
or disconnected polygons, and the only contributions to $m_l$ are from $\theta$-graphs
and dumbbells
with or without disconnected polygonal components.
For the disconnected graphs we can use the reduction theorem. It should be noted that in the
overlap decomposition new topologies of graphs appear, for example,
graphs with more than two odd vertices,
still rendering the higher-order computations non-trivial.

It is not difficult to derive the probability distribution relevant for the dual lattice
from the one of the original lattice.
The probability of having two horizontal links at a vertex, one to the right and one to the
left, is again 1.
The probability for an incoming or outgoing vertical link at a vertex of a given
chain of horizontal links is given by the probability of having a triangle or an upside-down
triangle in the corresponding strip of the original triangulation.
Since the two cases are mutually exclusive, the in- and out-probabilities are
not independent and each takes the value $1/2$,
which may also be thought of
as the weight associated with a triangle in the original lattice.
With these rules the lattice constant of an open chain of length 1 is $3/2$:
1 for the horizontal
link (associated to its left vertex, say, in order to avoid overcounting),
plus $1/2$ for the vertical link. Alternatively,
this easy example can be computed by use of the Euler relation.
For higher-order graphs it is often convenient to go back to the original lattice
and count -- with the appropriate
weights -- the configurations which can be associated to the dual graph under consideration.\\

\begin{figure}[ht]
\centering
\vspace*{12pt}
\includegraphics[width=9cm]{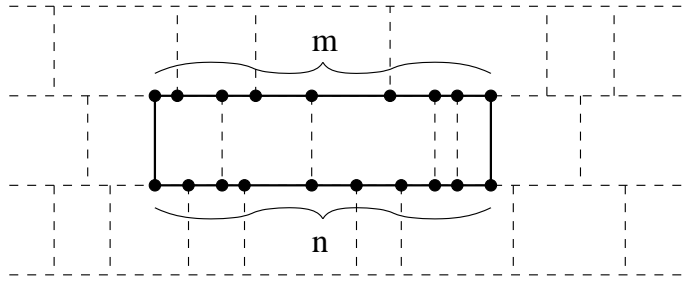}
\vspace*{13pt}
\caption{\footnotesize An embedding of the polygon considered in example 1.}
\label{dualpoly1}
\end{figure}

\noindent\textbf{Example 1.} Let us consider again a closed polygon with only two
vertical links, with $m$ links on the top and $n$ on the bottom, but this time
on the dual CDT lattice (see Fig.\ \ref{dualpoly1}).
A careful evaluation of the probabilities involved yields
\beq
\langle c_{m,n}\rangle^*=\frac{1}{2^{m+n+1}}\sum_{k=0}^{min(m,n)-1}\matrice m-1 \\
k \ematrice\matrice n-1 \\ k \ematrice
\label{bino}
\eeq
for its lattice constant, where $k$ is the number of vertical links of the lattice which
lie inside the polygon,
and $\langle\cdot\rangle^*$ denotes the average lattice constant per vertex of the
dual graph.
The binomials in (\ref{bino}) count the number of possible ways to arrange the
$k$ internal links, and the remaining probability -- expressed as a power of
$1/2$ -- turns out not to depend on $k$.

As already mentioned, the in-and out-probabilities at a vertex of the dual lattice are
not independent, which means that
we cannot apply the strip decomposition of the original lattice directly, but instead
have to proceed more carefully.\\

\begin{figure}[ht]
\centering
\vspace*{12pt}
\includegraphics[width=9cm]{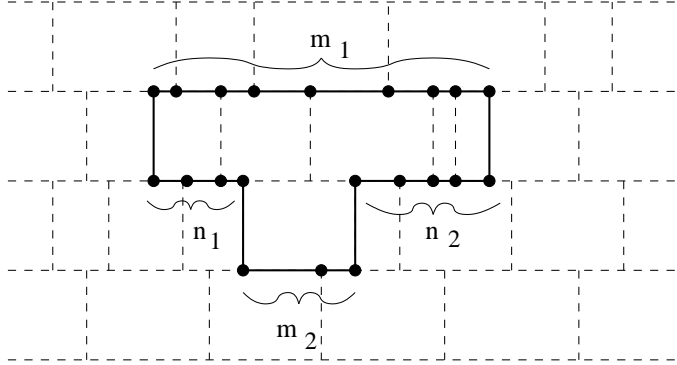}
\vspace*{13pt}
\caption{\footnotesize An embedding of the polygon considered in example 2.
}
\label{dualpoly2}
\end{figure}

\noindent\textbf{Example 2.} As an example of this, consider a polygon
embedding that extends over
two strips (Fig.\ \ref{dualpoly2}), a type of configuration that appears from order 8 onward. Its lattice
constant is given by
\beq
\begin{split}
\langle c_{m_1,m_2,n_1,n_2}\rangle^* &=\sum_{k_1=0}^{m_1-1}\matrice m_1-1 \\ k_1 \ematrice
\sum_{k_2=0}^{m_2-1}\matrice m_2-1 \\ k_2 \ematrice \times \\
& \times \sum_{i=0}^{min(k_1,n_1+n_2-4)} \matrice k_1-i+k_2 \\ k \ematrice
\matrice n_1+n_2-2 \\ i \ematrice \frac{1}{2^{m_1+m_2+n_1+n_2+k_1-i+k_2+2}}.
\end{split}
\label{twostrip}
\eeq
In eq.\ (\ref{twostrip}), $m_1$ and $m_2$ count the links on the top and bottom horizontal
lines, and $n_1$ and $n_2$ count the two sets of contiguous links on the central horizontal line.
The numbers of internal vertical links in the upper and lower strip are denoted by
$k_1$ and $k_2$, while $i$ counts how many links
out of the $k_1$ end on one of the two intermediate horizontal lines.
The logic behind the various counting factors appearing under the sums
is very similar to that of the previous example.

We have  found the method illustrated by examples 1 and 2 the most compact
one to deal with the counting of graph embeddings on the dual CDT lattice. It has
enabled us to push our ``counting by hand" to order 12, the results of which
are given in Table~\ref{tab-dual-coeff}. Because of the fixed, low coordination
number of the dual vertices, there are far fewer graphs at a given order than
there are on the original CDT lattice.
\begin{table}[hbtp]
\begin{center}
\begin{tabular}{|r||r|r|r|r|r|r|r|r|r|r|r|r|}
\hline
{\rule[-3mm]{0mm}{8mm} $n$} & 1 & 2 & 3 & 4 & 5 & 6 & 7 & 8 & 9 & 10 & 11 & 12\\
\hline\hline
{\rule[-3mm]{0mm}{8mm} $a_n$} & 3 & 6 & 12 & 23 & 42+$\frac{3}{4}$ & 78+$\frac{1}{2}$ & 142+$\frac{3}{4}$
& 258 & 461+$\frac{13}{16}$ & 820+$\frac{1}{8}$ & 1446+$\frac{13}{32}$ & 2532+$\frac{11}{16}$\\
\hline
\end{tabular}
\caption{\footnotesize Susceptibility coefficients for the dual CDT lattice up to order 12.}
\label{tab-dual-coeff}
\end{center}
\end{table}

\section{Series analysis} \label{series}

\subsection{Review of different methods}

Having computed the two high-temperature series to some order, we will now turn
to analyzing them, and try to extract information on the critical properties of the underlying
gravity-matter systems. The standard analytic methods are well-known from
the case of regular lattices and reviewed in
\cite{guttmann}. We will recall them briefly here in order to be self-contained.

The first and most straightforward tool is that of the {\it ratio method}, which works as follows.
Assuming a simple behaviour of the susceptibility of the form
\beq
\chi (u)\sim A(u) \Bigl( 1-\frac{u}{u_c}\Bigr)^{-\gamma}+B(u)
\eeq
near the critical point $u_c$, with analytic functions $A$ and $B$, its series expansion $\chi (u)=
1+\sum_{n>0} a_n u^n$ should yield (to first order in $1/n$)
\beq\label{ratio}
r_n\equiv\frac{a_n}{a_{n-1}}=
\frac{1}{u_c} \Bigl( 1+\frac{\gamma - 1}{n}\Bigr).
\eeq
One can then generate sequences of estimates of the critical parameters from
sequences of point pairs
$\{r_n,r_{n-1}\}$ \cite{guttmann}, namely,
\beq\label{seq1}
\gamma_n=\frac{n(2-n)r_n+(n-1)^2r_{n-1}}{n r_n-(n-1)r_{n-1}},
\eeq
\beq\label{seq2}
u_{c,n}=\frac{1}{n r_n-(n-1)r_{n-1}}.
\eeq
In general, these will converge very slowly. In addition, if the series has an
oscillatory behaviour the method can yield alternating over- and underestimates.
In this case a fit of a whole sequence $\{r_{min},...,r_{max}\}$, computed with
the help of (\ref{ratio}), may be better suited than the sequence of estimates
to find the straight line masked by the oscillations ($r_{min}$ is chosen
properly to exclude large deviations from (\ref{ratio}) at small $n$, and $r_{max}$ is the
ratio computed for $n=n_{max}$).

Oscillations and other irregularities in the expansion are caused by a more
complicated behaviour of the thermodynamic function, for example,
the presence of other singularities in the complex plane close to the physical
singularity, which in unfortunate cases may even lie closer to the origin than
the singularity of interest.
In case the behaviour near the physical singularity is like
\beq
\chi (u)\sim A(u) \Bigl( 1-\frac{u}{u_c}\Bigr)^{-\gamma},
\eeq
with $A(u)$ a {\it meromorphic} function with singularities close to $u_c$,
a method that should give
better results than the ratio method is that of the so-called {\it Pad\'e approximants}.
The method consists in the approximation of a function, known through its series expansion
to order $\cal N$, by a ratio of two polynomials of order $\cal L$ and $\cal M$,
subject to the condition $\cal L+M\leq N$,
\beq\label{pade}
\frac{P_{\cal L}(x)}{Q_{\cal M}(x)}\equiv\frac{\sum_{k=0}^{\cal L} p_k x^k}
{1+\sum_{k=1}^{\cal M} q_k x^k}=F_{\cal N}(x)+O(x^{{\cal L+M}+1})
\eeq
with $F_{\cal N}(x)=\sum_{k=0}^{\cal N} a_k x^k$. By common usage, the notation
$\cal [L/M]$ indicates
the order of the polynomials used.

In our specific case one takes as the function $F(x)$ to be approximated
the derivative of the logarithm of $\chi(u)$,
so that $u_c$ can be recognized as a pole and $\gamma$ as the associated residue
in
\beq
\frac{d}{du}\log \chi(u)=\frac{\gamma}{u_c-u}\left(1+O(u_c-u)\right).
\eeq
This is also referred to as the Dlog Pad\'e method.
Usually one only looks at the tridiagonal band $ [({\cal N}-1)/{\cal N}]$, $\cal [N/N]$ and
$[({\cal N}+1)/{\cal N}]$ because of the invariance of the diagonal Pad\'e approximants
under Euler transformations.

The Pad\'e approximants work well only when there is no additive term $B(u)$.
In presence of such a term there is
a generalization -- known as {\it differential approximants} -- to account for
functional behaviours of the form
\beq
\chi (u)\sim \prod_{i=1}^n A_i \Bigl( 1-\frac{u}{u_i}\Bigr)^{-\gamma_i} + B(u),
\eeq
with $B(u)$ and $A_1,...,A_n$ analytical functions and $u_1,...,u_n$ a set of singular points.
If we are considering the logarithmic derivative of a function $f(x)$
we can rewrite (\ref{pade}) as
\beq\label{dlog}
P_{\cal L}(x) f(x)-Q_{\cal M}(x) f'(x)=O(x^{{\cal L+M}+1}).
\eeq
The idea of the differential approximants method is to generalize this equation
according to
\beq\label{diff-approx}
R_{{\cal M}_2}(x)F''_{\cal N}(x)+Q_{{\cal M}_1}(x) F'_{\cal N}(x)-P_{\cal L}(x)
F_{\cal N}(x)=S_{\cal K}(x)+O(x^{{\cal K+L}+{\cal M}_1+{\cal M}_2+1}),
\eeq
where $R_{\mathcal{M}_2}(x)$ and $S_{\mathcal{K}}(x)$ are two other
polynomials
of order $\mathcal{M}_2$ and $\mathcal{K}$, and where
one can substitute the differential operator by $D=x\frac{d}{dx}$,
forcing the point at the origin to be a regular singular point.
In the following we will only consider the special case
$R_{\mathcal{M}_2}\equiv 0$,
giving rise to the so-called {\it inhomogeneous 1st-order
differential approximant}, denoted by
$[\mathcal{K}/\mathcal{L};\mathcal{M}_1]$.
With this method, the exponent $\gamma$ can be evaluated as
\beq
\gamma=\frac{P_{\mathcal{L}}(x_c)}{x_c Q'_{\mathcal{M}_1}(x_c)},
\label{gammada}
\eeq
where $x_c$ is a simple root\footnote{The case of multiple roots can also
be considered, but in our case never occurs.}
of $Q_{\mathcal{M}_1}(x)$, which gives an estimate
of the critical point.

\subsection{The CDT lattice series}

\begin{figure}
\centering
\includegraphics[width=11cm]{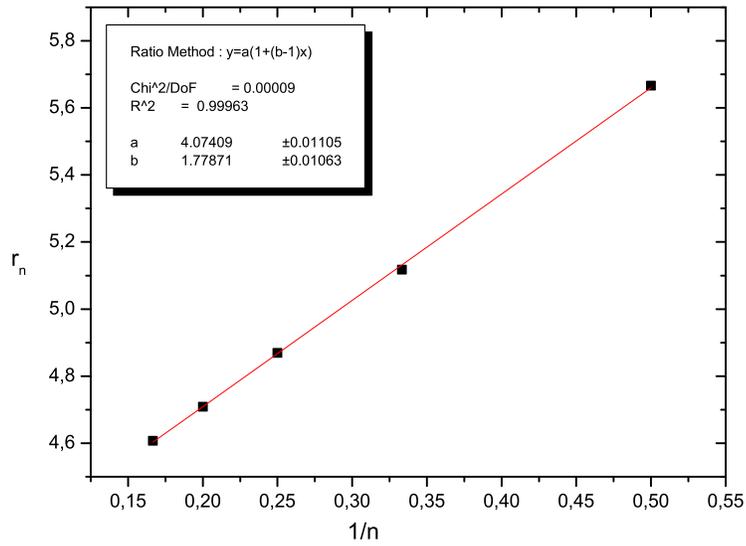}
\caption{\footnotesize Plot of the ratios (\ref{ratio}) for the case of the CDT lattice
(data from Table~\ref{tab-fits}).}
\label{ratioplot1}
\end{figure}

The high-temperature series expansion for the Ising susceptibility on the CDT model is
given by
\beq \label{h-series}
\chi_{\rm CDT}(u)=1+6u+34u^2+174u^3+(847+\frac{1}{3})u^4+3990u^5+
(18384+\frac{16}{27})u^6+O(u^7).
\eeq
Applying formulas (\ref{seq1}) and (\ref{seq2}) we get the sequence of estimates for
$\gamma$ and $u_c$ reported in Table~\ref{tab-2ptseq}.
\begin{table}[hbtp]
\begin{center}
\begin{tabular}{|c||c|c|}\hline
{\rule[-3mm]{0mm}{8mm} $n$ }
& \hspace{1cm} $u_c \hspace{1cm}$ & \hspace*{25pt} $\gamma$ \hspace*{25pt}\\ \hline\hline
{\rule[-3mm]{0mm}{8mm} 3} & 0.2489 &
1.819\\ \hline
{\rule[-3mm]{0mm}{8mm} 4} & 0.2424 &
1.721\\ \hline
{\rule[-3mm]{0mm}{8mm} 5} & 0.2459 &
1.791\\  \hline
{\rule[-3mm]{0mm}{8mm} 6} & 0.2438 &
1.740\\  \hline
\end{tabular}
\caption{\footnotesize Two-point linear extrapolations of the critical point and the critical exponent
of the series (\ref{h-series}) obtained using  (\ref{seq1}) and (\ref{seq2}).}\label{tab-2ptseq}
\end{center}
\end{table}
The results of linearly fitting the whole sequence of ratios $\{r_2,...,r_{max}\}$ instead
are listed in Table~\ref{tab-fits},
and a plot of the ratios $\{r_2,...,r_6\}$ versus $1/n$ together with their fit is shown in
Fig.\ \ref{ratioplot1}. Despite the shortness of the sequence, the linear fit is of very
good quality.
\begin{table}[hbtp]
\begin{center}
\begin{tabular}{|c||c|c|}\hline
{\rule[-3mm]{0mm}{8mm} $n_{\rm max}$ }
& \hspace{1cm} $u_c$  \hspace{1cm} & \hspace*{25pt} $\gamma$ \hspace*{25pt}\\
\hline\hline
{\rule[-3mm]{0mm}{8mm} 3} & \hspace{.2cm} 0.2488
\hspace{.2cm} &
1.820\\ \hline
{\rule[-3mm]{0mm}{8mm} 4} & 0.2462 &
1.789\\ \hline
{\rule[-3mm]{0mm}{8mm} 5} & 0.2458 &
1.783\\  \hline
{\rule[-3mm]{0mm}{8mm} 6} & 0.2454 &
1.779\\  \hline
\end{tabular}
\caption{\footnotesize Linear fits of the sequences $\{r_2,...,r_{max}\}$ for the series
(\ref{h-series}) with assumed functional form (\ref{ratio}).}\label{tab-fits}
\end{center}
\end{table}
It is worth noting that if we remove the point $r_2$, which of course is subject
to the largest deviations from a pure $1/n$-behaviour due to higher-order terms,
and fit the truncated sequence $\{r_3,...,r_6\}$, we obtain $\gamma=1.749$ and
$u_c=0.2441$ (see Table~\ref{tab-fits2}).

\begin{table}[hbtp]
\begin{center}
\begin{tabular}{|c||c|c|}\hline
{\rule[-3mm]{0mm}{8mm} $n_{\rm max}$ }
& \hspace{1cm} $u_c$  \hspace{1cm} & \hspace*{25pt} $\gamma$ \hspace*{25pt}\\
\hline\hline
{\rule[-3mm]{0mm}{8mm} 4} & 0.2424 &
1.721\\ \hline
{\rule[-3mm]{0mm}{8mm} 5} & 0.2439 &
1.745\\  \hline
{\rule[-3mm]{0mm}{8mm} 6} & 0.2441 &
1.749\\  \hline
\end{tabular}
\caption{\footnotesize Linear fits of the sequences $\{r_3,...,r_{max}\}$ for the
series (\ref{h-series}) with assumed
functional form (\ref{ratio}).}\label{tab-fits2}
\end{center}
\end{table}
It is surprising how fast the ratio sequence seems to converge to the result
$\gamma=1.75$ of the exact solution for the flat regular case,
as compared to the analogous one obtained from the series expansion on the plane
triangular lattice (see \cite{sykes-h} for the expansion coefficients to order $u^{16}$), even
though we cannot claim the result to be conclusive because of the limited number of
terms at our disposal (see Fig.\ \ref{gamma}).
\begin{figure}[ht]
\centering
\includegraphics[width=11cm]{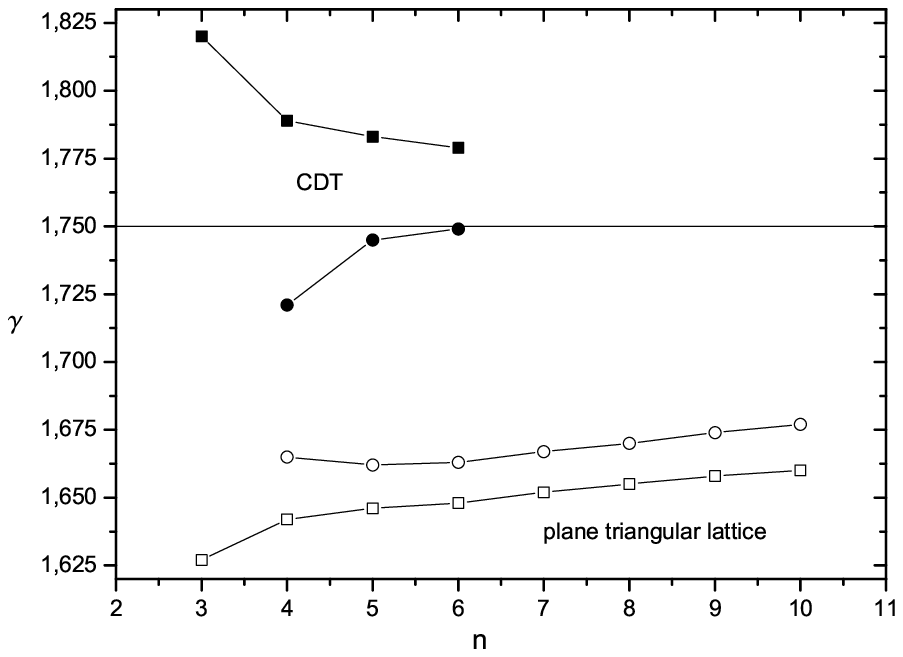}
\caption{\footnotesize A comparative plot of estimates of the critical exponent
$\gamma$ for the Ising model on the CDT versus that on a plane triangular lattice
(data taken from \cite{sykes-h}), obtained via the ratio method.
The filled shapes refer to the CDT model and the open
ones to the plane triangular lattice.
The squares indicate fits of $\{r_{n_{min}},...,r_n\}$ sequences, and the circles
those of sequences $\{r_{n_{min}+1},...,r_n\}$ with the first ratio eliminated.}
\label{gamma}
\end{figure}

What is the picture when we use one of the alternative methods to evaluate the
series expansions? Results from the Dlog Pad\'e and differential approximants
methods are somewhat inconclusive, most likely because of the small number of
estimates we can perform with our 6th-order series (see Table~\ref{tab-pade}).
To get a feel for what may be expected at this order, we report in Table~\ref{tab-pt}
the corresponding Dlog Pad\'e approximants $[({\cal N}+j)/\cal N]$ to the series for the
plane triangular lattice (see again \cite{sykes-h}). The estimates for the critical
exponent $\gamma$ in the latter case are clearly closer to the known exact value,
whereas for the CDT model, with only two values for the
diagonal $\cal [N/N]$, it is quite impossible to extrapolate the behaviour of $\gamma$.
(Note that we are reporting the estimates for the critical point $u_c$ only for
completeness; they are or course not expected to coincide for the two models.)
We have also computed the inhomogeneous first-order differential
approximants $[{\cal L}/({\cal N}+j);\cal N]$ for the CDT series, which was defined
just before formula (\ref{gammada}). Values for the critical
exponents are not completely off,  but there are simply too few of them to make
any statement about their convergence behaviour (see Table~\ref{tab-diff}).

\begin{table}[hbtp]
\begin{center}
\begin{tabular}{|c||c|c|c|c|c|c|}\hline   &
\multicolumn{2}{|c|}{{\rule[-3mm]{0mm}{8mm}
$[({\cal N}-1)/{\cal N}]$}} & \multicolumn{2}{|c|}{{\rule[-3mm]{0mm}{8mm}
$[{\cal N}/{\cal N}]$}} & \multicolumn{2}{|c|}{{\rule[-3mm]{0mm}{8mm}
$[({\cal N}+1)/{\cal N}]$}} \\
\hline
{\rule[-3mm]{0mm}{8mm} ${\cal N}$} & $u_c$ & $\gamma$ & $u_c$ & $\gamma$ & $u_c$ & $\gamma$ \\
\hline\hline
{\rule[-3mm]{0mm}{8mm} 1} & 0.1875 & 1.125 & 0.2540 & 2.064 & 0.2513 & 2.000 \\
\hline
{\rule[-3mm]{0mm}{8mm} 2} & 0.2514 & 2.003 & 0.2483 & 1.8810 & 0.2517 & 2.008 \\
\hline
{\rule[-3mm]{0mm}{8mm} 3} & 0.2519 & 2.012 & - & - & - & - \\
\hline
\end{tabular}
\caption{\footnotesize Dlog Pad\'e approximants method applied to the CDT
series (\ref{h-series}).}\label{tab-pade}
\end{center}
\end{table}
\begin{table}[hbtp]
\begin{center}
\begin{tabular}{|c||c|c|c|c|c|c|}\hline   &
\multicolumn{2}{|c|}{{\rule[-3mm]{0mm}{8mm}
$[({\cal N}-1)/{\cal N}]$}} & \multicolumn{2}{|c|}{{\rule[-3mm]{0mm}{8mm}
$[{\cal N}/{\cal N}]$}} & \multicolumn{2}{|c|}{{\rule[-3mm]{0mm}{8mm}
$[({\cal N}+1)/{\cal N}]$}} \\
\hline
{\rule[-3mm]{0mm}{8mm} ${\cal N}$} & $u_c$ & $\gamma$ & $u_c$ & $\gamma$ & $u_c$ & $\gamma$ \\
\hline\hline
{\rule[-3mm]{0mm}{8mm} 1} & 0.2500 & 1.500 & 0.2666 & 1.706 & 0.2678 & 1.730 \\
\hline
{\rule[-3mm]{0mm}{8mm} 2} & 0.2679 & 1.732 & 0.2670 & 1.712 & 0.2661 & 1.688 \\
\hline
{\rule[-3mm]{0mm}{8mm} 3} & 0.2662 & 1.692 & 0.2667 & 1.705 & 0.2672 & 1.722 \\
\hline
\end{tabular}
\caption{\footnotesize Dlog Pad\'e approximants method applied to the plane triangular
series in \cite{sykes-h}.}\label{tab-pt}
\end{center}
\end{table}

\begin{table}[hbtp]
\begin{center}
\begin{tabular}{|c||c||c|c|}\hline   & &
\multicolumn{2}{|l|}{{\rule[-3mm]{0mm}{8mm}
${\cal N}=$}}  \\
{\rule[-3mm]{0mm}{8mm} ${\cal L}$} & $j$ & 1 & 2 \\
\hline\hline
1 & -1 & $u_c=0.2143$ & \hspace*{3mm} 0.2454 \hspace*{3mm} \\
  &  & $\gamma=1.767$ & 1.815 \\
  & 0 & 0.2373 & 0.2461 \\
  &  & 1.568 & 1.832 \\
  & 1 & 0.2496 & - \\
  &  & 1.944 & - \\
\hline
2 & -1 & 0.2252 & 0.2463\\
  &  & 1.265 & 1.840 \\
  & 0 & 0.2529 & - \\
  &  & 2.087 & - \\
  & 1 & 0.2431 & - \\
  &  & 1.685 & - \\
\hline
3 & -1 & 0.2334 & - \\
  &  & 1.361 & - \\
  & 0 & 0.2402 & - \\
  &  & 1.577 & - \\
  & 1 & - & - \\
  &  & - & - \\
\hline
4 & -1 & 0.2357 & - \\
  &  & 1.396 & - \\
  & 0 & - & - \\
  &  & - & - \\
  & 1 & - & - \\
  &  & - & - \\
\hline
\end{tabular}
\caption{\footnotesize Inhomogeneous 1st-order differential approximants
$[{\cal L}/({\cal N}+j);{\cal N}]$ method for the  series (\ref{h-series}).}
\label{tab-diff}
\end{center}
\end{table}

\clearpage

\subsection{The dual CDT lattice series}

The high-temperature series expansion for the Ising susceptibility on the dual CDT model is
given by
\beq \label{h-dual-series}
\begin{split}
\chi_{\rm CDTd}(u) &=1+3u+6u^2+12u^3+23u^4+(42+\frac{3}{4})u^5+(78+\frac{1}{2})u^6+(142+\frac{3}{4})u^7\\
   +&258u^8+(461+\frac{13}{16})u^9+(820+\frac{1}{8})u^{10}+(1446+\frac{13}{32})u^{11}+(2532+\frac{11}{16})u^{12}+O(u^{13})
\end{split}
\eeq
Using expressions (\ref{seq1}) and (\ref{seq2}) we get the sequence of ratio-method
estimates for $\gamma$ and $u_c$ reported in Table~\ref{tab-dual-2ptseq}.
\begin{table}[hbtp]
\begin{center}
\begin{tabular}{|c||c|c|}\hline
{\rule[-3mm]{0mm}{8mm} $n_{\rm max}$ }
& \hspace{1cm} $u_c \hspace{1cm}$ & \hspace*{25pt} $\gamma$ \hspace*{25pt}  \\ \hline\hline
{\rule[-3mm]{0mm}{8mm} 3} & 0.5 &
1.\\ \hline
{\rule[-3mm]{0mm}{8mm} 4} & 0.6 &
1.6\\ \hline
{\rule[-3mm]{0mm}{8mm} 5} & 0.6147 &
1.713\\  \hline
{\rule[-3mm]{0mm}{8mm} 6} & 0.5800 &
1.390\\  \hline
{\rule[-3mm]{0mm}{8mm} 7} & 0.5842 &
1.436\\ \hline
{\rule[-3mm]{0mm}{8mm} 8} & 0.5782 &
1.360\\ \hline
{\rule[-3mm]{0mm}{8mm} 9} & 0.6057 &
1.758\\  \hline
{\rule[-3mm]{0mm}{8mm} 10} & 0.6064 &
1.769\\  \hline
{\rule[-3mm]{0mm}{8mm} 11} & 0.6093 &
1.820\\  \hline
{\rule[-3mm]{0mm}{8mm} 12} & 0.6202 &
2.033\\  \hline
\end{tabular}
\caption{\footnotesize Two-point linear extrapolations of the critical point and the critical exponent
of the dual series (\ref{h-dual-series}) obtained using (\ref{seq1}) and (\ref{seq2}).}
\label{tab-dual-2ptseq}
\end{center}
\end{table}

\begin{figure}
\centering
\includegraphics[width=11cm]{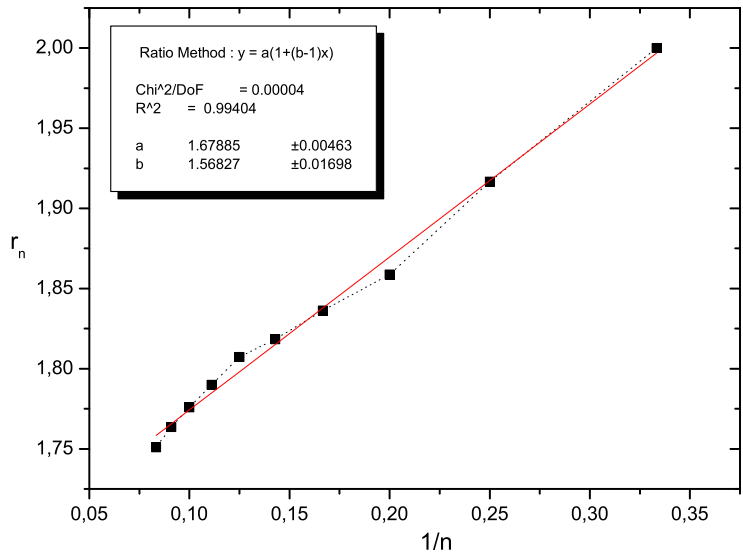}
\caption{\footnotesize Plot of the ratios (\ref{ratio}) for the case of the dual CDT lattice.}
\label{ratioplot2}
\end{figure}

\begin{figure}
\centering
\includegraphics[width=11cm]{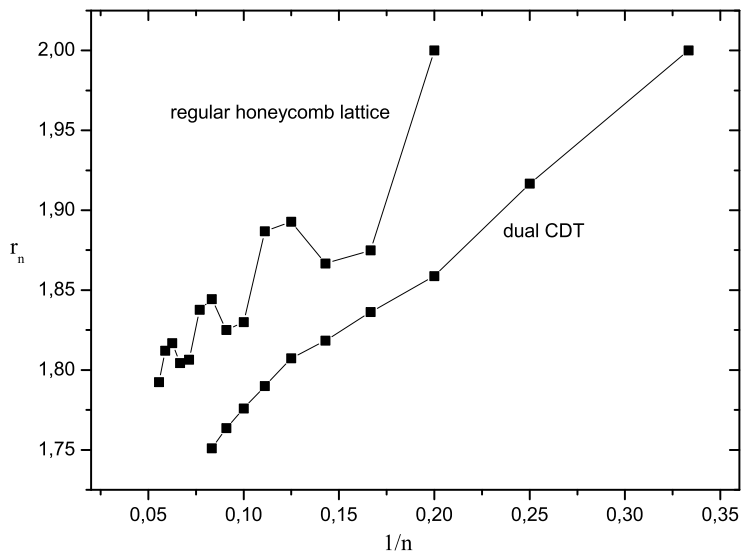}
\caption{\footnotesize A comparative plot of the ratios of the regular honeycomb
lattice (data taken from \cite{sykes-h})
versus the dual CDT model.}
\label{HCvsDual}
\end{figure}

\noindent The relative ratio plot is shown in Fig.\ \ref{ratioplot2}, and shows gentle
oscillations around the best linear fit of the sequence
$\{ r_3,...,r_{12}\}$, leading to the estimates $\gamma=1.568$ and $u_c=0.5956$.
However, these numbers do not tell it all. The strong qualitative differences with
the planar honeycomb lattice, the appropriate non-random
version of the dual CDT lattice, are illustrated in Fig.\ \ref{HCvsDual}. It shows a
comparative plot of the ratios extracted for the two models (the expansion coefficients
for the honeycomb lattice, known to order $u^{32}$, can be found in \cite{sykes-h}).
Obviously, the influence of interfering unphysical singularities present on the
regular honeycomb lattice is much reduced in the fluctuating dual CDT ensemble.


On the basis of the rather well-behaved results for the Ising susceptibility
on dynamically triangulated spacetimes we conjectured in \cite{mio-ising1}
that ``coupling to quantum gravity" may be an optimal method to learn
about the critical behaviour of a matter or spin system, in the sense that
the randomness of the underlying geometry eliminates spurious unphysical
singularities (due to the presence of discrete symmetries in the case of
regular lattices), but is not strong enough to change the universality
class of the matter system. This is a dynamical version of similar
conjectures made in the 1980s, which advocated the use of {\it fixed}
random lattices to improve the convergence behaviour of lattice gauge theory,
say (see, for example, \cite{random}), but ultimately did not succeed.

If it is indeed the case that the singularity structure of the thermodynamic
quantities in the complex plane is simplified, this could explain that the
simple ratio method does indeed give the best result, at least at the relatively
low orders we have been considering. In the case of the Ising model coupled
to Euclidean dynamical triangulations, an analogous result has already been
obtained with regard to the locus of the zeros of the partition function
in the complex-temperature plane, where unphysical singularities
of the flat regular lattices have been shown to be largely absent \cite{fat}.
However, as already mentioned, the
underlying geometries are too disordered to serve our present purpose, because
their critical matter behaviour is changed compared to the flat case.    \\

\begin{table}[hbtp]
\begin{center}
\begin{tabular}{|c||c|c|c|c|c|c|}\hline   &
\multicolumn{2}{|c|}{{\rule[-3mm]{0mm}{8mm}
$[({\cal N}-1)/{\cal N}]$}} & \multicolumn{2}{|c|}{{\rule[-3mm]{0mm}{8mm}
$[{\cal N}/{\cal N}]$}} & \multicolumn{2}{|c|}{{\rule[-3mm]{0mm}{8mm}
$[({\cal N}+1)/{\cal N}]$}} \\
\hline
{\rule[-3mm]{0mm}{8mm} ${\cal N}$} & $u_c$ & $\gamma$ & $u_c$ & $\gamma$ & $u_c$ & $\gamma$ \\
\hline\hline
{\rule[-3mm]{0mm}{8mm} 3} & 0.5952 & 1.554 & 0.5574 & 0.9738 & 0.5686 & 1.170 \\
\hline
{\rule[-3mm]{0mm}{8mm} 4} & 0.5691 & 1.180 & 0.5564 & 0.9567 & 0.5861 & 1.446 \\
\hline
{\rule[-3mm]{0mm}{8mm} 5} & 0.5890 & 1.492 & 0.5935 & 1.560 & 0.6359 & 2.996 \\
\hline
{\rule[-3mm]{0mm}{8mm} 6} & 0.5761 & 1.402 & - & - & - & - \\
\hline
\end{tabular}
\caption{\footnotesize Dlog Pad\'e approximants method for the dual CDT series (\ref{h-dual-series}).}\label{tab-dual-pade}
\end{center}
\end{table}
\begin{table}[hbtp]
\begin{center}
\begin{tabular}{|c||c|c|c|c|c|c|}\hline   &
\multicolumn{2}{|c|}{{\rule[-3mm]{0mm}{8mm}
$[({\cal N}-1)/{\cal N}]$}} & \multicolumn{2}{|c|}{{\rule[-3mm]{0mm}{8mm}
$[{\cal N}/{\cal N}]$}} & \multicolumn{2}{|c|}{{\rule[-3mm]{0mm}{8mm}
$[({\cal N}+1)/{\cal N}]$}} \\
\hline
{\rule[-3mm]{0mm}{8mm} ${\cal N}$} & $u_c$ & $\gamma$ & $u_c$ & $\gamma$ & $u_c$ & $\gamma$ \\
\hline\hline
{\rule[-3mm]{0mm}{8mm} 3} & - & - & - & - & 0.6063 & 2.153 \\
\hline
{\rule[-3mm]{0mm}{8mm} 4} & 0.5589 & 1.416 & 0.5680 & 1.531 & 0.5676 & 1.525 \\
\hline
{\rule[-3mm]{0mm}{8mm} 5} & 0.5676 & 1.525 & 0.5679 & 1.530 & 0.5589 & 1.459 \\
\hline
{\rule[-3mm]{0mm}{8mm} 6} & 0.5645 & 1.490 & 0.5730 & 1.618 & 0.5709 & 1.571 \\
\hline
\end{tabular}
\caption{\footnotesize Dlog Pad\'e approximants method for the honeycomb lattice series
in \cite{sykes-h}.}
\label{tab-hc}
\end{center}
\end{table}

\begin{table}[hbtp]
\begin{center}
\begin{tabular}{|c||c||c|c|c|}\hline   & &
\multicolumn{3}{|l|}{{\rule[-3mm]{0mm}{8mm}
${\cal N}=$}}  \\
{\rule[-3mm]{0mm}{8mm} $\cal L$} & $j$ & 2 & 3 & 4 \\
\hline\hline
1 & -1 & $u_c=0.5000$ & \hspace*{3mm} 0.6246 \hspace*{3mm}& \hspace*{3mm} 0.5873 \hspace*{3mm}\\
  &  & $\gamma=1.000$ & 2.1435 & 1.503\\
  & 0 & 0.4974 & 0.5722 & 0.4797\\
  &  & 1.018 & 1.205 & 2.688\\
  & 1 & 0.5782 & 0.5759 & 0.6024\\
  &  & 1.298 & 1.234 & 1.828\\
\hline
2 & -1 & 0.4967 & 0.5767 & 0.6094\\
  &  & 1.023 & 1.312 & 1.932\\
  & 0 & 0.5735 & 0.5802 & 0.5768\\
  &  & 1.265 & 1.395 & 1.029\\
  & 1 & 0.5753 & 0.5620 & 0.4902\\
  &  & 1.276 & 0.9503 & 2.201\\
\hline
3 & -1 & 0.5722 & 0.5810 & 0.5966\\
  &  & 1.216 & 1.414 & 1.622\\
  & 0 & 0.5715 & 5692 & 0.5010\\
  &  & 1.195 & 1.208 & 2.031\\
  & 1 & 0.5820 & 0.5606 & - \\
  &  & 1.436 & 0.9257 & - \\
\hline
4 & -1 & 0.5735 & 0.5502 & 0.4711\\
  &  & 1.237 & 1.062 & 2.966\\
  & 0 & 0.5867 & 0.5100 & - \\
  &  & 1.571 & 1.880 & - \\
  & 1 & 0.5376 & 0.6467 & - \\
  &  & 1.202 & 3.062 & - \\
\hline
5 & -1 & 0.5882 & complex & - \\
  &  & 1.572 & complex & - \\
  & 0 & complex & - & - \\
  &  & complex & - & - \\
  & 1 & complex & - & - \\
  &  & complex & - & - \\
\hline
\end{tabular}
\caption{\footnotesize Inhomogeneous 1st order differential approximants
$[{\cal L}/{\cal N}+j;{\cal N}]$ method for dual CDT
series (\ref{h-dual-series}).}
\label{tab-dual-diff}
\end{center}
\end{table}
\noindent Returning to the evaluation of the
dual CDT series, we have used both the Dlog Pad\'e and the differential
approximants methods, in addition to the ratio method already described.
Even with the larger number of terms compared to the original CDT case
the results are reluctant to show any clear sign of convergence. Our results
for the Dlog Pad\'e approximant are summarized in Table~\ref{tab-dual-pade}.
The corresponding data for the regular honeycomb lattice \cite{sykes-h},
which we are including for comparison in Table~\ref{tab-hc}, seem to show more
consistency at the same order of approximation.
The result from using the
inhomogeneous first-order differential approximants for the dual CDT lattice
(see Table~\ref{tab-dual-diff}) is similarly inconclusive. It would clearly be
desirable to understand in more analytic terms the influence of the
fluctuating geometry of the underlying quantum gravitating lattice on the
singularity structure of spin models like the one we have been
considering, and thus determine which of the approximation methods is
best suited for extracting their critical behaviour.

\section{Comments on the low-temperature expansion} \label{Ising-lowT}

Having dealt with the high-temperature expansion, it is natural to ask whether similar
expansion techniques can be applied in a low-temperature expansion of our
coupled models of matter and gravity, in the same way this is possible for
spin systems on regular lattices. The first thing to notice is that at the point of
zero temperature around which one expands, the spins are all frozen
to point into the same direction, and effectively play no role. The
quantum-gravitational ensemble is therefore
again characterized by the probability distribution (\ref{vertprob}) for the coordination number
of pure gravity, as was also the case in the limit of infinite temperature.

In the low-temperature expansion on the regular lattice
one starts from a ground state with all spins
aligned, and then includes perturbations with $s$ overturned spins,
obtaining \cite{sykes-l}
\beq\label{low-exp}
  \ln Z_N(K,H)=l K + v H + \sum_{s,r} [ s;r;G ] z^p \mu^s,
\eeq
in terms of the expansion parameters $z=e^{-2K}$ and $\mu=e^{-2H}$, where
$K$, $H$, $l$ and $v$ were all defined at the beginning of Sec.\ 3 above.
The part linear in $N$ of the (strong) lattice constant (see \cite{domb1} for a
definition) for a graph
with $s$ vertices and $r$ links embedded in $G$ is denoted by
$[s;r;G]$, and $p$ is the number of lattice links
which are incident on any of the vertices of the graph, but do not themselves belong to
the embedded graph.
On a lattice of fixed coordination number $q$ it is easy to prove that $p=qs-2r$.
Graphs with given values of $s$ and $r$ therefore contribute only at a specific order.

The low-temperature expansion on CDT lattices via graph counting
represents a slight complication. It is easy to see that already by overturning just
a single spin the number of links whose interaction changes sign as a result depends on the
spin's position, so that
each site will contribute to a different order in the perturbative
expansion.\footnote{By contrast, for the low-temperature expansion on {\it dual} CDT
lattices, where the coordination number is fixed to 3, the expansion parallels
that on flat lattices, with the (strong) lattice constants replaced by their
ensemble averages.}
In the case at hand, this difficulty
can be overcome by counting  the number of elementary
polygons (i.e. faces) with given boundary length $p$ {\it on the dual lattice}.
In the case of several overturned spins we can proceed similarly by associating
certain patterns on the dual lattice with a given order of $z$.
More precisely, to obtain the coefficient at order $z^p \mu^s$ we need to count
polygons on the dual lattice which have
total boundary length $p$ and are constructed out of $s$ faces.
Denoting the number of such patterns
by $\mathcal{P}(p,s)$, we can write the free energy per unit volume as
\beq\label{low-exp2}
  \frac{1}{N}\ln Z_N(K,H)=\frac{3}{2} K + \frac{1}{2} H + \sum_{s,p} \mathcal{P}(p,s) z^p \mu^s
\eeq
and the susceptibility at vanishing magnetic field as
\beq \label{low-chi}
  \chi(z)=\sum_{s,p} 4 s^2 \mathcal{P}(p,s) z^p.
\eeq
Although the procedure is slightly more involved, and cannot be seen as a
computation of (averaged) strong lattice constants, the expansion is still doable
using the counting techniques introduced earlier in this chapter.
Table~\ref{tab-low-coeff} gives the results for $\mathcal{P}(p,s)$ up to order 10.
\begin{table}[hbtp]
\begin{center}
\begin{tabular}{|c||c|c|c|c|}\hline
{\rule[-3mm]{0mm}{8mm}  } &  $s=1$ & $s=2$ & $s=3$ & $s=4$ \\ \hline\hline
{\rule[-3mm]{0mm}{8mm} $p=4$} & $\frac{1}{8}$ & - & - & - \\ \hline
{\rule[-3mm]{0mm}{8mm} $p=5$} & $\frac{1}{8}$ & - & - & - \\ \hline
{\rule[-3mm]{0mm}{8mm} $p=6$} & $\frac{3}{32}$ & $\frac{1}{32}$ & - & - \\ \hline
{\rule[-3mm]{0mm}{8mm} $p=7$} & $\frac{1}{16}$ & $\frac{1}{16}$ & - & - \\ \hline
{\rule[-3mm]{0mm}{8mm} $p=8$} & $\frac{5}{128}$ & $\frac{3}{64}$ & $\frac{1}{128}$ & - \\ \hline
{\rule[-3mm]{0mm}{8mm} $p=9$} & $\frac{3}{128}$ & $\frac{9}{64}$ & $\frac{7}{128}$ & - \\ \hline
{\rule[-3mm]{0mm}{8mm} $p=10$} & $\frac{7}{512}$ & $\frac{19}{512}$ & $\frac{29}{512}$ & $\frac{9}{512}$ \\ \hline
\end{tabular}
\caption{\footnotesize Free energy coefficients $\mathcal{P}(p,s)$ as defined in (\ref{low-exp2})
for the CDT lattice.}\label{tab-low-coeff}
\end{center}
\end{table}
The susceptibility low-temperature series (\ref{low-chi}) turns out to be very irregular
and none of the methods of analysis considered seems to give
a reasonable indication of the critical exponent, at least not from the relatively few terms
we have computed. Although this is somewhat disappointing, in view of the fact that
low-temperature expansions for regular lattices are notoriously ill-behaved,
it does not come as a total surprise.

What helps in the evaluation of the low-temperature series for
the Ising model on {\it flat} triangular, square or honeycomb lattices is the fact that
their critical temperatures are known exactly from duality arguments
and thus can be used as an input
in so-called biased approximants to considerably improve the estimates of
critical exponents. Namely, for
$H=0$ one can apply the standard duality transformation \cite{baxter} which maps
the low-temperature expansion (\ref{low-exp}) of the triangular lattice model to
the high-temperature
expansion (\ref{high-exp}) of its dual and vice versa.
This can be combined with the star-triangle transformation to obtain
the critical point. The analogous deriviation for the simple square lattice
is even easier, because it is self-dual.
Unfortunately, the star-triangle transformation is not applicable in the CDT case,
which prevents us from making a similar argument in the coupled Ising-gravity
model. In order to pursue the analysis of the low-temperature series further,
we would therefore have to rely on the evaluation of higher-order terms in
the expansion.

\section{Summary and outlook} \label{Ch2-conlusion}

In this chapter we have given a concrete example of a method which can
be used to extract physical properties of a strongly coupled model of
gravity and matter.
We showed how the method -- estimating critical matter exponents from
a series expansion of suitable thermodynamic functions of the system --
can be adapted successfully from the case of a fixed, flat lattice to that
of a fluctuating ensemble of geometries, as is relevant in studies of
non-perturbative quantum gravity. For the ensemble of two-dimensional
causal dynamical triangulations, we have formulated an explicit algorithm for
counting embedded graphs which allows us to do the counting
recursively for increasing order. The method can in principle be
applied to other matter and spin systems which admit a similar
diagrammatic expansion in terms of weak or strong lattice constants
around infinite or zero
temperature, and where one has sufficient information about the
probability distribution of local geometric lattice properties like the
vertex coordination number. Even in cases where these lattice
properties are not available explicitly, the lattice constants could still be
extracted from simulating the pure gravity ensemble.

As a potential spin-off, we noticed that the explicit expansion for the
susceptibility in vanishing external field of the Ising model on CDT (up
to order 6) or
dual CDT lattices (up to order 12) indicates a more regular behaviour of
this function
in the complex-temperature plane than that of the Ising model on the
corresponding fixed triangular or honeycomb lattice. This may also
explain why the simple ratio method, applied to CDT lattice results from
only six orders, gives an excellent approximation to the Onsager
susceptibility exponent $\gamma=1.75$. As we saw in some detail, other
approximation methods, namely, Dlog Pad\'e and differential approximants,
do not produce results of a similar quality. We believe that for the case of
the dual CDT lattice, the series computed is simply too short to yield
reliable estimates with any of the approximation methods
(even at order 12, the number of terms contributing
is much smaller than the number of terms contributing at order 6 on
the original CDT lattices). At any rate, if one aims to make an argument of
improved convergence of matter behaviour on {\it fluctuating} lattices
(see also \cite{mio-ising1}),
it is plausible that this will be achieved optimally with the triangulated
CDT geometries
whose vertex coordination number can vary dynamically from 4 all
the way to infinity, rather than with the dual tesselations which have
a fixed, low coordination number of 3. To investigate the convergence
issue in more detail will
require going to higher order than 6 on CDT lattices, which is not really
feasible `by hand', as we have been doing so far, but will require
the setting up of a computer algorithm to perform (at least part of)
the graph counting.

Our work should be seen as contributing to the study of non-perturbative
systems of quantum gravity coupled to matter, which is only just beginning.
One challenging task will be
to establish {\it computable} criteria characterizing and quantifying
the influence of geometry on matter and vice versa, in the physically
relevant case of four spacetime dimensions, something about which
we currently know close to nothing. Obvious quantities of interest
are critical exponents pertaining to geometry (like, for instance, the
Hausdorff and spectral dimensions of spacetime already measured for
the ground state of four-dimensional CDT \cite{ajl-rec}), and critical
matter exponents, like that of the susceptibility investigated in the present
work. One would like to have a classification of possible universality classes
of gravity-matter models as a function of the characteristics of the
ensemble of quantum-fluctuating geometries. A necessary condition
for viable quantum gravity models is that at sufficiently large distances
and for sufficiently weak matter fields the correct classical limits for
these critical parameters must be recovered.

Two-dimensional toy models like the one we have been considering
can serve as a blueprint for what phenomena one might expect to
find. In two dimensions, there have been several investigations of the effect
of random geometric disorder on the critical properties of matter systems,
and attempts to formulate general criteria for when a particular type of
disorder is relevant, i.e. will lead to a matter behaviour different from that
on fixed, regular lattices. Good examples are the Harris \cite{harris} and the
Harris-Luck \cite{harrisluck} criteria,
which tie the relevance of random disorder to the
value of the specific-heat exponent $\alpha$ and to correlations among
the disorder degrees of freedom. However, not all models fit the predictions,
and open problems remain (see \cite{luck} and references therein).
Causal dynamical triangulations coupled to matter add another class of
models inspired by quantum gravity,
whose randomness lies in between that of Poissonian Voronoi-Delaunay
triangulations and the highly fractal Euclidean dynamical triangulations,
and over which
one has some analytic control. It will be interesting to see to what degree
the robustness of the matter behaviour with respect to the geometric
fluctuations observed so far \cite{aal1,aal2,mio-ising1,mio-ising2} will persist for
different types of spin and matter systems and in higher dimensions.


\chapter{Moving to three dimensions}

{\small
In this chapter I will present work on a three-dimensional model of CDT.
The model I will consider is a constrained version of the general one
reviewed in the first chapter. Rather than allowing all possible triangulations
of the spatial slices and of the spacetime in between two of them, only
a particular class of triangulations will be allowed, in a similar fashion
to the construction leading to two-dimensional CDT from Euclidean DT.
The restriction will permit an analytic evaluation of the partition
function for the one-step propagator and of its continuum limit.
A quantum Hamiltonian for this model will be derived.
This chapter is based on \cite{mio-prm}.
}

\section{The challenge of higher dimensions}

Although the CDT model has been solved exactly in (1+1)-dimensions,
it turns out to be very difficult to extend this treatment to higher
dimensions,
for which most results so far have come from numerical simulations.
An exception to this is the matrix-model formulation of the
(2+1)-dimensional
model \cite{ABAB1}. An exact solution of the relevant matrix model has
been
given \cite{zinn-just-1}, but its rather complicated and implicit form has
so far
been an obstacle to performing its continuum limit analytically, and
extracting the quantum Hamiltonian.

This situation is not particularly surprising, as most of the known
solvable statistical
models are only one- or two-dimensional.
Even a simple spin model like the Ising model has been solved only in two
dimensions,
and for zero external field.
One can therefore already anticipate that an extension of analytical
methods and
results to the case of higher-dimensional CDT models will be a challenging
task.
One possible way to approach this problem is to try to modify the model
by making simplifying assumptions to improve its solubility, but without
changing its
physical content in the continuum.
The main aim of the present work is to study a reduced model of
three-dimensional CDT, which is a particular case of a class of models
introduced in \cite{bianca-bh}. This model possesses some additional
``order",
which will enable us to bring a number of technical tools to bear on its
solution. As we will see, this leads to a non-trivial dynamics for the
three-dimensional quantum universe, justifying our original ``truncated"
ansatz.

This chapter is organized as follows. In Sec.~\ref{sec-model} we describe
the geometric configurations whose partition function we are going
to study and use as a transfer matrix. The configurations are from an
ensemble of
triangulations of product type,
interpolating between successive constant-time slices, with free boundary
conditions and
boundary cosmological constants, conjugate to the areas of the slices.
In Sec.~\ref{sec-inversion} we introduce the first tool that will help us
in solving the
three-dimensional CDT model,
the inversion formula for {\it heaps of pieces}.
In Sec.~\ref{sec-matrices} we reformulate the partition function in terms
of the transfer
matrices for a set of one-dimensional hard bi-coloured dimer models.
In Sec.~\ref{sec-zeros} we study the location of the singularities of the
partition function
and define a critical surface in the parameter space, which is the
boundary of the
region where the model is convergent and well defined.
In Sec.~\ref{sec-replica} we introduce the second tool we will need for
the solution,
the replica trick for products of random matrices.
This enables us to identify a critical point on the critical surface
where the mean volume goes to infinity, thus making it a potential
candidate for
taking a non-trivial continuum limit.
We can give an exact solution to the model at the critical point, and
approximate it to the desired order in the displacement parameter when
moving away from it.
In Sec.~\ref{sec-continuum} we start exploring the continuum properties of
the
model by deriving expectation values for the volume and curvature of
spacetime.
In Sec.~\ref{sec-gluing} we construct the area-to-area transfer matrix by
summing
over unphysical degrees of freedom, and
explain the procedure for extracting information about the spatial volume
of the universe.
In Sec.~\ref{sec-hamiltonian} we show that for a vanishing bare inverse
gravitational constant,
one obtains a well-defined quantum Hamiltonian acting on the Hilbert space
of the
area eigenstates. By contrast, a canonical scaling ansatz for the
gravitational constant
does not seem to lead to a meaningful result.
In Sec.~\ref{microcan} we consider a possible extension of these
calculations,
encoding further global metric information in the form of an additional
Teichm\"uller parameter. Finally, our conclusions are presented in
Sec.~\ref{sec-conclusions}.
The four appendices at the end of the thesis collect technical results and
discussions needed in the
main text of this chapter.

\section{Introducing the model} \label{sec-model}

\subsection{The product type triangulations}

Part of our strategy for trying to solve the non-perturbative three-dimensional
model of quantum gravity defined by causal dynamical triangulations (CDT) is to
identify a suitable (sub-)class of all CDT configurations whose superposition
can be tackled {\it analytically}. The aim of such a reduction is to simplify the
analytic treatment, without
eliminating relevant degrees of freedom to such a degree that the universality
class is changed, compared to that of the full CDT model.\footnote{This assumes
that the full CDT model leads to an essentially unique three-dimensional quantum gravity theory.
Strictly speaking, rather little is known in dimension three about the universality
classes of statistical models of random geometry like the one we are using.
There are certainly more, but we are only interested in those which possess a
continuum interpretation in terms
of quantum gravity.} To achieve this, we propose to
work with a set of triangulated, causal geometries which have an additional
``order" imposed on them. This order is mild in the sense of
not implying ``isometries" of the triangulations: both the local spacetime curvature
and the local curvature of spatial slices can still vary arbitrarily. Our model
therefore has less order than the hexagon model considered in \cite{bianca-hex}, which has
flat spatial slices.

The way in which we will introduce more structure or ``order" on triangulated
spacetimes -- inspired by similar earlier ideas \cite{difra-hardobj,bianca-bh} -- is that of applying
the building principle inherent in {\it causal} dynamical triangulations twice over.
This ``causality principle" recognizes that in a metric spacetime of Lorentzian
signature not all directions
are equivalent, but there is a distinguished (class of) time direction(s), in line with
the existence of light cones and causal relations (none of which are present
in spaces of purely Euclidean signature). As we have seen in Chapter~1, in CDT this causal structure is
implemented via a discrete global time slicing.
This simply means that each simplicial
building block of a triangulated spacetime must be contained in exactly one
spacetime ``sandwich" (the region between discrete proper times $t$ and $t+1$).
This implies a causal ordering on the simplices, without constraining
the local curvature degrees of freedom.

In addition to this physically motivated choice of a distinguished time direction,
which is one of the key ingredients of the approach of causal dynamical
triangulations,
we will introduce here a second distinguished direction, but this time purely
for computational convenience, and under the assumption that it will not
affect the universal properties of the gravitational model\footnote{An assumption
that obviously will have to be borne out by the final result}. This will roughly
speaking correspond to an additional slicing, but this time in one of the spatial
directions. As explained in \cite{bianca-bh}, the resulting structure can be thought of
as a staggered fibration in a piecewise flat setting.

\begin{figure}[ht]
\centering
\vspace*{13pt}
\includegraphics[width=9cm]{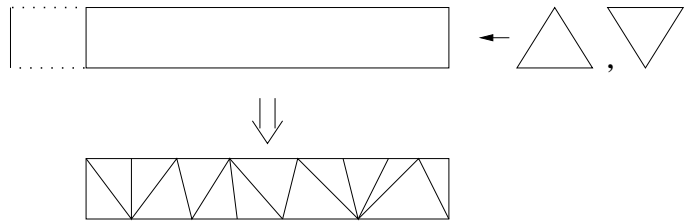} 
\vspace*{13pt}
\caption{\footnotesize A triangulated strip constructed as a tower over a one-dimensional link.}
\label{segment-tower}
\end{figure}

To illustrate the main idea behind this type of triangulation let us start
from the easier (1+1)-dimensional CDT model. A spacetime contributing
to the sum over geometries is usually described as a sequence of
triangulated strips glued together, where each strip represents a
piece of spacetime between proper times $t$ and $t+1$, and the
lengths of the spatial boundaries of adjacent strips must match pairwise.
For our purposes, it is useful to think of a strip as being constructed
``sideways" (see Fig.\ref{segment-tower}): starting from a segment (a time-like link),
build a frame on it and then fill the frame with some sequence of up- and down-triangles.
We will refer to this construction as
``building a two-dimensional {\it tower} over a link". The entire two-dimensional
triangulated spacetime can then be regarded as fibration over a one-dimensional
chain of links.

\begin{figure}[ht]
\centering
\vspace*{13pt}
\includegraphics[width=9cm]{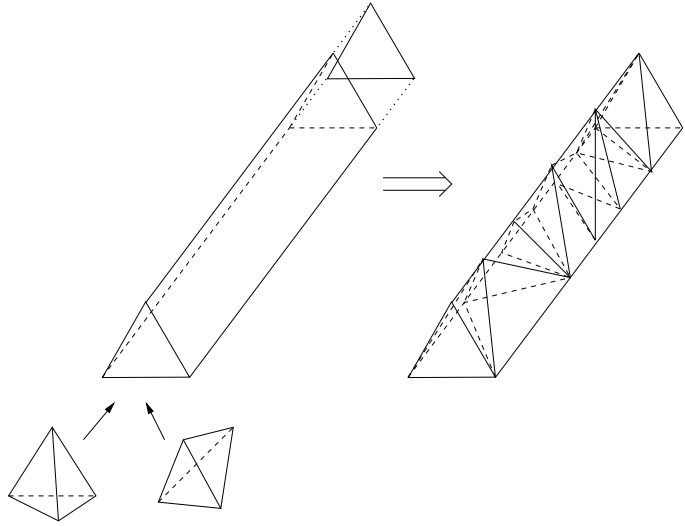} 
\vspace*{13pt}
\caption{\footnotesize A triangulated prism constructed as a tower over a
two-dimensional triangle.}
\label{triangle-tower}
\end{figure}

More generally, {\it product triangulations} \cite{bianca-bh} are obtained by building towers
over (arrays of) higher-dimensional simplices, such as triangles. How a three-dimensional
tower is built over a triangle is illustrated in Fig.\ref{triangle-tower}.\footnote{It should be noted that
the figures do not represent faithfully the {\it metric} properties of the building blocks, that is, their edge
lengths. By construction, up to a relative factor between space- and time-like links, the edge
lengths in a causal dynamical triangulation are all identical.}
If the base space is not just a single triangle, but a two-dimensional triangulation
consisting of $n$ triangles, we can construct a three-dimensional product
triangulation by erecting a tower over each of them, in such a way that the boundary
triangulations of the resulting prisms again match pairwise.
A general $(n+m)$-dimensional product triangulation is a simplicial manifold
constructed by consistently building $(k+m)$-dimensional towers (for all $k\leq n$) over the
$k$-dimensional sub-simplices of an $n$-dimensional {\it base} simplicial manifold.

In the present work we will use this construction in the context of (2+1)-dimensional
causal dynamical triangulations. By definition, these spacetimes can be regarded
as product triangulations whose base space -- like in (1+1) dimensions -- is a
one-dimensional triangulation in the time direction, and the fibres over each
link of the base space are the sandwiches between integer values of the discrete
time $t$ introduced earlier. Now, instead of allowing sandwich geometries which
are arbitrary triangulations of thickness $\Delta t=1$, we will impose an additional
product structure on them. Namely, they should themselves have the form of a
sequence of up- and down prisms, as depicted in Fig.\ \ref{tower-sandwich} below.
That is, in addition to the slicing of the entire spacetime
corresponding to the global discrete proper time, each
sandwich possesses a discrete slicing in a given spatial direction. The number of
slices (that is, the number of prisms in a sandwich) is allowed to vary from
sandwich to sandwich.

An equivalent way of characterizing such spacetimes is to regard them as fibrations
over a two-dimensional base space (with one space- and one time-dimension)
which is itself an arbitrary (1+1)-dimensional CDT, and where over each triangle of
the base space we have erected a tower ``filled" with a sequence of tetrahedra, in such
a way that the triangulations on the faces of neighbouring towers match and can be
glued together consistently. In this way, we may think of our triangulations as being
(1+1+1)-dimensional.

\subsection{The partition function}\label{sec-part}

We will concentrate on the dynamics of a ``sandwich geometry",
given by the transition amplitude from a spatial geometry at time $t$ to
one at time $t+1$, in other words, the transfer matrix $\hat T$ of the
causal dynamical triangulation (CDT) model, which in turn contains
information about its quantum Hamiltonian $\hat H$
by virtue of the relation $\hat T=e^{-a \hat H}$ (see \cite{ajl-def} for details).

Since pure gravity in (2+1) dimensions does not possess any local,
propagating degrees of freedom, we expect that most details of the
spatial geometries will be dynamically irrelevant, leaving only
the spatial two-volume $A(t)$ and the Teichm\"{u}ller parameters (together
with their canonically conjugate momenta) as
physical degrees of freedom. Therefore, rather than calculating the
matrix elements
\beq\label{firsttransfer}
G(g_1,g_2,\Delta t=1)=\langle g_2|\hat T |g_1\rangle
\eeq
of the transfer matrix
from an arbitrary spatial triangulated geometry $|g_1\rangle$ at time $t$ to an
arbitrary $|g_2\rangle$ at time $t+1$, we will only keep track of the two
boundary areas $A_1$ and $A_2$ and evaluate the reduced matrix elements
\beq\label{area-area}
  G(A_1,A_2,\Delta t=1)=
  \langle A_2|\hat T |A_1\rangle, 
\eeq
where $|A_i\rangle$ is a normalized linear combination of
states $|g_i\rangle$ with given area $A_i$ (see
\cite{ABAB1} for a detailed discussion). Note that one is still summing over the
{\it same} sandwich geometries as in (\ref{firsttransfer}), so the reduced matrix
elements still capture the effective dynamics of {\it all} geometric excitations of
the sandwich geometry.\footnote{The presence of physical degrees of freedom
beyond the spatial area depends on the topology of the spatial slices. There are
none for the case of spherical slices considered in \cite{ajl-3d,ABAB1}, and there is one
Teichm\"uller parameter
in the present model, where we choose boundary conditions corresponding to
cylindrical spatial slices. Work is under way to determine whether the
calculation of matrix elements presented here is still feasible when the
dependence on this parameter is kept explicitly.}
Once these matrix elements are known, a next step is to try to extract
the continuum Hamiltonian operator $\hat{H}$ of the system from an expansion
in the short-distance cutoff $a$ as $a\rightarrow 0$ according to
\beq \label{HfromT}
  \langle A_2|\hat T |A_1\rangle =
  \langle A_2|e^{-a \hat{H}} |A_1\rangle =
  \langle A_2|\left(\hat{1}-a\hat{H}+O(a^2)\right)|A_1\rangle.
\eeq

We will denote by $N_{ij}$ the number of simplices having $i$ vertices
on the initial boundary spatial slice at time $t$ and $j$ on the final one at
$t+1$.
In this way $A_1$ (resp. $A_2$) is given by $N_{31}$ (resp. $N_{13}$).
The prescription for evaluating the discrete one-step propagator (\ref{area-area}) is
given by
\beq\label{area-area-cdt}
  G(N_{31},N_{13},\Delta t=1)=\sum_{\TT_{|N_{31},N_{13}}}\frac{1}{C_{\TT}}\ e^{-S_{EH}},
\eeq
where the sum is over all sandwich triangulations $\TT$
with fixed boundary areas $N_{31}$ and $N_{13}$,
$S_{EH}$ the Wick-rotated discrete Einstein-Hilbert
action, and $C_{\TT}$ the order of the automorphism group of $\TT$.
Taking into account boundary terms in the action to ensure the
correct propagator behaviour, we find
\beq\label{}
  S_{EH}=\alpha (N_{13}+N_{31}) + \beta N_{22} +\gamma N
\eeq
as an explicit expression for the gravitational action, where $N$ is the number of triangle
towers in the sandwich, and where we have introduced the parameters
\beq\label{alphabeta}
\begin{split}
  &\alpha =-c_1 k + b_1 \lambda \\
  &\beta = c_2 k + b_2 \lambda \\
  &\gamma = c_3 k
\end{split}
\eeq
depending on the dimensionless bare cosmological and inverse Newton constants
$\lambda$ and $k$, and on positive numerical constants $c_i$ and $b_i$ characterizing
the geometric construction (see appendix \ref{App-action}).

As one might have anticipated from previous investigations of related
three-dimensional quantum gravity models,
the evaluation of (\ref{area-area-cdt}) remains a challenging task, also for
our specific choice of ensemble of triangulations.
Our main aim will be to calculate its discrete Laplace transform
\beq\label{laplace}
\begin{split}
  Z(x,y,\Delta t=1)&=\sum_{N_{31}}\sum_{N_{13}}x^{N_{31}}y^{N_{13}}G(N_{31},N_{13},\Delta t=1)\\
  &=\sum_{N_{31}}\sum_{N_{13}}(x e^{-\alpha})^{N_{31}}(y e^{-\alpha})^{N_{13}}\sum_{\TT_{|N_{13},N_{31}}}
  e^{-\beta N_{22}-\gamma N}
\end{split}
\eeq
which can be thought of either as the generating function of $G(N_{31},N_{13},\Delta t=1)$
or as the partition function of the sandwich geometry with free boundary conditions
(thus summing over all values of the boundary volumes) and with additional
boundary cosmological
terms $x=e^{-\lambda_{in}}$ and $y=e^{-\lambda_{out}}$ in the action.

To get back the transfer matrix from $Z(x,y,\Delta t=1)$, one needs to keep $x$
and $y$ distinct and variable. However, as shown by other examples
(see the end of section \ref{2d-solution} for the two-dimensional case, and \cite{ajl-3d}
for a numerical illustration in three dimensions), one may be able to extract
non-trivial information about the phase structure of the model by considering
the special values $x=y=1$, which simplifies the evaluation of (\ref{laplace}).

\begin{figure}[ht]
\centering
\vspace*{13pt}
\includegraphics[width=9cm]{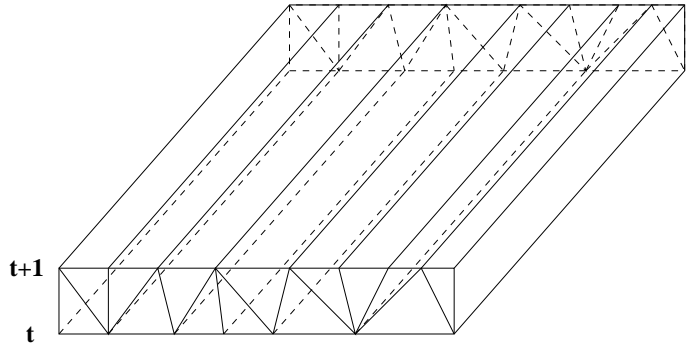} 
\vspace*{13pt}
\caption{\footnotesize A ``sandwich geometry" of product type in the
(1+1+1)-dimensional CDT model, built over a
given triangulated base strip. We split the calculation of its partition function into
two parts, calculating first the partition function for such a geometry
and then summing over all possible sequences of towers. By ``towers of type
(2,1)" we will mean the towers built on a triangle with two vertices at time $t$
and one at time $t+1$, and similarly for ``towers of type (1,2)".
Each of the towers is filled up with a sequence of tetrahedra,
as illustrated in Fig. \ref{triangle-tower}. There are three different possibilities
for how a tetrahedron can be oriented inside one of the prisms (see \cite{bianca-bh}
for more geometric details).}
\label{tower-sandwich}
\end{figure}

At this stage we will exploit the special product structure of our chosen
spacetime geometries to split the sum over all sandwich triangulations into two simpler
sums. The idea is to first perform the sum over all fillings (i.e. triangulations)
of towers for a {\it fixed} triangulated base strip, and then
to perform the sum over all possible triangulations of the base strip (see
Fig.\ref{tower-sandwich}), schematically
\beq \label{sum-split}
  \sum_{\TT_{sandwich}}=\sum_{\TT_{base \ strip}}\hspace {.3cm}\sum_{\TT_{towers}}.
\eeq

We are thus naturally led to studying the partition function $Z_N$ for
sandwich geometries of product type with $N$ towers, and investigating its
properties in the large-$N$ limit. In this sense we are treating the second
distinguished direction, which defines the slicing within the sandwich
geometry, in the same way as we treat the time direction in (1+1)-dimensional CDT.
It is related to (\ref{laplace}) via
\beq \label{zndefine}
 Z(x,y,\Delta t=1)=\sum_{N}  e^{-\gamma N} Z_N(u,v,w,\Delta t=1).
 \eeq
Introducing the weights $\sqrt{u}=x e^{-\alpha}$, $\sqrt{v}=y e^{-\alpha}$
and $w= e^{-\beta}$, we can express $Z_N$ as
\beq \label{Z_N}
  Z_N(u,v,w,\Delta t=1)=\sum_{S_N}\hspace {.3cm}
  \sum_{\TT_{|N_{13},N_{31},N}}
  u^{\frac{N_{31}}{2}} v^{\frac{N_{13}}{2}} w^{N_{22}},
\eeq
where the sequences $S_N$ consist of $N$ prism towers,
$N-R$ of them of type (2,1) and $R$ of type (1,2), for any $R<N$.

\section{Inversion formula} \label{sec-inversion}

The special product form of our sandwich geometries will enable us to make
use of an {\it inversion formula},
relating its partition function to that of a
model of one dimension less. As we have seen in Chapter~1, this
technique essentially provides a complete solution to the CDT model in
dimension (1+1). This is not true in the much more involved case of
CDT in dimension (1+1+1) we are addressing here, but -- as we will
demonstrate -- it will nevertheless allow
us to make substantial progress in the evaluation of the partition function.

In order to apply the inversion formula to the case of three-dimensional
causal triangulations, we will use the fact that the propagator of this
model can be characterized in terms of geometric data which are ``almost"
two-dimensional. We have seen this characterization in the first chapter,
when talking about the  ABAB-matrix model. The procedure to be used here
in order to obtain the two-dimensional dual graph describing the propagator
is exactly the same, but the class of graphs we obtain is different, reflecting
the difference in the class of triangulations.

\begin{figure}[ht] 
\centering
\vspace*{13pt}
\includegraphics[width=12cm]{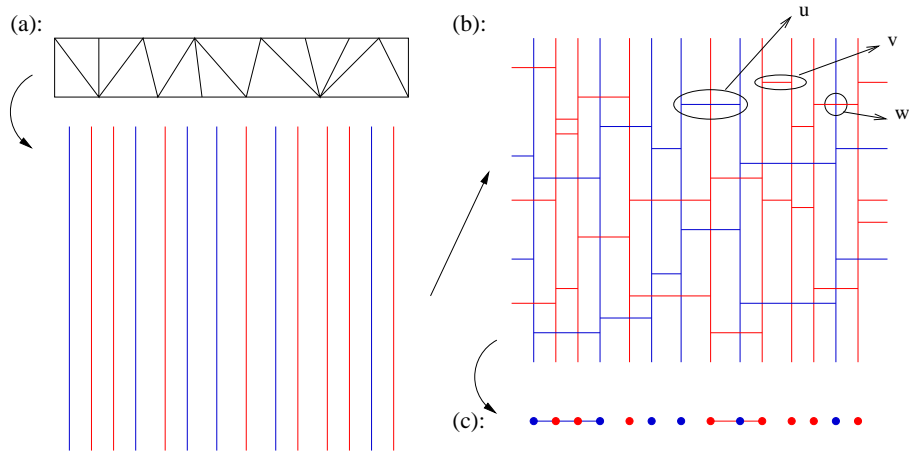} 
\vspace*{13pt}
\caption{\footnotesize  From sandwich geometries with product structure to hard dimers:
(a) vertical lines representing towers over a triangulated base strip; (b) horizontal links
encode how the tetrahedra are connected, the corresponding weights can be read off
as indicated; (c) projection to a sequence of hard-dimer configurations, as
explained in appendix \ref{App-inversion}.}
\label{3d-dual}
\end{figure}

In the case of our product triangulations the dual graphs obtained
 have a special form.
Namely, part of the dual graph can be drawn as
a fixed sequence of straight lines (which we will call
``vertical lines", and which are drawn vertically in Fig.\ \ref{3d-dual})
coming from the prism towers, blue for a tower over
a type-(2,1) triangle and red for a tower over a type-(1,2) triangle.
The sequence of blue and red towers
depends on the triangulation of the base strip.
In addition, the dual graph contains ``horizontal links", which
start on a vertical line of the same colour and end on the next
(to the right, say) vertical line of the same
colour. The position of the horizontal links encodes the three-dimensional
triangulation of the original sandwich geometries.

Looking at the subgraphs of one colour in the dual graph example of
Fig.\ \ref{3d-dual}, one notes that they are particular examples of
dual graphs for (1+1)-dimensional CDT spacetimes. It is indeed the
case that the blue graph is precisely the graph dual to the triangulation
of the initial two-dimensional surface of the sandwich geometry, and
the red graph the dual to its final surface. We may therefore think of
the bi-coloured graph as a particular superposition of these two
uni-coloured graphs. Bi-coloured graphs of this type are again of the
form of {\it heaps of pieces} in the sense of \cite{viennot},
which means that we can map them to dimer configurations as
in the (1+1)-dimensional situation (c.f. appendix \ref{App-inversion}),
but with the difference that now the one-dimensional
dimer model will also be bi-coloured, with a fixed sequence of
blue and red sites and with the blue (red) dimers linking a blue
(red) site with the next one of the same colour, as depicted in Fig.\ \ref{3d-dual}.

The upshot of our considerations so far is that for a sandwich
geometry associated with a fixed base strip $S$
an analogue of formula (\ref{fund2}) holds, namely,
\beq\label{fund3}
  Z_{S_{|R_1,R_2}}(u,v,w)= \frac{ 1}{Z_{S_{|R_1,R_2}}^{hcd}(-u,-v,w)},
\eeq
where
\beq\label{hcd}
  Z_{S_{|R_1,R_2}}^{hcd}(u,v,w)= \sum_{ {\rm coloured}\ {\rm hard-dimer}\ {\rm config.}\ D_{|S}}
  u^{|D|_b}v^{|D|_r}w^{|\cap D|}
\eeq
is the partition function of the coloured dimer problem for a given
sequence S of $R_1$ blue and $R_2$ red sites, $|D|_b$ ($|D|_r$) is the
number of blue (red) dimers in the configuration $D$,
and $|\cap D|$ is the number of crossings between dimers and sites of different colour.
We should emphasize that our application of the inversion formula
relates only to the {\it second} sum in (\ref{sum-split}), since we sum over
the triangulation of the towers for a {\it fixed} triangulation of the base strip.
This allows us to rewrite (\ref{Z_N}) as
\beq
\label{Zinv}
  Z_N(u,v,w,\Delta t=1) =\sum_R\sum_{S_{|R,N-R}}\hspace {.3cm}
  \frac{ 1}{Z_{S_{|R,N-R}}^{hcd}(-u,-v,w)}, \hspace {.2cm}
\eeq
whose further evaluation will be the subject of the remainder of this chapter.

\section{Random matrix formulation} \label{sec-matrices}

The coloured-dimer partition function can
be written in terms of transfer matrices, by introducing
three vectors (1,0,0), (0,1,0) and (0,0,1) corresponding
to an empty link, a blue link and a red link between neighbouring sites.
A transfer matrix can be associated with transitions
between these states, taking into account that hard dimers
are not allowed to touch each other, and that a dimer of a given
colour connects only sites of the same colour which are first
neighbours (among the sites with same colour), and may cross
sites of the opposite colour in between. The transfer matrices
can therefore be associated with the sites or vertices of the
dimer model, and their explicit form depends on the colour label
of the site.
The transition empty-empty gets a weight $1$, the transitions
dimer-empty or empty-dimer get a weight $\sqrt{u}$ or $\sqrt{v}$,
depending on the colour, and the crossing of a site by a
dimer of any colour gets a weight $w$; all other possibilities have
weight $0$.
In this way we can associate a matrix
\beq\label{A}
  \tilde A=\matrice 1 & \sqrt{u} & 0 \cr
                    \sqrt{u} & 0 & 0 \cr
                    0 & 0 & w \ematrice
\eeq
with every blue site, and a matrix
\beq\label{B}
  \tilde B=\matrice 1 & 0 & \sqrt{v} \cr
                    0 & w & 0 \cr
                    \sqrt{v} & 0 & 0 \ematrice
\eeq
with every red site.
Consequently, the partition function $Z_{S_{|R_1,R_2}}^{hcd}(u,v,w)$
can be expressed as a product of $R_1$ matrices $\tilde A$
and $R_2$ matrices $\tilde B$, ordered according to the sequence
specified by $S_{|R_1,R_2}$. (Note that the matrices
$\tilde A$ and $\tilde B$ do not commute.)
We can now write the partition function (\ref{fund3}) associated with a fixed
sequence $S_{|R_1,R_2}$ as
\footnote{Our choice of taking the trace in the denominator implies periodic
boundary conditions in one of the spatial directions, by gluing together the first
and last prism tower of the sandwich. Since we are leaving the other spatial
direction open, the spatial slices of our model have the topology of a cylinder.
Instead of compactifying one of the directions, we could also have left it
open, in which case we would have to specify
the boundary lengths as boundary conditions, or introduce conjugate boundary variables
analogous to the $x$- and $y$-variables in (1+1) dimensions, eqs.\ (\ref{gen2}) and (\ref{gen2bis}),
and contract the product of matrices with the vectors
$(1\;ix_1\;ix_2)$ and $(1\;iy_1\;iy_2)$.
This would complicate the treatment of the partition function considerably.}
\beq\label{fund3-AB}
  Z_{S_{|R_1,R_2}}(u,v,w)= \frac{ 1}{Tr(ABAABABBB...)_{S_{|R_1,R_2}}}
\eeq
where the matrices $A$ and $B$ are the same as $\tilde A$ and $\tilde B$, but
with the substitution $\sqrt{u}\rightarrow {\rm i}\sqrt{u}$,  $\sqrt{v}\rightarrow {\rm i}\sqrt{v}$.
In order to express the sum over sequences $S$ in eq.\ (\ref{Z_N}), it
is convenient to combine the two matrices and define a general transfer matrix
\beq\label{M_q}
  M_{q_j}(u,v,w)=q_j A + (1-q_j) B =\matrice 1 & q_j {\rm i}\sqrt{u} & (1-q_j){\rm i}\sqrt{v} \\
                       q_j{\rm i}\sqrt{u} & (1-q_j)w & 0 \\
                       (1-q_j){\rm i}\sqrt{v} & 0 & q_j w \ematrice,
\eeq
with $q_j$ taking the values 1 ($M=A$) or 0 ($M=B$). The sum over sequences $S$
is then replaced by a sum over sequences of elements
$\{0, 1\}$. In this formulation, the calculation of the partition
function assumes the form of a problem of products of random matrices
\cite{crisanti}, and we can rewrite the full partition function (\ref{Z_N}) for
a fixed number $N$ of prism towers (equivalently, a total number of matrices
$A$, $B$ in the sequence) as
\beq
\label{ZZZ}
  Z_N(u,v,w,\Delta t=1) =\sum_{R<N} \sum_{\{q_j\}_{R,N}} \frac{ 1}{Tr\prod_{j=1}^N M_{q_j}(u,v,w)}.
\eeq
This in turn can be thought of as the average of (\ref{fund3-AB}) over
all possible configurations $\{q_j\}_N$ (with any R), with
$q_j=1$ and $q_j=0$ each having probability $p=1/2$.
Introducing the notation
\beq
|P_{N,\{q\} }(u,v,w)|=Tr\prod_{j=1}^N M_{q_j}(u,v,w)
\eeq
we can finally write
\beq\label{Z}
  Z_N(u,v,w,\Delta t=1) =2^N\langle \frac{1}{|P_{N,\{q\} }(u,v,w)|}
  \rangle_{p=\frac{1}{2},1-p=\frac{1}{2}}.
\eeq

\section{The zeros of the denominators} \label{sec-zeros}

An important feature to be noted about the inversion formula is that it maps an infinite
series with positive coefficients to the inverse of a finite sum with alternating sign
coefficients. The latter will have a real root corresponding to the radius of convergence
of the infinite series. This is evident in the (1+1)-dimensional case, formula (\ref{gen2bis}).
The same is clearly true for the (1+1+1)-dimensional case (\ref{fund3}) too, which will
have a two-dimensional {\it locus} of zeros in the three-dimensional parameter space
spanned by $u$, $v$ and $w$.
However, the location of this locus will depend on the sequence $\{ q_j\}$, and
we must keep in mind that we still have to sum over all the sequences $\{ q_j\}$.
Consequently, we will be interested in determining the envelope of all the loci.
In the absence of an analytical solution, which appears difficult to come by,
we will determine its location in the limit of infinite $N$ by studying
particular classes of sequences and by numerical computations, in conjunction
with a bit of (well-motivated) conjecture.

\subsection{u = v}

Let us temporarily assume that $x=y$ (and eventually $=1$) in the Laplace transform of
the one-step propagator, (\ref{laplace}), and therefore $u=v$.
This clearly makes things easier, but should still permit us to extract information
about the phase structure of the model, as is the case in both two and
three dimensions (see \cite{ABAB1} and \cite{ABAB2} for illustrations of the latter).
This is not an implausible proposition in the sense that we want to find the
critical line in the $k$-$\lambda$ plane, which concerns the ``bulk" behaviour
of the spacetime and should be insensitive to details of the boundary
data, like those encoded in $x$ and $y$.

In order to determine the zeros of the polynomials
$|P_N(u,v=u,w)|=Tr\prod_{j=1}^N M_{q_j}(u,v=u,w)$,
we write this product as $A^{n_1}B^{n_2}...A^{n_{M-1}}B^{n_M}$ for some sequence
of $M\leq N$ positive integers $\{n_1,...,n_M\}$ which sum up to $N$, $\sum_{i=1}^Mn_i=N$.
Note that from $u=v$ follows $B=J A J$, where the matrix
\beq
  J=\matrice 1 & 0 & 0\\ 0 & 0 & 1\\ 0 & 1 & 0\ematrice
\eeq
is a projector ($i.e.$ $J^2=Id$), so that we can write
\beq \label{AJ^M}
  |P_{N,\{q\} }|=Tr(A^{n_1}J A^{n_2}J...A^{n_{M-1}}J A^{n_M}J).
\eeq
Note furthermore that
\beq
  A^n=\matrice (\alpha^n)_{11} & (\alpha^n)_{12} & 0 \cr
                    (\alpha^n)_{12} & (\alpha^n)_{22} & 0 \cr
                    0 & 0 & w^n \ematrice ,
\eeq
where $\alpha^n$ is the n-th power of the matrix
\beq
\alpha :=
\matrice 1 & {\rm i} \sqrt{u} \\
{\rm i} \sqrt{u} & 0\ematrice
\eeq
found in (\ref{Z-2d-1}) above, and which has the following properties:
\beq \label{a^n_11}
  (\alpha^n)_{11}=\frac{\lambda_+^{n+1}-\lambda_-^{n+1}}{\lambda_+-\lambda_-}
  =u^{\frac{n}{2}} U_n(\frac{1}{2 \sqrt{u}})
\eeq
\beq
  (\alpha^n)_{12}=i\sqrt{u}(\alpha^{n-1})_{11}
\eeq
\beq
  (\alpha^n)_{22}=(\alpha^n)_{11}-(\alpha^{n-1})_{11}
\eeq
where $\lambda_{\pm}=(1\pm\sqrt{1-4u})/2$ are the eigenvalues
of the matrix $\alpha$ ($w$ is the third eigenvalue of the matrix $A$),
and $U_n(x)$ is the n-th Chebyshev polynomial of the second kind.

\subsubsection{u=0}
For the subcase $u=0$ everything is trivial since $|P_N(0,0,w)|=1$ for every sequence.
(We exclude degenerate sequences with $N$ matrices of one type and none of the other, which
would yield $|P_N(0,0,w)|=1+w^N$. At any rate, they would give a negligible contribution
in the average.) We thus have $Z_N(0,w)=1$.

\subsubsection{w=0}
For the subcase $w=0$ the trace (\ref{AJ^M}) reduces to
\beq\label{w=0}
  |P_{N,\{q\} }|=z_0=\prod_{i=1}^{M} (\alpha^{n_i})_{11}\equiv
 \prod_{i=1}^{M}
  u^{\frac{n_i}{2}} U_{n_i}(\frac{1}{2 \sqrt{u}})
\eeq
by virtue of (\ref{a^n_11}), which implies
that (\ref{AJ^M}) becomes the product of $M$ uncoupled
partition functions of the simple (uncoloured) hard-dimer problem.

Note first of all that the term $u^{\frac{n}{2}}$ in (\ref{a^n_11}) does not vanish at $u=0$,
because its product with $U_n(\frac{1}{2 \sqrt{u}})$ gives a polynomial of order $[\frac{n}{2}]$
in $u$ whose zeroth order term is $1$.
From mathematical handbooks we know that $U_n(x)$ has zeros only in the interval $[-1,1]$,
more precisely, at the values $x_i=\cos{(\frac{i}{n+1}\pi)}$, for $i=1,...,n$. It follows
that no zeros can occur for $u<1/4$. By contrast, for fixed $u>1/4$ there is always
a sequence, for some N, such that $|P_{N,\{q\} }|$ has zeros between $1/4$ and $u$.
For positive $\epsilon=u-\frac{1}{4}$, the occurrence of the first root
is when $\frac{1}{2\sqrt{\frac{1}{4}+\epsilon}}\leq x_1=\cos{(\frac{1}{n+1}\pi)}$.
For small $\epsilon$ this requires a large $n$. And this in turn means
that (some but not necessarily all of) the sequences contributing to the critical
behaviour at $w=0$ and
$u=\frac{1}{4}$ possess a large group of consecutive matrices of
the same kind; particular examples of this are configurations like $A B^{N-1}$.
A predominance of such configurations would signal a decoupling of
neighbouring slices in the three-dimensional geometry, and a geometric
degeneracy of the model there.

\subsubsection{u,w$\neq$0}

\begin{figure}[h]
\begin{center}
$\begin{array}{c@{\hspace{1in}}c}
\multicolumn{1}{l}{} &
    \multicolumn{1}{l}{} \\ [-0.53cm]
\includegraphics[width=4cm]{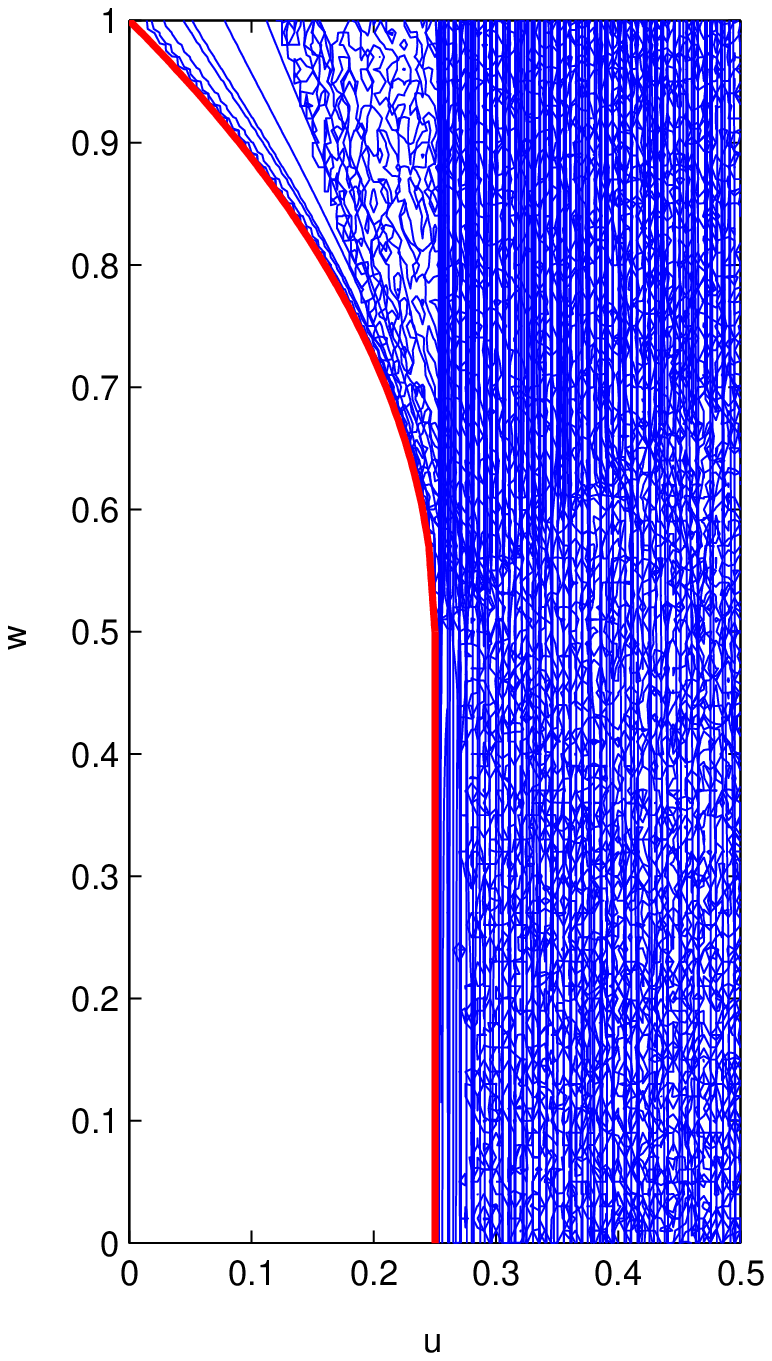} &
\includegraphics[width=7cm]{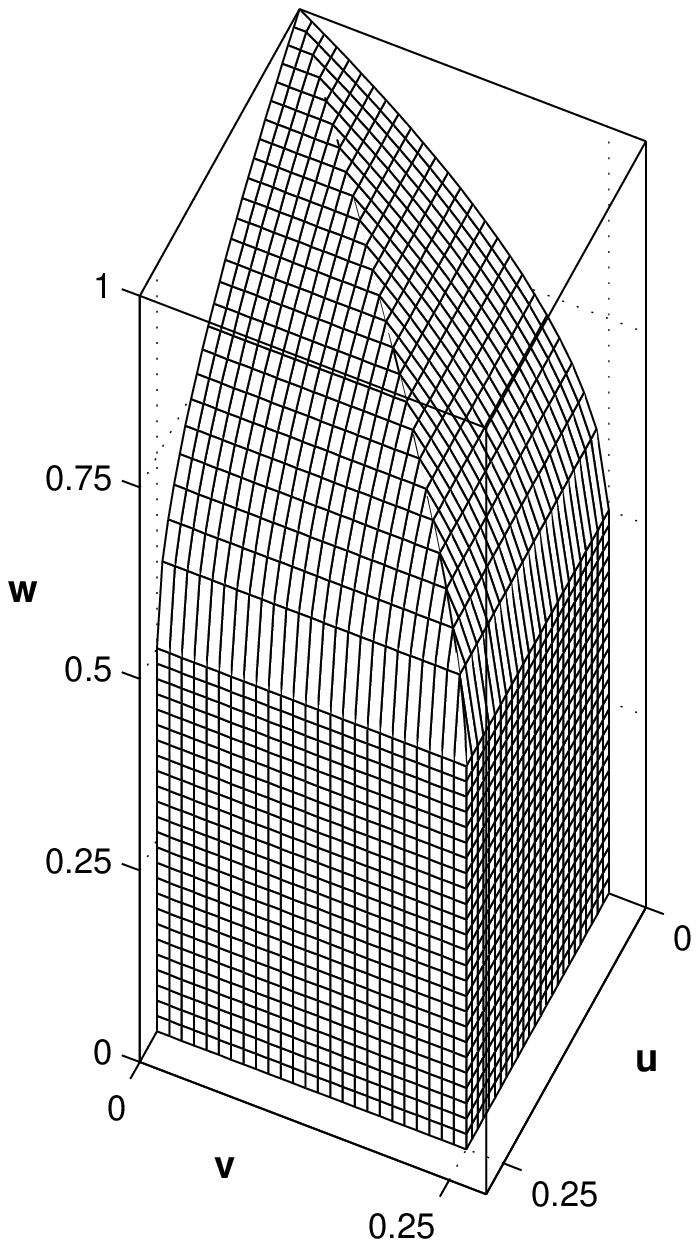} \\ [0.4cm]
\mbox{\bf (a)} & \mbox{\bf (b)}
\end{array}$
\end{center}
\caption{\footnotesize \mbox{\bf (a):} The critical
line, combined from $u=1/4$, $w=\lambda_+(u)$, together with plots of the zeros
of $|P_{N,\{q\} }|$ for random sequences $\{q\}$ and for sequences $AB^{N-1}$
and $A^{\frac{N}{2}}B^{\frac{N}{2}}$
at different values of N, in the $u=v$ plane.
\mbox{\bf (b):} The plot of the critical surface in the full parameter space of the
$( u,v,w)$.}
\label{zeros}
\end{figure}

For both $u$ and $w$ different from zero the general expression for
(\ref{AJ^M}) becomes highly non-trivial.
To get a better idea of the distribution of zeros, we used the program
{\sc Matlab} to plot some of the roots of $|P_N(u,w)|$.
This is in principle straightforward, but for increasing $N$ requires
considerable computing power, because the number
of sequences grows like $2^N$.
A superposition of several such plots is shown in Fig.~\ref{zeros}a.\footnote{Typical
$N$-values for the sequences ranged between 50 and 200,
and the total number of sequences analyzed was on the order of 50.}
From this picture it appears that the singularity-free region for $u<1/4$ found
for $w=0$ is also present for $w>0$, up to some value of about 0.5,
and then starts to shrink toward zero. Because of the limit on $N$ in our numerical
study, we could not determine whether the value $u=0$ is reached
at $w=1$ or rather some small $u=\epsilon >0$.


What we can do analytically to put more stringent bounds on the envelope
is to study specific sequences and find their
locus of zeros asymptotically for $N\rightarrow\infty$.
We find that $(u,w)=(\frac{1}{4},w)$ continues to be an
accumulation point for the zeros of
the sequences $A B^{N-1}$ for all values of $w\leq 1$.
For sequences of the form
$A^{\frac{N}{2}}B^{\frac{N}{2}}$ it can easily be shown analytically that
a solution of $Tr(A^{\frac{N}{2}}B^{\frac{N}{2}})=0$ for
$N\rightarrow\infty$ is given by $w=\lambda_+(u)=(1+\sqrt{1-4u})/2$.
For the alternating sequences $(AB)^{N/2}$ we find instead a locus
given by $uw=4/27$, which is tangent to the curve
$w=\lambda_+(u)$ in the point $(u,w)=(2/9,2/3)$, and otherwise lies completely to
the right of it (in a $u$-$w$-plot like that of Fig.\ \ref{zeros}a).
Many other sequences with regular distribution patterns for $A$- and
$B$-matrices whose large-$N$ limit we have studied seem to have an
asymptotic locus of zeros which is a smooth curve lying in between
$w=\lambda_+(u)$ and $w=4/(27 u)$, and passing through their common
point $(2/9,2/3)$.

Based on these findings, and in the absence of any evidence to the contrary,
we conjecture that the $(u,w)$-region free of singularities is defined by the
simultaneous conditions
$u< 1/4$ and $w<\lambda_+(u)$, whose boundary is given by the red
curve in Fig.\ \ref{zeros}a. To its right we have plotted the locations of
zeros, both of random sequences corresponding to some $q_i$'s, and of
special sequences we have been able to treat exactly.
Interestingly, this singular curve can be characterized in terms of the eigenvalues
$(\lambda_+,\;\lambda_-,\;w)$ of the matrix $A$, as the curve along which
the largest two of the eigenvalues are degenerate. For $u>1/4$, the $\lambda_\pm$
are distinct and complex, and thus cannot be equal to $w$ (which is real).
For $u=1/4$, $\lambda_\pm$ are both equal to $1/2$ and larger than $w$,
as long as $w<1/2$. For $w\geq 1/2$, the two largest eigenvalues are
exactly defined by $w$ being equal to the larger one of the pair $\lambda_{\pm}$.
This suggests the existence of an analytical argument for why
such a degeneracy gives rise to roots of $|P_N|$, but we have not been
able to find it.

Following this line of argument further, we can give yet another
characterization of the singular line.
It can be checked that the trace $|P_N|$ is always real.
Denoting the eigenvalues of $P_N$ by $\lambda_i$,  $i=1,2,3$,
their sum must therefore be real. Their product is the determinant of $P_N$,
which from (\ref{AJ^M}) is easily computed as the product of the determinants,
$\lambda_1\lambda_2\lambda_3=\lambda_+^N\lambda_-^Nw^N=(uw)^N$,
which is always real and positive for $u,w>0$.
This leaves only three possibilities for the signature/character of the $\lambda_i$:
$(+++)$, $(+--)$ or $(+\;c\;\bar{c})$, where $c$ and $\bar{c}$ denotes complex
conjugates.
Of course with signature $(+++)$ the trace can never be zero, and this is
precisely the
case in the region $u<1/4$ and $w<\lambda_+(u)$. From this we cannot
go directly to
signature $(+--)$ because two of the eigenvalues would have to pass
through zero, in which case the determinant would become zero,
leading to a contradiction. The transition occurring at the singular line
should be to the region with $(+\;c\;\bar{c})$. From there, once two of the eigenvalues
are complex, their real part can become negative and also the trace can
become zero. This scenario is confirmed by numerical computations, but
a general, algebraic proof is at this stage still missing.

\subsection{u $\neq$ v}

Using the same mixture of analytical and numerical methods as for the case $u=v$,
and with a comparable computational effort\footnote{amongst other things, by
systematically ``scanning" the three-dimensional coupling space in $(v,w)$-planes
for various fixed values of $u$}
we have found a similar picture in the full three-dimensional parameter space with
$u\neq v$. The critical surface is determined by the same relations as for
$u=v$ above, but with $u$ substituted by $\max (u,v)$. That is, for $u<v$ the
critical surface is given by $v=1/4$ for $w\leq 1/2$
and $w=\lambda_+(v)$ for $w>1/2$, while for $u>v$ it is
given by $u=1/4$ for $w\leq 1/2$
and $w=\lambda_+(u)$ for $w>1/2$.
Since our extensive numerical and analytical checks have turned up no
contradictions to this picture, we will in the
following assume it to be correct. The critical surface is depicted in
Fig.\ \ref{zeros}b.

\section{The large-$N$ limit} \label{sec-replica}

We will now assume that we are inside the region free of singularities. In this region
the partition sum is convergent and its large-$N$ limit well defined.
The boundary of this region is the critical line (or critical surface for $u\neq v$).
We want to characterize the (non-)analytic character of the partition function
$Z_N(u,v,w,\Delta t=1)$ on this boundary
{\it after} the limit for large $N$ has been taken (see the discussion in the previous section).
Initially the evaluation of this function seems an insurmountable task, because
according to (\ref{Zinv}) and (\ref{ZZZ}) we have to sum over the {\it inverses}
of the matrix traces resulting from the application of the inversion formula.
However, it was exactly with this difficulty in mind that we introduced
the reformulation of the partition function in terms of random matrix products
in Sec.\ \ref{sec-matrices} above. Techniques available for such products of
random matrices will help us to estimate precisely the limit we are interested in,
namely
\beq
L_{-1}(u,v,w)  =
\lim_{N\rightarrow \infty} \frac{1}{N}\ln \left\langle\frac{1}{|P_N(u,v,w)|}\right\rangle
= -\ln 2 +\lim_{N\rightarrow \infty} \frac{1}{N}\ln Z_N(u,v,w,\Delta t=1),
\eeq
where, as in (\ref{Z}) above, the average has been taken over all random
sequences $\{q_j\}$.
(The reason for the notation $L_{-1}$ will become clear soon.)
The aim of this section is to compute  $L_{-1}(u,v,w)$, at least in a perturbative expansion
around the critical point (which still needs to be identified).

\subsection{Generalized Lyapunov exponents}

\begin{figure}[h]
\centering
\includegraphics[width=12cm]{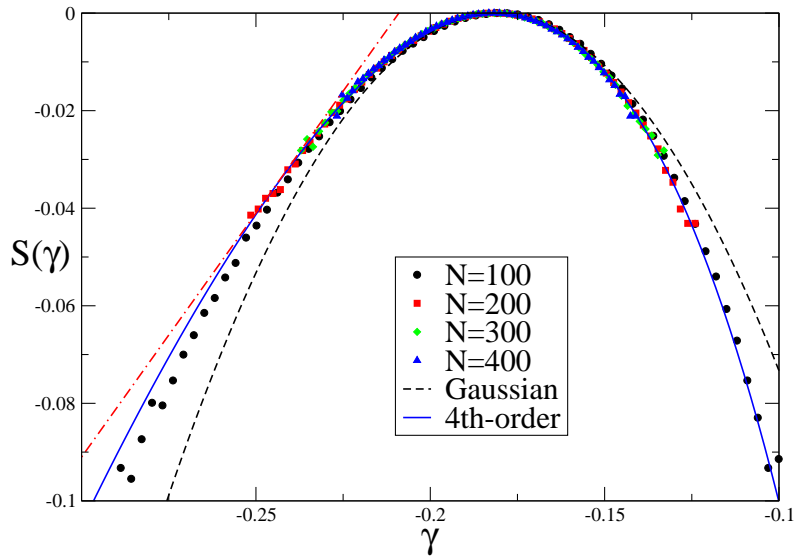}
\caption{Determining the function $S(\g)$ numerically for $u=v=0.24$ and $w=0.1$.
The Gaussian approximation
is shown as a dashed line. The blue continuous line is a fourth-order fit. The straight
(dashed-dotted) line
on the left has
slope 1, and is tangent to $S(\g)$ in $\g^*_{-1}$. In this case, the Gaussian approximation
is not sufficient to compute $L_{-1}$ from (\ref{lns}),
because in the point $\g = \g^*_{-1}$, $S(\g)$ is very different
from the Gaussian fit. The figure illustrates a ``worst-case scenario" in that for
the chosen parameter values the deviation of the actual curve from the Gaussian
approximation is maximal, see also Fig.\ \ref{replichefig}.
See appendix A-III for details on the numerical method.}
\label{Sgamma}
\end{figure}

For a given sequence $\{q_j\}$, define a real number $\gamma$ by
\beq
\gamma=\frac{1}{N}\ln |P_N|  .
\eeq
Since $\{q_j\}$ is random, $\gamma$ will be a random variable as well;
its probability distribution has, for large $N$, the general form~\cite{crisanti}
\beq
\pi_N(\gamma)\propto e^{N S(\gamma)} .
\eeq
The function $S(\g)$ is called {\it large deviations function}; it is a convex
function that has a maximum in $\g=\bar\g$, where $\bar\g$ is defined as
\beq\label{barg}
\bar\g=\lim_{N\rightarrow\infty}\frac{1}{N}\ln |P_N|
\eeq
and called the \emph{maximum Lyapunov characteristic exponent}.\footnote{In general,
the Furstenberg theorem
\cite{crisanti} guarantees that $\bar\g$
exists with probability 1 and is a non-random quantity,
$i.e.$ $\bar\g=\lim_{N\rightarrow\infty}\frac{1}{N}\langle\ln |P_N|\rangle$.
This means that $\gamma$ is a \emph{self-averaging} quantity, and that
$S(\gamma)$ must be a function peaked around $\bar\g$.
Note that instead $|P_N|=e^{N\gamma}$ is not in general a
self-averaging quantity: it will be shown later in this
section that
$\la |P_N| \ra = e^{N L_1}$ and $\la |P_N|^2 \ra = e^{N L_2}$ where the $L_n$ are $O(1)$, so that
$\la (|P_N| - \la |P_N| \ra)^2 \ra \sim e^{N L_2}$ is of the same order as $\la |P_N| \ra^2$, as long as $L_2 > 2 L_1$.
}
We will choose $S(\bar\g)=0$, so that the distribution $\pi_N(\g)$ is
normalized at the leading order for $N\to \io$:
\beq
\int d\g \, e^{N S(\g)} \sim e^{N S(\bar\g)} = 1 .
\eeq
We can then write
\beq
  \left\langle\frac{1}{|P_N(u,v,w)|}\right\rangle=\langle e^{-N\gamma}\rangle =
  \int d\gamma \, e^{N[S(\gamma)-\gamma]}\sim e^{N[S(\gamma^{\ast})-\gamma^{\ast}]}  ,
\eeq
where $\gamma^{\ast}$ is the solution to the saddle point equation\footnote{We are assuming that $S(\g)$ is
analytic and convex, so that the solution to the saddle point equation exists and is unique. Even if we cannot prove
this assumption, it is strongly supported by numerical simulations, see Fig.~\ref{Sgamma}. The same assumption guarantees
that $L_n$ is an analytic function of $n$ and is at the basis of the replica method used in
the next subsection.}
\beq
  \frac{\partial S(\gamma)}{\partial\gamma}=1  .
\eeq
Similarly, one can define the Legendre transform of $S(\g)$ (or {\it generalized
Lyapunov exponent}) by\footnote{Obviously if $\pi_N(\gamma)=\delta(\g-\bar\g)$,
then $L_n=n \bar\g$.}
\beq
\label{lns}
L_n = \lim_{N \to \io} \frac{1}{N} \ln \langle |P_N|^n \rangle =
\lim_{N \to \io} \frac{1}{N} \ln \langle e^{n\g N} \rangle = S(\g^*_n) + n\g^*_n ,
\eeq
where $\g^*_n$ is the solution of
\beq
\label{gn}
\frac{dS}{d\g}=- n .
\eeq
Clearly, for integer $n$, $e^{N L_n}=\la |P_N|^n \ra$ is the $n$-th moment of $|P_N|$.
If $S(\gamma)$ is Gaussian, $S(\g)=-\frac{(\g-\bar\g)^2}{2\s^2}$, we easily find
\beq
\label{gauss}
\gamma^{\ast}_n=\bar{\gamma} + n \sigma^2  ,
\hskip30pt
L_n = n \bar\g + \frac{1}{2} n^2 \s^2  .
\eeq
In this case the knowledge of the first two moments $L_1$ and $L_2$ is
equivalent to the knowledge of
$\bar \g$ and $\s$, and determines the full curve $L_n$, now regarded
as an analytic function for all values of $n$.
If $S(\g)$ is not exactly Gaussian, a systematic expansion of $L_n$ in powers of $n$,
\beq\label{serieL}
L_n = \sum_{k=1}^\io \frac{l_k}{k!} n^k ,
\eeq
can be obtained starting from the expansion of $S(\g)$ in powers
of $\g - \bar\g$ and solving eq.~(\ref{gn})
order by order, given that $n=O(\g - \bar\g)$.
The knowledge of the first $k$ integer moments $L_k$ is equivalent to the
knowledge of the first $k$ derivatives of $S(\g)$ in $\g=\bar\g$ and
allows us to reconstruct $L_n$ up to a given order of approximation.
Therefore, one possible way of investigation is to measure $S(\g)$ numerically,
use a Gaussian
fit to estimate $\bar{\gamma}$ and $\sigma^2$, and eventually compute the
corrections coming from the cubic, quartic, ... terms of $S(\g)$. An illustrative numerical result
for the function $S(\g)$ is plotted in Fig.~\ref{Sgamma} for $u=v=0.24$, $w=0.1$,
where the deviation from Gaussianity is maximal. The Gaussian approximation becomes
better as the critical point is approached. The insights gleaned from this numerical
analysis become most powerful when combined with a different tool for computing
the moments $L_n$, based on the so-called replica trick, to which we will turn in
the next subsection.

\subsection{Replica trick}\label{replica-sec}

A more efficient strategy is to compute the integer moments $L_n$ directly
with the help of the replica trick.
For $n$ a positive integer and for any sequence
of matrices $M_j$, $j=1,\cdots,N$, it is easy to show that
\beq
\left(\Tr\prod_j M_{q_j} \right)^n = \Tr \prod_j M_{q_j}^{\otimes n}  ,
\eeq
where $\otimes$ is the tensor product. Then, if the $M_j$ are independently
and identically distributed,
\beq
\langle |P_N|^n\rangle=\left\langle \left(\Tr\prod_jM_{q_j} \right)^n\right\rangle
=\Tr\left\langle \prod_j M_{q_j}^{\otimes n}\right\rangle =
\Tr \prod_j \la M_{q_j}^{\otimes n} \ra = \Tr \la M^{\otimes n} \ra^N \sim \nu_n^N .
\eeq
where $\nu_n$ is the largest eigenvalue of the matrix $\la M^{\otimes n} \ra$, which can
be easily evaluated for $n$ small.
Thus one has, for positive integer $n$,
\beq
L_n = \ln \nu_n .
\eeq
Knowing the function $L_n$ for integers $0 < n\leq k$ allows one to compute the first
$k$ coefficients $l_1,\cdots,l_k$ of the $n$-expansion (\ref{serieL}) simply by
solving a linear system.
This yields an approximate expression for
$\bar\g = \lim_{n\to 0} \frac{d L_n}{d n} = l_1$ and for
$L_{-1}$, which is the quantity we want to compute.
Obviously the method assumes that $L_n$ is an analytic function of $n$ and
would break down if there were a singularity at some value of $n$.

\begin{figure}[t]
\centering
\includegraphics[width=10cm]{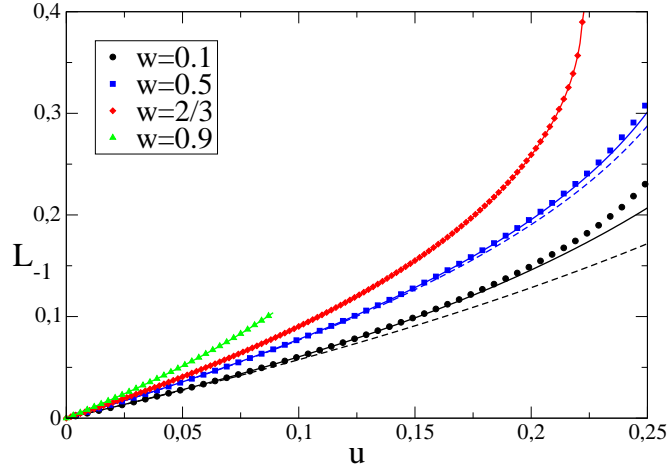}
\caption{
The numerical result for $L_{-1}$ as a function of $u(=v)$ for $w=0.1,0.5,w_c,0.9$.
For $w \geq w_c=2/3$, the simplest approximation $L_{-1}=-L_1$ (full line) works well. For $w=0.1$
and $0.5$
the Gaussian approximation, $L_{-1} = L_2 - 3 L_1$ (continuous line), is needed to approximate
the data better. For small $w$ and $u\to 1/4$ it is clear that additional
higher-order corrections are necessary, see also Fig.~\ref{Sgamma}.
}
\label{replichefig}
\end{figure}

\subsubsection{First moment}

The simplest approximation is $L_n = l_1 n$. The coefficient $l_1$ is then
equal to $L_1$ and is simply the logarithm of the largest eigenvalue of the matrix
$\langle M \rangle = (A+B)/2$, that is,
\beq\begin{split}
&\nu_1(u,v,w) = \frac{1}{4} \left( 2 + w + \sqrt{(2-w)^2 - 4(u+v)} \right) , \\
&L_1(u,v,w) = \ln \n_1(u,v,w)  .
\end{split}\eeq
This approximation is equivalent to neglecting the fluctuations of $\g$ since
\beq
e^{NL_n} = \left\langle |P_N|^n \right\rangle = e^{nNL_1}= \la |P_N|\ra ^n   .
\eeq
The function $\nu_1(u,v,w)$ has a singularity at
\beq
u+v=\left(\frac{2-w}{2}\right)^2,
\eeq
which intersects the critical surface only at the {\it critical point} $(u_c,v_c,w_c)
\equiv ( 2/9,2/9,2/3)$.
This point is therefore a good candidate for a singularity
of $L_{-1}$, if it does not cancel at the next order.

\subsubsection{Second moment: Gaussian approximation}

Including the first correction is equivalent to the Gaussian approximation and is obtained
by considering $L_n = l_1  n + \frac{1}{2} l_2 n^2$ as in (\ref{gauss}) above. One has
\beq
l_1 = \bar\g = 2 L_1 - \frac{1}{2} L_2 , \hskip.5cm
l_2 = \s^2 = L_2 - 2 L_1  ,
\eeq
with $L_1=\ln \n_1$ as before and $L_2=\ln \n_2$, the logarithm of the
largest eigenvalue of the $9\times 9$ matrix
$\langle M^{\otimes 2} \rangle = (A^{\otimes 2}+B^{\otimes 2})/2$. For our searched-for
expression this implies $L_{-1}=L_2 - 3L_1$.
We have calculated the eigenvalue $\n_2$ with the help of {\sc Mathematica} for
$u=v$, but refrain from reporting it here, because it is long and not particularly
illuminating.
However, the expressions simplify considerably on the critical line $u=v=w-w^2$,
where one has
\beq\label{nue2}
\n_2(w-w^2,w) =
\begin{cases} &w^2  \hskip200pt  w > \frac{2}{3}  , \\
&\frac{1}{4}\left[2-2w+w^2 + \sqrt{4(1-w)^2 + w^2 (2-3w)^2}\right] \hskip15pt w <\frac{2}{3}  .
\end{cases}
\eeq
What is rather remarkable about this result is that for $w>2/3$ we have
$\n_2 = \n_1^2$, \ie $L_2 = 2L_1$.
It means
that the approximation $L_n=nL_1$ is {\it exact}\footnote{To prove this statement one should
prove that $\n_n = \n_1^n$ for all $n$. Although this ought to be possible, we have
limited ourselves here to checking, by sampling random values on the critical
line above $w>2/3$, that it is
true for $n$ large but finite.} on the critical line for $w>2/3$.
On this line we then obtain $L_{-1} = -\ln \n_1 = -\ln w$.

More generally, we can write $L_{-1}=-L_1 + (L_2 - 2L_1) = -L_1 + \D L_2$ and
study the behaviour of $\D L_2$ close
to the critical line.
For later use in our scaling analysis, we
consider two linearly independent types of perturbation away from a
given point on the critical line, labelled by its $w$-coordinate $w_0$.
The first one is for $w = w_0 - \d$
and $u=v=w-w^2$, thus describing motion
along the critical line, for which we find
\beq\label{deltacase}
\begin{split}
&\n_2 \sim \left(\frac{2-w}{2}\right)^2 + \frac{1}{2} \d^2 , \\
&\D L_2 = \begin{cases}
0 \hskip98pt \text{ for } w_0 > 2/3,\\
\kappa_1 \delta^2+O(\delta^3) \hskip45pt \text{ for } w_0 = 2/3,\\
\kappa_2+\kappa_3\delta +O(\delta^2) \hskip27pt \text{ for } w_0 < 2/3.
\end{cases}
\end{split}\eeq
The second one moves away from the critical line inside the diagonal
plane $u=v$, and toward the interior of the singularity-free region
according to $u = w_0-w_0^2-\ee$, yielding
\beq\label{epscase}
\D L_2 = \begin{cases}
\kappa_4 \ee^4 + O(\ee^5) \hskip62pt \text{ for } w_0 > 2/3  , \\
\kappa_5 \ee + \kappa_6 \ee^{3/2} + O(\ee^2) \hskip29pt \text{ for } w_0 = 2/3  , \\
\kappa_7 + \kappa_8 \ee + O(\ee^2) \hskip45pt \text{ for } w_0 < 2/3
       \end{cases}
\eeq
where the $\kappa_i$ in (\ref{deltacase}) and (\ref{epscase}) are numerical constants.
Interestingly, these corrections
do not change the divergent term in the first derivatives of
$L_{-1}$ with respect to $u$ and $v$ at the critical point, which govern
the presence or otherwise of an infinite-volume limit. Neither, since $\nu_2$
according to (\ref{nue2}) has no further singularities, can singularities appear at
second order at points with $w_0\not= 2/3$.
In summary, since all correction terms give only finite contributions to the
derivatives, it is clear that our first-order analysis already correctly identified
$(u_c,v_c,w_c)= ( 2/9,2/9,2/3)$ as the only point at which an
infinite-volume limit exists.

Since $\D L_2$ is proportional to $\ee$ at the critical
point, we expect it to contribute to the calculation of
the Hamiltonian. This would at first seem to necessitate an analytic determination for
the eigenvalues of $(A^{\otimes 2}+B^{\otimes 2})/2$ for  $u\neq v$,
which is currently out of reach. Fortunately,
all that is needed is a perturbative evaluation after inserting an ansatz for
the scaling of $u$, $v$ and $w$ near
the critical point, which is a perfectly feasible task we will perform in Sec. 9 below.

\subsubsection{Higher moments: beyond the Gaussian approximation}

The precision in determining $L_{-1}$ can be improved further by
computing $L_3$, $L_4$, etc. In Fig.~\ref{replichefig} we
report our numerical findings for $L_{-1}$ (see appendix \ref{App-numeric}),
together with the first- and second-moment approximation just described.
For $w>2/3$ the first-order approximation is excellent, and the numerical
deviations from $\D L_2 \sim 0$
for $w>2/3$ and any value of $u$ are very small.
Around $w=1/2$, the Gaussian correction starts to
be observable. For $w<1/2$ and $u \sim 1/4$, the correction is large, c.f.
Fig.~\ref{Sgamma}, and higher-order corrections have to be taken into account
to correctly reproduce the numerical result.

For approaches to the critical point, we have found that $\D L_r$, the incremental correction
for the calculation of $L_{-1}$ which results from adding the $r$'th one to
that of the first $r-1$ moments, scales like $\ee^{r/2}$ and thus
is not expected to contribute to the calculation of the Hamiltonian. We have confirmed
this by perturbative calculations for $L_3$.


\section{Continuum limit, canonical scaling and properties of the slices} \label{sec-continuum}

As usual in dynamically triangulated models of quantum gravity, our next step will be to search
for a continuum limit in which the details of the discretization procedure will be ``washed out",
and only universal, physical properties will remain. A necessary part of this limit is to take our
short-distance cutoff, the edge length $a$, to zero, while keeping the {\it physical} spacetime
volume $\mathcal V$ finite. This is only possible if the number of building blocks
in the simplicial manifold (the {\it discrete} spacetime volume $V$) simultaneously goes to
infinity. More precisely, we have to take the continuum limit
following a trajectory (parameterized by $a$) in the coupling-constant space which ends up
(for $a=0$) in a point
where the expectation value of the number of building blocks diverges, and which approaches this point
in such a way that the physical volume stays finite. In other words, we have to renormalize the bare
cosmological (and possibly other coupling) constants so that relevant
physical quantities remain finite.

Our first step, the identification of a critical point with suitable properties, follows from the
discussion of the previous section: the only point where the derivatives of the partition
function\footnote{Remember that the average volume
is given by the derivative of the logarithm of the partition function with respect to the bare cosmological
constant, which translates into a linear combination of $u\frac{\partial}{\partial u}$,
$v\frac{\partial}{\partial v}$ and $w\frac{\partial}{\partial w}$, see (\ref{sandwich-volume}).}
diverge is $(u_c,v_c,w_c)= ( 2/9,2/9,2/3)$.
As a second step we will assume a canonical scaling, where the critical point is approached according
to the canonical dimensions of the corresponding continuum coupling constants. To lowest
non-trivial order in $a$, these are by definition
\beq \label{canonical}
k\simeq k_c+\frac{a}{G}, \hspace{.3cm} \ln x\simeq -\l_{b,c}-a^2 X, \hspace{.3cm} \ln y\simeq -\l_{b,c}-a^2 Y,
 \hspace{.3cm} \l\simeq \l_c+a^3\L
\eeq
for the inverse Newton, the two boundary cosmological and the bulk cosmological constants.
In addition, a canonical scaling would require $T=a t$ and ${\cal N}=aN$ for the time and
spatial extensions.

Translated into the variables $u$, $v$ and $w$ (which were defined in the discussion
preceding (\ref{Z_N})), the canonical scaling becomes
\beq \label{canonical2}
 u=\frac{2}{9}\ e^{2 a c_1/G-2a^2 X-2 a^3 b_1\L}, \hspace{.3cm}
 v=\frac{2}{9}\ e^{2 a c_1/G-2a^2 Y-2 a^3 b_1\L}, \hspace{.3cm}
 w=\frac{2}{3}\ e^{-a c_2/G-a^3 b_2\L} .
\eeq
Already without detailed inspection we can anticipate difficulties with the standard
interpretation of the bulk and boundary cosmological constants as the couplings
conjugate to the bulk volume and boundary areas.
Since $k$ scales with the lowest power of $a$ when approaching the critical point,
terms proportional to $1/G$ will be the ones which survive in the lowest-order expressions
of the continuum limit,
unless unexpected cancellations occur.

We can check this immediately by looking at what this particular choice of scaling
implies for geometrical quantitites like the volume of sandwich geometries.
As argued in the previous section, the second moment in the replica method only contributes to the
next-to-leading order, which therefore we will need for the Hamiltonian,
but not to recover the continuum expression for the volume.
This means that as a starting point for our computation we can take
the simple expression
\beq
Z_N\sim e^{-N\ln \n_1(u,v,w)}
\eeq
as the partition function of a sandwich geometry. Assuming canonical scaling for
the three-volume and using previous definitions, we have
\beq \label{sandwich-volume}
\begin{split}
\langle {\mathcal V} \rangle &= \lim_{a\rightarrow 0} a^3 \langle V \rangle,\\
\langle V \rangle &= \left ( -\frac{\partial}{\partial \l} \ln Z_N\right )=
\left ( 2 b_1 u\frac{\partial}{\partial u}+ 2 b_1 v\frac{\partial}{\partial v} +
 b_2 w\frac{\partial}{\partial w} \right ) \ln Z_N ,
\end{split}
\eeq
which after inserting the scalings (\ref{canonical2}) and taking the limit $a\rightarrow 0$ yields
\beq
\langle {\mathcal V} \rangle = a^{5/2} N \frac{2b_1+b_2}{4\sqrt{(c_2-2c_1)/G}}
\eeq
to lowest order in $a$.
As expected, only the renormalized Newton constant $G$ appears in the continuum expression
for the volume. In addition, this expression
indicates an anomalous scaling of either the time variable $t$ or the variable $N$ measuring the
linear spatial extension. For example, if we insist on the canonical scaling
${\cal N} = a N$ for the finite continuum counterpart of $N$, we find that the volume
goes to zero like $a^{3/2}$ instead of $a$ as would be required for a canonical scaling $T=at$
of the time variable.

If instead we set $k=0$ identically, corresponding to working with a bare action which only
contains a cosmological term
the sandwich volume scales canonically according to
\beq
\langle {\mathcal V} \rangle = a^2 N \frac{2b_1+b_2}{4\sqrt{X+Y}}=a {\cal N }\frac{2b_1+b_2}{4\sqrt{X+Y}}.
\eeq

Since we do not want to touch the canonical scaling of the three-volume, which would affect
the interpretation of the model in a fundamental way, possible solutions are
to either assume a non-canonical scaling only for Newton's constant,
for example, $k\simeq k_c+\left(\frac{a}{G}\right)^2$, or
to insist on canonical scaling but choose the constants $c_i$ such that $c_2=2 c_1$ and the
order-$a$ terms cancel each other.
In the first case we would obtain
\beq
\langle {\mathcal V} \rangle = a^2 N \frac{2b_1+b_2}{4\sqrt{X+Y-(2c_1-c_2)/G^2}},
\eeq
and in the second
\beq
\langle {\mathcal V} \rangle = a^2 N \frac{2b_1+b_2}{4\sqrt{X+Y-\frac{3}{4}c_2^2 /G^2}},
\eeq
which are the same, up to a finite rescaling of $G$.
The scaling now {\it is} canonical, and the sandwich volume is governed by both the
boundary cosmological constants and the Newton constant. We will see below that
related issues appear when we try to derive the continuum Hamiltonian.

Another quantity one can consider is the total, integrated scalar curvature
${\cal R}_{tot}\equiv\int d^3 x \sqrt{g}\ {\cal R}$.
Its counterpart at the discrete level is exactly the term $R_{tot}$
multiplying $k$ in the gravitational action, for which we have used the standard
Regge prescription. We can express the average total curvature
as a derivative with respect to $k$ of the partition function, and then take the continuum limit.
The analogues of relations (\ref{sandwich-volume}) are given by
\beq
\begin{split}
\langle {\cal R}_{tot} \rangle &= \lim_{a\rightarrow 0} a\langle R_{tot}\rangle,\\
\langle R_{tot}\rangle &= \frac{\partial \ln Z_N}{\partial k}=
 -c_3 N+\left( c_1\left( 2u\frac{\partial}{\partial u}+ 2v\frac{\partial}{\partial v}\right)
 -c_2w\frac{\partial}{\partial w}\right)\ln Z_N .
\end{split}
\eeq

Since our spacetime sandwiches in the continuum limit have only infinitesimal
thickness and thus infinitesimal volume, it is more appropriate to work with
the average curvature per volume,
\beq
\bar{\cal R} \equiv \frac{\langle {\cal R}_{tot} \rangle }{\langle {\cal V}\rangle } .
\eeq
It is a well-known feature of dynamically triangulated models of gravity that
this quantity generically diverges like $a^{-2}$, unless specific
cancellations occur.
We find here not only the same behaviour, but even the same algebraic expression
to leading order, namely,
\beq
\bar{\cal R} =  \frac{1}{a^2}\left(\frac{2c_1-c_2}{2b_1+b_2}\right),
\eeq
regardless of
whether we assume the scaling $k=0$, the canonical $k\simeq k_c+a/G$, or the
non-canonical $k\simeq k_c+(a/G)^2$.
However, in line with our earlier remarks, {\it if} one makes the choice\footnote{
This can be achieved by fixing the finite relative factor $r$ between the
time- and space-like (squared) edge lengths of the triangulations (c.f. appendix A-I).
Using eq.\ (\ref{alphabeta2}), it corresponds to taking $r\simeq 0.724$.} $2 c_1=c_2$,
a different, less divergent, behaviour is found.
Combining this condition and canonical scaling, we obtain
\beq
\label{rav1}
\bar{\cal R} =  \frac{1}{a}\left(\frac{3c_2^2/G}{4b_1+2b_2}
+\left( \frac{c_2-4c_3}{2b_1+b_2} \right)\sqrt{X+Y--\frac{3}{4}c_2^2 /G^2}\
\right) .
\eeq
For $k=0$ we find instead
\beq
\label{rav2}
\bar{\cal R} =  \frac{1}{a} (c_2-4c_3)\frac{\sqrt{X+Y}}{2b_1+b_2}
-\frac{3}{2}\left(\frac{c_2}{2b_1+b_2}\right)(X+Y).
\eeq
We will encounter somewhat similar scaling relations in the discussion of
the quantum Hamiltonian in Sec.\ \ref{sec-hamiltonian} below.


\section{Transfer matrix for areas}\label{sec-gluing}

One way to construct the propagator for finite times $t$ of a statistical model is
by iteration of the transfer matrix $\hat T$, which in our case would take the
form  $\langle g_2|\hat T^t|g_1\rangle$. Although the determination of the complete
transfer matrix for our three-dimensional model is out of reach, we have already
argued in Sec.\ \ref{sec-part} that most of the detailed dependence of the propagator
on the boundary geometries $g_i$ will be dynamically irrelevant, because
of the absence of local degrees of freedom in the three-dimensional gravity theory.
Following \cite{ABAB1}, keeping track only of the boundary areas $A_i$ in the one-step propagator
as we have been doing throughout this work, should be enough\footnote{possibly up to an additional
dependence on global Teichm\"uller
parameters}  to determine the finite-time
behaviour of the propagator in the continuum limit.
Introducing the ``area states"
\beq
|A\rangle = \frac{1}{\sqrt{\NN(A)}}\sum_{g_{|A}}|g_{|A}\rangle,
\eeq
where $\NN(A)$ is the number of triangulations of a given area $A$ and $g_{|A}$ any triangulation
with a given total number of triangles $A$, we obtain that
\beq \label{Z-Sqrt}
\begin{split}
Z(x,y,\Delta t=1)&=\sum_{A_1,A_2}x^{A_1}y^{A_2}\sum_{g_{|A_1},g_{|A_2}}\langle g_{|A_2}|\hat T|g_{|A_1}\rangle=\\
&=\sum_{A_1,A_2}x^{A_1}y^{A_2}\langle A_2|\hat T|A_1\rangle\sqrt{\NN(A_1)\NN(A_2)}.
\end{split}
\eeq
Assuming now in line with our earlier reasoning (see also \cite{ABAB1}) that
\beq \label{conjecture}
\langle A_2|\hat T|g_{|A_1}\rangle-\langle A_2|\hat T|g'_{|A_1}\rangle\rightarrow 0 \hskip29pt \text{for}
\hskip15pt A_2,A_1\rightarrow\infty,
\eeq
for all pairs $g$, $g'$ of boundary geometries of the same area, the completeness relation
\beq
\int dA |A\rangle\langle A|=\mathbb{I}
\eeq
holds in the large-area limit and we can use $\langle A_2|\hat T|A_1\rangle$
as our transfer matrix.
One remaining problem is the appearance of the square-root term $\sqrt{\NN(A_t)\NN(A_{t+1})}$
in (\ref{Z-Sqrt}), which
we will have to deal with either before or after performing the inverse Laplace transform of $Z(x,y,\Delta t=1)$.
Given that the number of triangulations of a given area $A$ scales like
\beq \label{entropy-A}
\NN(A)\sim A^{-\g} e^{\l_0 A},
\eeq
with $\gamma =1/2$ as we will show at the end of this section,
one possibility is to apply to  $Z(x,y,\Delta t=1)$ a fractional derivative
operator
\beq
\sum_{A_1,A_2}x^{A_1}y^{A_2}\langle A_2|\hat T|A_1\rangle=\left(\frac{\partial}{\partial\ln x}\right)^{\g /2}
\left(\frac{\partial}{\partial\ln y}\right)^{\g /2} Z(x,y,\Delta t=1).
\eeq
There are several definitions of fractional derivatives in the literature; what we need here is an operator
$D^{\a}_x$ such that $D^{\a}_x e^{A x}=A^{\a} e^{A x}$, so that applying $D^{\g/2}_{\ln x}D^{\g/2}_{\ln y}$
term by term to (\ref{Z-Sqrt}) we get rid of the entropy factors (the remaining exponential term
$e^{\l_0 A}$ is unproblematic since it only shifts the location of the critical point).
We need yet another property for our fractional derivative, namely, a ``chain rule", since the
final expression for $Z(x,y,\Delta t=1)$ will involve some function of $x$ and $y$.
In particular, since the final expression is expanded in powers of $\ln x$, it would be nice if
the derivative would act on powers with the simple rule $D^{\a}_x x^n\propto x^{n-\a}$.
An explicit representation of a fractional derivative operator with the desired properties exists
and is reviewed in appendix \ref{App-fracder}.

Unfortunately, at the relevant value $\g=1/2$, for some terms in the final expression for
$Z(x,y,\Delta t=1)$ the fractional derivatives suffer from convergence problems.
These can be circumvented in a somewhat {\it ad-hoc} fashion by integrating by parts or
by introducing a regularization in the fractional integrals,
which should be removed only after the ordinary derivative has been applied.
(As explained in appendix \ref{App-fracder},
the fractional derivative is actually defined as an ordinary derivative acting on a fractional integral.)

A cleaner alternative to get the correct Hamiltonian is to first work out the inverse
Laplace transform and then identify the contribution coming from the entropy factor
and remove it.
This method on the other hand requires some special care for the following reason.
Starting from (\ref{Z-Sqrt}), keeping only the two lowest orders in $a$ (see (\ref{HfromT})), and doing an inverse
Laplace transform will produce $\NN(A_1)\d(A_1-A_2)$ at the lowest order and some differential operator acting
on $\d(A_1-A_2)$ at the next order. If we divided this operator by $\NN(A_t)$ to obtain the Hamiltonian,
we would make a mistake since at next-to-leading order $A_2\simeq A_1 +a \d A$, which needs to be
taken into account.
To do so let us rewrite the coefficients of the power series (\ref{Z-Sqrt}) as follows:
\beq \label{H-comm}
\begin{split}
\langle A_2|\hat T|A_1\rangle &\sqrt{\NN(A_1)\NN(A_2)} =
\langle A_2|\sqrt{\NN(\hat A)}\ \hat T\ \sqrt{\NN(\hat A)}\ |A_1\rangle = \\
 &=\langle A_2|\NN(\hat A)\hat T|A_1\rangle
 + \langle A_2|\sqrt{\NN(\hat A)}\left[\hat T,\sqrt{\NN(\hat A)}\right] |A_1\rangle =\\
 &=\NN(A_2)\langle A_2|(1-a\hat H)|A_1\rangle
 -a\sqrt{\NN(A_2)}\langle A_2|\left[\hat H,\sqrt{\NN(\hat A)}\right]|A_1\rangle +O(a^2) ,
\end{split}
\eeq
where we have introduced the area operator $\hat A|A_1\rangle=A_1|A_1\rangle$.
If we define the auxiliary Hamiltonian $\hat H'$ by
\beq \label{H-aux}
\langle A_2|\hat T|A_1\rangle \sqrt{\NN(A_1)\NN(A_2)} =: \NN(A_2)\langle A_2|(1-a\hat H')|A_1\rangle
 +O(a^2) ,
\eeq
we can invert (\ref{H-aux}) by using (\ref{H-comm}) to find the real Hamiltonian
\beq \label{H-inversion}
\hat H = \hat H' + \sqrt{\NN(\hat A)} \left[\hat H',\frac{1}{\sqrt{\NN(\hat A)}}\right] .
\eeq
It turns out that also with this method convergence problems appear in some of the
inverse Laplace transforms to be performed. Fortunately, since the divergences
in the fractional derivative method and in the auxiliary Hamiltonian method appear
in different terms, they can be regularized consistently by requiring the end result
in both methods to agree.

We conclude this section by showing that $\g=1/2$ in the scaling relation
(\ref{entropy-A}) for the number of boundary geometries.
The entropy of this class of triangulations can be deduced from its leading critical behaviour.
Summing over all the triangulations with a cosmological weight -- as is appropriate for evaluating
the gravitational partition function in two dimensions -- one has
\beq \label{gamma}
\sum_{\TT} e^{-\l A} \sim \sum_A A^{-\g} e^{-(\l-\l_0)A} \sim (\l-\l_0)^{\g-1} .
\eeq
Of course, the boundaries of our model are nothing but standard (1+1)-dimensional causal
dynamical triangulations with periodically identified time. For precisely this ensemble
we have already established earlier the relation (\ref{Z-2d-1}), with whose help we will now
be able to find $\g$ analytically. Summing over $t$ to include all triangulations (and
including a factor of $1/2^t$, which is necessary for convergence),
we simply get
\beq
\sum_t \frac{1}{2^t} Z^{2d}_t = \frac{1}{\ln 2 + \ln \l_+(u)} \sim \frac{1}{\sqrt{1-4 u}},
\eeq
since there is no power-like subleading behaviour in $t$.
Comparing with (\ref{gamma}), we deduce that for the (1+1)-dimensional boundaries we have $\g=1/2$.


\section{The Hamiltonian} \label{sec-hamiltonian}

The continuum dynamics of the model is encoded in the Hamiltonian operator, which
can be derived by taking the continuum limit of the transfer matrix, as we explained before.
To obtain the area-to-area transfer matrix we will first remove
the $N$-dependence from the partition function, to avoid that the propagator will be dressed with
the exponential decay of the one-dimensional boundaries correlation.
To do so we normalize\footnote{The normalization is needed in order to take
the continuum limit of the
partition function at fixed ${\cal N}$. An analogous situation is found in
(1+1) dimensions (see, for example, \cite{difra-calogero}). There, if we start
with the generating function (\ref{gen2bis}) for the transfer matrix, we find
to leading order
{\it twice} the identity operator in the continuum limit. Consequently, if we iterate (\ref{gen2bis})
$t$ times, we will get a multiplicative factor $2^t$, which has to be removed
if we want to obtain the continuum $T$-propagator (the analogue of our
partition function at fixed ${\cal N}$).} $Z_N$
and sum\footnote{Since we are evaluating the continuum
limit, it is perfectly legitimate to sum over $N$ after having performed the large-$N$ limit.
Performing the continuum limit first and then integrating over the continuous variable associated to $N$
would lead to the same result.
What is more important is that we are not including any power-like corrections
to the behaviour of $Z_N$, which could in principle occur. This is justified by the fact that it can be
shown with the replica method that
they are absent for the positive integer $n$ analogues of $Z_N$.} over $N$, yielding
\beq
 Z(u,v,w)=\sum_N e^{-N (L_{-1}(u_c,v_c,w_c)-L_{-1}(u,v,w))} 
 \sim \frac{1}{L_{-1}(u_c,v_c,w_c)-L_{-1}(u,v,w)}
\eeq
(note that $L_{-1}(u_c,v_c,w_c)-L_{-1}(u,v,w)>0$).
To derive the quantum Hamiltonian in the ``$X$-representation",
we substitute the known function $L_{-1}(u,v,w)$ as well as an ansatz for
the scaling relations into the evolution equation
\beq
\psi(x,t+1)=\oint\frac{dy}{2\p i y} Z(x,y^{-1},\Delta t=1)\psi(y,t)
\eeq
for the wave function
(remember that $x$ and $y$ were absorbed in $u$ and $v$),
and evaluate its continuum limit to order $a$, namely,
\beq \label{H-formula}
 \left(1- a\hat H_X+{\cal O}(a^2)\right)\psi(X,T)=-a^2\int_{-i\infty+\m}^{+i\infty+\m}\frac{dY}{2\p i}
 Z(X,-Y;a)\psi(Y,T).
\eeq
In (\ref{H-formula}), $\m$ is chosen such that the integration contour lies to the right of the
singularities of $\psi(Y,T)$ and
to the left of those of $Z(X,-Y;a)$, around which we have to close it.

To show that to lowest order the identity operator is reproduced and to extract
the Hamiltonian at the next order,
we clearly need the first two orders in $a$ of $L_{-1}(u,v,w)-L_{-1}(u_c,v_c,w_c)$.
As discussed in Sec.\ \ref{replica-sec} above, to get the order-$a^2$ terms
right we need to use the replica trick
up to the second moment approximation, $i.e.$ $L_{-1}=L_2-3 L_1$.
It does not matter that we do not have a closed analytic expression available, since we will
only need the expansion
up to order $a^2$. This can be found perturbatively around the solution at the critical point
(where we {\it can} solve for the eigenvalues of $(A^{\otimes 2}+B^{\otimes 2})/2$), with the
eigenvalue problem perturbed in accordance
with the chosen scaling of coupling constants.

In anticipation of difficulties with the scaling of $k$ we first discuss the results for $k=0$.
Given the analytic expression of $L_1$ and after computing
\beq
L_2(X,Y,G,\L)=\ln \left( \frac{4}{9}+a\frac{4}{9}\sqrt{X+Y}+ \frac{a^2}{27}\left(
\frac{4 X^2 + 5 X Y + 4 Y^2}{X+Y}+\frac{(4b_1+2b_2)\L}{\sqrt{X+Y}} \right)  \right)+O(a^3),
\eeq
we have all we need to calculate the partition function, resulting in
\beq \label{Z-k=0}
a^2 Z(X,-Y;\L)=\frac{2 a}{\sqrt{X-Y}}+a^2\left( \frac{5}{6}+\frac{XY}{(X-Y)^2}-\frac{\L}{(X-Y)^{3/2}}
 \right)+O(a^3),
\eeq
where we have absorbed a finite numerical factor in $\Lambda$.
This expression is in perfect agreement with our previous discussion, since due to the entropy factor
$\NN(A)\sim A^{-1/2} e^{\l_0 A}$ from the boundaries we expect the leading term to have an inverse
square-root singularity if the transfer matrix is to reduce to the identity at lowest order.
Using the fractional derivative method before the integration in (\ref{H-formula}) or, alternatively,
the method of the auxiliary Hamiltonian one finds
\beq \label{H}
\hat H_{\cal A}=-{\cal A}^{\frac{3}{2}}\frac{\partial^2}{\partial {\cal A}^2}
-\frac{3}{2}{\cal A}^{\frac{1}{2}}\frac{\partial}{\partial {\cal A}}
-\frac{1}{16}\frac{1}{{\cal A}^{\frac{1}{2}}}+\L \ {\cal A}
\eeq
for the quantum Hamiltonian in the ``${\cal A}$-representation" (with $\cal A$ denoting
the (finite) continuum area). This looks like a bona-fide Hamiltonian, with a second-order
kinetic term and a potential term depending on the cosmological constant.
In the absence of a dimensionful
Newton coupling $G$, the factor ${\cal A}^{\frac{1}{2}}$ multiplying the kinetic term must be
present for dimensional reasons.
Note also the appearance of the factor $3/2$ in front of the first-order derivative, ensuring
the self-adjointness of the Hamiltonian with respect to the trivial $d{\cal A}$-measure.
By introducing a new variable
\beq \label{L-variable}
L=4\sqrt{{\cal A}}
\eeq
and simultaneously defining new wave functions\footnote{With such a change of wave functions
we make sure that the measure is preserved, in the sense that
\beq
\int d{\cal A}\ \psi_1({\cal A})\psi_2({\cal A}) = \int dL\ \phi_1(L) \phi_2(L) \nonumber.
\eeq
Note that we also require
\beq
\int d{\cal A}\ \psi_1({\cal A})\hat H_{\cal A} \psi_2({\cal A}) = \int dL\ \phi_1(L)\hat H_L \phi_2(L) ,\nonumber
\eeq
which implies
\beq
\hat H_L=\hat H_{{\cal A}\rightarrow L}+\sqrt{L}\left[\hat H_{{\cal A}\rightarrow L},\frac{1}{\sqrt{L}}\right],
 \nonumber
\eeq
with the $\cal A$ in the Hamiltonian substituted according to (\ref{L-variable}).          }
\beq \label{L-function}
\phi(L)=\frac{\sqrt{2 L}}{4} \psi(\frac{L^2}{16}),
\eeq
the Hamiltonian becomes
\beq \label{H-L}
\hat H_L=-L \frac{\partial^2}{\partial L^2}-\frac{\partial}{\partial L}+\frac{\L}{16} L^2 ,
\eeq
with the $1/L$-term having disappeared from the potential.
Apart from the cosmological term, which has the appropriate dimension for a
(2+1)-dimensional Hamiltonian,
this is exactly the quantum Hamiltonian of the (1+1)-dimensional CDT
model \cite{2d-ldt} (see also \cite{difra-calogero}
for a similar transformation of variables).

Since by setting $k=0$ we have not included any spacetime-curvature
term in the discrete action, it is clear that the kinetic terms in (\ref{H}) and (\ref{H-L})
have their origin in the ``entropy" of configurations, or, using a continuum language,
in the non-trivial path integral measure underlying the dynamically triangulated model.
This is also underlined by the absence of such a kinetic term from a related (1+2)-dimensional
model considered and solved in \cite{difra-hardobj}, which uses the same product structure as our model,
but works instead with a {\it fixed}, flat base manifold.
Rather intriguingly, this gravity-inspired model (for finite $t$) can be related through
an inversion formula to a problem of hard hexagons on a regular triangulation.
Its one-step propagator can be seen to resurface in our model as contributing just
one of the terms in the partition function, namely, $\tilde Z_N=1/Tr(AB)^{N/2}$.
Solving the model in the large-$N$ limit does not require the replica trick, and simply leads to
$\tilde Z_N\sim 1/\l_{max}^{N/2}$, where $\l_{max}$ is the largest eigenvalue of the matrix $AB$.
It is straightforward to extract the Hamiltonian, which only contains
a term proportional to the area, and no derivatives, which implies that the area is not a dynamical
quantity in the continuum
limit\footnote{The mapping to the hard hexagon model of \cite{difra-hardobj} suggested
a fractal dimension $d_f=12/5$ for the (1+2)-dimensional simplicial complexes.
The relation with our results for the associated one-step propagator
is currently unclear. We conjecture that in our model, where we sum over all base triangulations,
the fluctuations of the base serve as a stabilizer for the geometry and lead to an effective,
non-anomalous dimension $d_f=3$ for finite times.
}.

Returning to the analysis of our model,
the case with $k\neq 0$ raises issues similar to those already encountered when
computing the spacetime volume and curvature.
The $a/G$-term dominates the continuum limit, giving rise to
\beq
Z(X,-Y;G,\L)=\frac{2 a^{3/2}}{3\sqrt{(c_2-2c_1)/G}}+
 a^2 \frac{(6c_1-5c_2+8c_3)/G-4\sqrt{4X-4Y-3(c_2/G)^2}}{(36c_1-18c_2)/G}+O(a^{5/2}),
\eeq
which is difficult to interpret within the scheme of the previous section (and to this order does not
even contain any reference to the cosmological constant $\Lambda$!)
Things simplify somewhat once we choose $c_2=2c_1$,
in which case we obtain
\beq
\label{zxyg}
Z(X,-Y;G,\L)=\frac{2 a}{4
c_3-c_2+\sqrt{X-Y-\frac{3}{4}\left(\frac{c_2}{G}\right)^2}}+
 O(a^2) .
\eeq
We have not bothered to include the next term in the expansion, because already
the lowest-order expression gives nothing like the
desired form ${\cal N}({\cal A})\delta({\cal A}-{\cal A'})$ upon performing an inverse
Laplace transform. It is not totally inconceivable that the offensive terms in
(\ref{zxyg}) could still be ``argued away". Note that we have encountered the numerical
term $(4 c_3-c_2)$ previously in the expressions (\ref{rav1}) and (\ref{rav2}) for the average
curvature. In particular, {\it if} we had a good reason for why $(4 c_3-c_2)$ could be set
to zero (which we currently do not), it would imply that in the case $k=0$ the average curvature was
turned into a finite expression. However, even if  $(4 c_3-c_2)$ could be made to vanish, it would still
leave us with the problem of having to absorb the shift in the location of the singularity
from $Y=X$ to $Y=X-\frac{3}{4}\left(\frac{c_2}{G}\right)^2$. In terms of the ${\cal A}$-representation,
the shift can be related to an overall multiplicative operator, which replaces the simple
transfer matrix $e^{-a\hat H_{\cal A}}$
by $e^{-a\hat H_{\cal A}}e^{\frac{3}{4}\left(\frac{c_2}{G}\right)^2\hat{\cal A}}$.
This might be related to a redundancy among the renormalized couplings of
the model inherent in the ansatz (\ref{canonical}), (\ref{canonical2}), but we have not yet
been able to make the relation precise.



\section{Going beyond the area representation} \label{microcan}

Up to now we have (through the boundary cosmological constants) only taken the area information
of the two-dimensional boundary geometries into account, but we expect from the
canonical analysis for cylindrical topology
also a Teichm\"uller parameter as a dynamical degree of freedom (see, for instance, \cite{polchinski}).
Although it does not seem possible to define
the Teichm\"uller parameter for a general triangulation exactly, a suitable discrete
implementation for the triangulated cylinder should be the ratio between the area of the cylinder
and the length of the cycle (the boundary of the cylinder).
The latter variable in our model is given by the number of matrices $A$ or $B$ (for the incoming
and outgoing cylinder respectively) appearing in the traces.
The problem of keeping track of the Teichm\"uller parameter is then reduced to that of keeping track
of the number of matrices $A$ or $B$ (as opposed to their sum $N=A+B$ as we have
been doing up to now).
This task is in principle solvable with the help of the so-called {\it microcanonical method}
\cite{paladin}.

The microcanonical method is based on a trick similar to the replica one.
We consider the ensemble of sequences $\{q_j\}$ such that $N_p = pN$
variables are equal to $0$, with $p\in [0,1]$,
\ie there are $N_p$ matrices $B$ and $N_{1-p}$ matrices
$A$.
One can define generalized Lyapunov exponents in the microcanonical ensemble
according to
\beq
\begin{split}
&L_n(p)=\lim_{N\to\io} \frac{1}{N} \ln \langle |P_N|^n \rangle_{pN} \ , \\
&\langle |P_N|^n \rangle_{pN} = \frac{1}{\matrice N \\ pN \ematrice} \sum_{\{q_j\}_{pN}}
\left( \Tr \prod_{j=1}^N M_{q_j} \right)^n \ ,
\end{split}
\eeq
where the sum is only over sequences which contain $pN$ matrices $B$.
The microcanonical method is based on the identity
\beq
\begin{split}
\sum_{\{q_j\}_{pN}} \left( \Tr \prod_{j=1}^N M_{q_j} \right)^n &=
\sum_{\{q_j\}_{pN}} \Tr \prod_{j=1}^N M^{\otimes n}_{q_j} =
\Tr \frac{1}{pN!} \left. \frac{d^{pN}}{dx^{pN}} (A^{\otimes n}+xB^{\otimes n})^N
\right|_{x=0} \\
&=\frac{1}{2\pi i} \oint_C dz \frac{\Tr(A^{\otimes n}+zB^{\otimes n})^N}{z^{pN+1}} \ ,
\end{split}
\eeq
where $C$ is a contour in the complex plane containing the point $z=0$.
Calling $\nu_n(z)$ the largest eigenvalue of the matrix
$(A^{\otimes n}+zB^{\otimes n})/2$, we have
\beq
\sum_{\{q_j\}_{pN}} \left( \Tr \prod_{j=1}^N M_{q_j} \right)^n \sim
\frac{1}{2\pi i} \oint_C dz \ e^{N [\ln 2 \nu_n(z) - p \ln z]} \sim
e^{N [\ln 2 \nu_n(\zeta_p) - p \ln \zeta_p]} \ ,
\eeq
where the saddle point $\zeta_p$ is the solution of
\beq \label{saddle/z}
\frac{d\ln \nu_n(z)}{d\ln z} = p \ .
\eeq
Expanding the binomial by means of the Stirling formula we have
\beq
\matrice N \\ pN \ematrice \sim e^{-N[p\ln p + (1-p) \ln (1-p)]} \= e^{N\s(p)} \ ,
\eeq
and can finally write
\beq \label{Lnp}
L_n(p) = \ln 2 \nu_n(\zeta_p) - p \ln \zeta_p - \s(p) \ .
\eeq
This shows that the generalized Lyapunov exponents can be evaluated analytically also in
the microcanonical ensemble.

We would now like to use this property to compute the partition function
\beq
\begin{split}
Z(\m_1,\m_2) &=\sum_{R_1,R_2} e^{-\m_1 R_1-\m_2 R_2} Z_{R_1,R_2}=\sum_N \sum_{R<N}
e^{-\m_1 (N-R)-\m_2 R} Z_{N,R}\sim\\
 & \sim\sum_N \int_0^1 dp\ e^{-\m_1 (1-p)N -\m_2 pN} Z_{N,p}\sim \sum_N
 \frac{1}{2^N}e^{-\m_1 N}\int_0^1 dp\ e^{-(\m_2-\m_1)N} e^{N(L_{-1}(p)+\s(p))}\ ,
\end{split}
\eeq
where we have introduced the lengths $R_1\equiv N-R\equiv (1-p)N$ and $R_2\equiv R\equiv pN$
of the boundaries of the incoming and outgoing cylinders,
and the associated conjugate variables $\m_1$ and $\m_2$.
We may again evaluate the integral by the saddle point method, namely,
\beq
Z(\m_1,\m_2)\sim \sum_N e^{-\m_1 N} e^{-N((\m_2-\m_1)\bar p - L_{-1}(\bar p )-\s(\bar p)+\ln 2)}\ ,
\eeq
where $\bar p$ is the solution of the saddle point equation
\beq \label{saddle/p}
(\m_2-\m_1) - \frac{\partial L_{-1}(p )}{\partial p} -\frac{\partial \s(p)}{\partial p} =0.
\eeq
Like in the replica trick, we are going to infer $L_{-1}(p)$ from knowledge of the $L_n(p)$.
If we take $L_{-1}(p)=-L_1(p)$ as lowest-order approximation, we find
(using (\ref{saddle/z}) and (\ref{Lnp}))
that the saddle point equation reduces to
\beq \label{saddle/p(-1)}
\zeta_{p}=e^{\m_2-\m_1} \left(\frac{p}{1-p}\right)^2\ .
\eeq
Since we do not have an analytic expression for $\zeta_p$,
we will make an educated guess at its solution.
If we replace $L_{-1}(p)$ by $L_n(p)$ in the above expressions,
we obtain for the saddle point equation
(using again (\ref{saddle/z}) and (\ref{Lnp}))
\beq \label{z-saddle/p}
\zeta_{\bar p}=e^{-(\m_2-\m_1)},
\eeq
and for the partition function
\beq
Z^{(n)}(\m_1,\m_2)\sim \sum_N e^{-\m_1 N} e^{+N \ln \n_n(e^{-(\m_2-\m_1)})}
 \sim \frac{1}{\m_1-\ln \n_n(e^{-(\m_2-\m_1)})}\ .
\eeq
We therefore conjecture that the solution of (\ref{saddle/p(-1)}) is given by
\beq
p=\frac{e^{-\m_2}}{e^{-\m_1}+e^{-\m_2}}\ ,
\eeq
which reproduces the value (\ref{z-saddle/p}) for $\zeta_{\bar p}$.
With this conjecture we obtain
\beq
Z(\m_1,\m_2)\sim \frac{1}{\m_1+\ln 4\n_1(e^{-(\m_2-\m_1)})+2(\m_2-\m_1)
\frac{e^{-\m_2}}{e^{-\m_1}+e^{-\m_2}}
 -2\s(\frac{e^{-\m_2}}{e^{-\m_1}+e^{-\m_2}})}\ .
\eeq
If we choose the scaling
\beq
\m_1\simeq \m_c+a M_1\ ,\ \ \m_2\simeq \m_c-a M_2\ ,
\eeq
together with the canonical scaling (\ref{canonical}) for the other variables
(with $k=0$, otherwise we run into the same problems as before), we find
\beq \label{mc-id}
Z(X,Y,M_1,M_2)=\frac{2 a^2}{M_1-M_2+\sqrt{X-Y}}+
 O(a^3)
\eeq
in the continuum limit.
That this expression is compatible with the presence of the identity operator
at lowest order can be seen as follows.
Since we are now dealing with a new set of (orthonormal) states depending
on two labels,
\beq
|A,R\rangle= \frac{1}{\sqrt{\NN(A,R)}}\sum_{g_{|A,R}}|g_{|A,R}\rangle\ ,
\eeq
with $\NN(A,R)$ denoting the number of triangulations of fixed area $A$ and length $R$,
we expect to lowest order in $a$ that
\beq
\sum_{g_{|A,R},g_{|A',R'}}\langle g_{|A',R'}|\hat T|g_{|A,R}\rangle
\sim \d(A-A')\d(R-R')\NN(A,R)\ .
\eeq
Performing an inverse Laplace transform of (\ref{mc-id}), we find that
\beq \label{entropy-AR}
\NN(A,R)\simeq\frac{R\ e^{-\frac{R^2}{4A}}}{2 A^{3/2}}\ e^{\l_0 A}\ ,
\eeq
where we have reinserted by hand the exponential term $e^{\l_0 A}$, which is otherwise absorbed
in the critical value of the coupling constants, and
which -- as it should -- satisfies (see (\ref{entropy-A}))
\beq
\int_0^{+\infty} dR\ \NN(A,R)=\NN(A)\ .
\eeq
Work is in progress on including the next-order approximation,
$L_{-1}(p)=L_2(p)-3L_1(p)$, and determining the resulting continuum Hamiltonian.

\section{Conclusions} \label{sec-conclusions}

In this chapter, we have for the first time derived a continuum Hamiltonian from a
three-dimensional quantum gravity model in terms of Causal Dynamical Triangulations.
This was made possible by restricting ourselves to a subclass of all triangulations,
which possess a product structure in the spatial direction, in addition
to the usual product structure in the time direction underlying the causal nature of
the model. This restriction enabled us to apply a number of analytical methods,
including the inversion formula, a random matrix formulation and the replica trick, to
solve the one-step propagator with free boundary conditions and then perform its
continuum limit. This led to an explicit expression for the quantum Hamiltonian
(\ref{H}) in ${\cal A}$-space, with wave functions depending on the area ${\cal A}$
of two-dimensional spatial slices.

Taking into account the special, topological character of three-dimensional pure
gravity, we have argued that for obtaining the full dynamics of the theory,
it is sufficient to keep track of only a finite number of
boundary data when deriving the model's transfer matrix from the
one-step propagator, and sum over everything else. This should not be
confused with an approach where the remaining degrees of freedom
(essentially the conformal mode of the metric) are fixed at the outset.
In our model, all of these are still present, but are summed over in the
discrete path integral. For simplicity, what we have done in the present piece of work is
to keep track of only a single variable, the volume (area) of the two-dimensional
universe. Because of the cylindrical topology of our spatial slices, we
expect there to be one additional Teichm\"uller parameter, corresponding to
the ratio between the area of the cylinder and the length of the cycle (the
boundary of the cylinder). To include this parameter in our model
requires to separately keep track of the number of matrices $A$ and $B$
(for the incoming and outgoing cylinder) in the traces, instead of only their
sum $N$, as we have done in the main part of our work. This task is in principle solvable
with the help of the so-called microcanonical method,
as explained in the previous section.

The fact that we have derived a non-trivial three-dimensional dynamics does
a posteriori justify our restriction to a subclass of all three-dimensional Lorentzian
triangulations. Since the continuum limit we have identified was obtained
by setting the bare inverse Newton constant to zero, it is clear that the
resulting continuum Hamiltonian describes the ``collective" effect of the
quantum fluctuations of we have summed over in the path integral, and
which elsewhere we have called the entropy of the model. In particular,
the net effect of these fluctuations is to generate a kinetic term for
the area which makes the Hamiltonian {\it bounded below}, and therefore
has the opposite sign from the corresponding term for the ``global conformal
mode" in the gravitational action.
This further supports a mechanism
already observed elsewhere in CDT models in three and four dimensions,
namely, the presence of contributions from the path integral measure
compensating for the divergence due to the conformal mode of the Euclidean
gravitational action.

There is a (small) price we are paying for using the additional constraint
of a double product structure for the triangulations, which after all was not
motivated by physics, but simply by the desire to be able to apply a number
of special solution techniques. The application of the inversion formula
implies that we are evaluating first the limit of infinitely many building blocks in
{\it one} of the spatial directions (the ``height" of the towers), and only then
the corresponding limit in the complementary direction, corresponding to
what we have called the large-$N$ limit. The effect of taking these two
limits in sequence, and not simultaneously, is that the $u$-$w$-diagram
(Fig.\ \ref{zeros}), which describes the behaviour {\it after} taking the first
of the infinite sums, has only a single point at which an infinite-volume
limit can be defined. This is in contrast to the corresponding phase diagram
for three-dimensional CDT, which has an entire critical line $\lambda_c (k)$
along which the infinite-volume limit can be taken.

The analytic solution found here has also highlighted a problem with the
scaling of Newton's constant, which has already been encountered in
previous analyses of the phase structure of three-dimensional CDT
models \cite{ABAB3}. The issue is that a standard canonical, additive
renormalization of Newton's constant as in (\ref{canonical}) does not seem
to lead to a good
continuum limit, simply because the term containing the renormalized Newton
constant $G$ has the lowest order in $a$, and dominates everything else.
The only way so far in which we have managed to obtain a nontrivial continuum limit
was by setting $k=0$, which implies that the
Newton constant does not play a dynamical role, which is of course
compatible with the absence of local gravitational excitations in three
dimensions. We believe a deeper understanding of the relation between
certain scalings of the coupling constants and that of physically relevant
quantities, like the curvature discussed in Sec.\ \ref{sec-continuum} above
would be useful, and potentially relevant for a better understanding of
the analogous issues in four dimensions. --
Even if this and other issues still remain open, we think that the work
presented here represents an important
step in the analytical understanding of CDT models in $d>2$ dimensions.

\clearpage

\renewcommand{\thesection}{A-\Roman{section}}
\setcounter{section}{0}  
\renewcommand{\theequation}{A-\arabic{equation}}
\setcounter{equation}{0}  
\section{The discrete action} \label{App-action}

We derive in this appendix the exact expression for the action of our model.
All the definitions and necessary ingredients were already introduced in \cite{ajl-def},
here we just recall and use them to derive the precise expression for our particular case.
We start from the Einstein-Hilbert action plus a Gibbons-Hawking boundary, that is,
\beq
{\cal S}=\int_M d^3x\sqrt{g(x)}\left(\L-\frac{R(x)}{2G}\right)+\frac{1}{G}\int_{\partial M}
d^2x\sqrt{h(x)} K(x),
\eeq
and then use Regge's prescription for the corresponding quantities on a simplicial manifold
\cite{regge} to obtain the discrete action
\beq \label{S-Regge}
{\cal S}\rightarrow S=\L_b\sum_{\s_3} V_{\s_3}-\frac{1}{G_b}\sum_{\s_1\in
\dot M} V_{\s_1}
(2\pi-\sum_{\s_3\supset\s_1}\th_{\s_3\rhd\s_1})-\frac{1}{G_b}\sum_{\s_1\in
\partial M} V_{\s_1}
    (\pi-\sum_{\s_3\supset\s_1}\th_{\s_3\rhd\s_1}),
\eeq
where $\s_n$ is an $n$-dimensional simplex, $V_{\s_n}$ its volume, $\dot M$ is the interior
of the simplicial manifold and $\partial M$ its boundary, $\th_{\s_3\rhd\s_1}$ is the dihedral
angle of $\s_3$ at $\s_1$, and the subscript $b$ on the coupling constants stands for {\it bare}.

In Dynamical Triangulations things simplify considerably because the edge lengths are held fixed.
In the old, Euclidean models one would have only a single type of building block in any dimension,
since all the edge lengths were chosen equal.
In CDT the Lorentzian nature of the model allows us to have two different types of edge lengths,
corresponding to time- and space-like directions.
Because of the way the causal gluing rules are implemented, what used to be a space-like and what
used to be a time-like link can still be distinguished after having mapped a Lorentzian triangulation
to its Euclidean counterpart.
Let us define the ratio $r=l_t^2/l_s^2$, where $l_t^2$ denotes the squared edge length for all time-like
links, and $l_s^2$ that for all space-like ones.
Its allowed range in the Euclidean signature is $r>\ee>0$, where $\ee$ is a positive constant, which is
the lower bound of the triangular inequalities and depends on the number of dimensions $d$ ($\ee=1/4$
for $d=2$, $\ee=1/2$ for $d=3$ and $\ee=7/12$ for $d=4$). An analytic continuation of $r$ in the complex
plane to negative values defines the (inverse) ``Wick rotation" in CDT.
Throughout this work, we will stick to positive values for $\epsilon$, corresponding to a Euclideanized,
real partition function.

In (2+1) dimensions we have precisely two different kinds of building blocks, tetrahedra with three space-like
and three time-like edges, and tetrahedra with two space-like and four time-like edges.
We will use the notation $N_{ij}$ for the number of $(i,j)$-simplices of dimension $i+j-1$ having $i$ vertices
in a constant-time slice $t$ and $j$ vertices in the subsequent one at time $t+1$. The Regge action (\ref{S-Regge})
in this case becomes
\beq
\begin{split}
S= &\L_b a^3 \left( V_{2,2} N_{2,2} + V_{3,1} (N_{3,1}+N_{1,3})\right)
 -\frac{a}{G_b}\bigg(2\pi\sqrt{r} N_{1,1}^{(\dot M)}+\pi \sqrt{r} N_{1,1}^{(\partial M)}+\\
  &-\sqrt{r}\left[4 \th_{2,2}^t N_{2,2} +3 \th_{1,3}^t (N_{1,3}+N_{3,1}) \right]
   +\pi (N_{2,0}+N_{0,2})+\\
  &-\left[2 \th_{2,2}^s N_{2,2} +3 \th_{1,3}^s (N_{1,3}+N_{3,1}) \right]\bigg)\ ,
\end{split}
\eeq
where we have distinguished the dihedral angles at time-like and space-like links by the superscripts
$t$ and $s$. Relevant volumes and angles are easily found to be
\beq
V_{2,2} = \frac{\sqrt{2r-1}}{6\sqrt{2}}\ ,\  \ \ \  V_{3,1} = \frac{\sqrt{3r-1}}{12}\ ,
\eeq
\beq
\begin{split}
&\cos \th_{1,3}^s = \frac{1}{\sqrt{3}\sqrt{4r-1}}\ , \ \ \ \cos \th_{1,3}^t = \frac{2r-1}{4r-1}\ , \\
&\cos \th_{2,2}^s =  \frac{4r-3}{4r-1}\ , \ \ \ \cos \th_{2,2}^t = \frac{1}{4r-1}\ .
\end{split}
\eeq
Finally, using Euler's formula and the Dehn-Sommerville relations for triangulated manifolds with
boundary\footnote{Note that the boundary of our ``sandwiches" of three-dimensional spacetime
(the product of a finite cylinder with an interval), which has space-like as well as time-like components,
is connected and has the topology of a torus.}, we arrive at the expression
\beq\label{}
   S=\alpha (N_{13}+N_{31}) + \beta N_{22} +\gamma N
\eeq
for the discrete action, where $\alpha$, $\beta$ and $\gamma$ depend on the dimensionless bare coupling
constants $\l$ and $k$ according to
\beq\label{alphabeta2}
\begin{split}
   &\alpha = \left(-\pi\sqrt{r}+3 \sqrt{r}\arccos \frac{2r-1}{4r-1}
             -\frac{3}{2}\pi+3\arccos \frac{1}{\sqrt{3}\sqrt{4r-1}}\right)k
    +\frac{\sqrt{3r-1}}{12}\lambda =-c_1 k + b_1 \lambda, \\
   &\beta = \left(-2\pi\sqrt{r} +4 \sqrt{r}\arccos \frac{1}{4r-1}+2\arccos \frac{4r-3}{4r-1}\right)k
    +\frac{\sqrt{2r-1}}{6\sqrt{2}}\lambda = c_2 k + b_2 \lambda, \\
   &\gamma = \left(6 \pi \sqrt{r}-\pi\right) k = c_3 k .
\end{split}
\eeq

\section{The inversion formula} \label{App-inversion}

\begin{figure}[h] 
\centering 
\includegraphics[width=12cm]{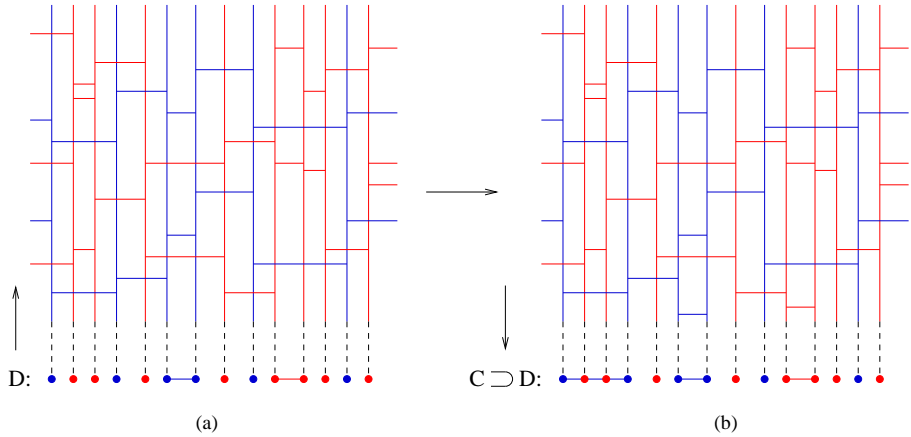} 
\vspace*{13pt} \vspace*{13pt}
\caption{\footnotesize By completing an arbitrary triangulation of our sandwich geometry with a hard-dimer
configuration $D$ (a), we build a larger triangulation (b) whose projection $C$ contains $D$. With this procedure
(with fixed $D$) we build all triangulations whose projection $C$ contains $D$ exactly once.}
\label{color-completion}
\end{figure}

The proof of the inversion formula in (2+1) dimension (for a fixed sequence $S_{|R_1,R_2}$ of $R_1$
blue and $R_2$ red towers) proceeds very similarly to that in (1+1) dimensions given in \cite{difra-hardobj}.
We first switch to the dual picture of a triangulation, and assign a weight $u$ or $v$ to each horizontal edge,
depending on its colour, and a weight $w$ to each red-blue intersection, as depicted in Fig.
\ref{3d-dual} above.
\begin{figure}[h] 
\centering 
\includegraphics[width=6cm]{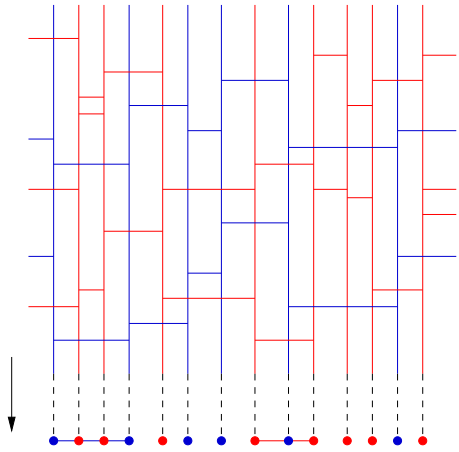} 
\vspace*{13pt}
\caption{\footnotesize Decomposing the dual picture of a sandwich geometry into a sequence of projections
onto its one-dimensional base line.
The projection is obtained by letting the ``bottom layer" of the heap of pieces, consisting of its lowest-lying edges,
drop down to the horizontal line at the base.
Each such projection defines a hard-dimer configuration on the horizontal line.}
\label{color-projection}
\end{figure}
Next, we decompose a given configuration into a sequence of projections onto the base, as illustrated in
Fig.\ref{color-projection}.
Such a projection gives rise to a hard-dimer configuration, if by a dimer we simply mean an edge linking two nearest
vertices of the same colour.
The ``hardness" refers to the fact that any vertex of the base can be occupied by at most one dimer, incorporating
a mutual avoidance of dimers.
Such a construction is possible because the dual graph is a heap of pieces.
One may think of the vertical lines in Fig.\ \ref{color-projection} as tracks along which the horizontal edges can
slide up and down, with the only restriction of not being allowed to touch or pass each other.
Using this decomposition, we can write the partition function $Z_T(t)$ as
\beq \label{Zc}
  Z_{S_{|R_1,R_2}}(u,v,w)=\sum_{{\rm hard}\ {\rm dimer}\ {\rm config.}\ C} u^{|C|_b} v^{|C|_r} w^{|\cap C|}
   Z_{S_{|R_1,R_2}}^{(C)}(u,v,w)\ ,
\eeq
where the sum extends over all hard-dimer configurations $C$ on the one-dimensional lattice (including the empty
configuration).
The numbers $|C|_b$ and $|C|_r$ count the blue and red dimers in $C$, and $|\cap C|$ the crossings between dimers
and sites of different colour in the configuration $C$. For fixed $C$, $Z^{(C)}(u,v,w)$ is the restricted partition
function involving those configurations having projection $C$, and from which we have factored out the weight
$u^{|C|_b} v^{|C|_r} w^{|\cap C|}$ of the projected part. More generally, we have the relations
\beq \label{Zd}
    u^{|D|_b} v^{|D|_r} w^{|\cap D|} Z_{S_{|R_1,R_2}}(u,v,w)=
    \sum_{C \supset D} u^{|C|_b} v^{|C|_r} w^{|\cap C|} Z_{S_{|R_1,R_2}}^{(C)}(u,v,w)\ ,
\eeq
valid for any hard-dimer configuration $D$ (eq.\ (\ref{Zc}) corresponding to $D=\emptyset$).
This expresses the fact that by completing any dual geometric configuration under consideration by a given row
of horizontal edges (corresponding to a hard-dimer configuration $D$), one builds each configuration having
a projection containing $D$, i.e. having $D$ as a sub-configuration exactly once, see Fig.\ref{color-completion}.
Let us now rewrite (\ref{Zd}) as
\beq \label{g=f.zeta}
   g(D)=\sum_C f(C)\zeta(C,D)
\eeq
with
\beq
   g(D)=u^{|D|_b} v^{|D|_r} w^{|\cap D|} Z_{S_{|R_1,R_2}}(u,v,w), \eeq \beq
   f(C)=u^{|C|_b} v^{|C|_r} w^{|\cap C|} Z_{S_{|R_1,R_2}}^{(C)}(u,v,w), \eeq and \beq
   \zeta(C,D)=
   \begin{cases}
   1 & \text{if $C\supset D$},\\
   0 & \text{otherwise}.
   \end{cases}
\eeq
Thinking of the space of configurations as a vector space in this manner,
(\ref{Zd}) can be thought of as a vector-matrix multiplication, with an upper-triangular matrix $\zeta(C,D)$
and with all diagonal elements being equal to 1. Then $\zeta$ has an inverse matrix $\mu(D,C)$ of the form
\beq
   \mu(D,C)=
   \begin{cases}
   (-1)^{|D|-|C|} & \text{if $D\supset C$},\\
   0 & \text{otherwise}.
   \end{cases}
\eeq
This property can be verified by noting that \beq
   \sum_D \zeta(C,D)\mu(D,C')=
   \begin{cases}
   \sum_{D\supset C'\atop D\subset C}(-1)^{|D|-|C|} & \text{if $C\supset D\supset C'$},\\
   0 & \text{otherwise},
   \end{cases}
\eeq
and
\beq
   \sum_{D\supset C'\atop D\subset C}(-1)^{|D|-|C|}=
   \begin{cases}
   1 & \text{if $C=C'$},\\
   \sum_{i=0}^{|C|-|C'|}(-1)^i\matrice |C|-|C'|\\i\ematrice = (1-1)^{|C|-|C'|}=0 & \text{otherwise}.
   \end{cases}
\eeq
We can now invert (\ref{Zd}) as\footnote{This is the famous M\"{o}bius inversion formula of the theory
of partially ordered sets, and $\mu$ is the associated M\"{o}bius function.}
\beq
   u^{|C|_b} v^{|C|_r} w^{|\cap C|} Z_{S_{|R_1,R_2}}^{(C)}(u,v,w) = \sum_{D \supset C} (-1)^{|D|-|C|}
   u^{|D|_b} v^{|D|_r} w^{|\cap D|} Z_{S_{|R_1,R_2}}(u,v,w).
\eeq
Noting that $Z_{S_{|R_1,R_2}}(u,v,w)$ factors out of the sum on the right-hand side, we finally get
\beq
   Z_{S_{|R_1,R_2}}(u,v,w)=
    \frac{(-u)^{|C|_b} (-v)^{|C|_r} w^{|\cap C|} Z_{S_{|R_1,R_2}}^{(C)}(u,v,w)}
    {\sum\limits_{D \supset C} (-u)^{|D|_b} (-v)^{|D|_r} w^{|\cap D|}},
\eeq
where we have used that  $|D|=|D|_b+|D|_r$, leading to the minus sign in front of both $u$ and $v$.
Picking $C=\emptyset$, we arrive at the fundamental inversion relation (eq.\ (\ref{fund3}) above)
\beq
   Z_{S_{|R_1,R_2}}(u,v,w)= \frac{ 1}{Z_{S_{|R_1,R_2}}^{hcd}(-u,-v,w)},
\eeq
where
\beq
   Z_{S_{|R_1,R_2}}^{hcd}(u,v,w)= \sum_{{\rm hard}\ {\rm dimer}\ {\rm config.}\ D}
   u^{|D|_b} v^{|D|_r} w^{|\cap D|}
\eeq
denotes the partition function for hard coloured dimers with fugacity $u$
($v$) per blue (red) dimer
and weight $w$ per crossing.

\section{Numerical computations} \label{App-numeric}

\begin{figure}[t]
\centering
\includegraphics[width=10cm]{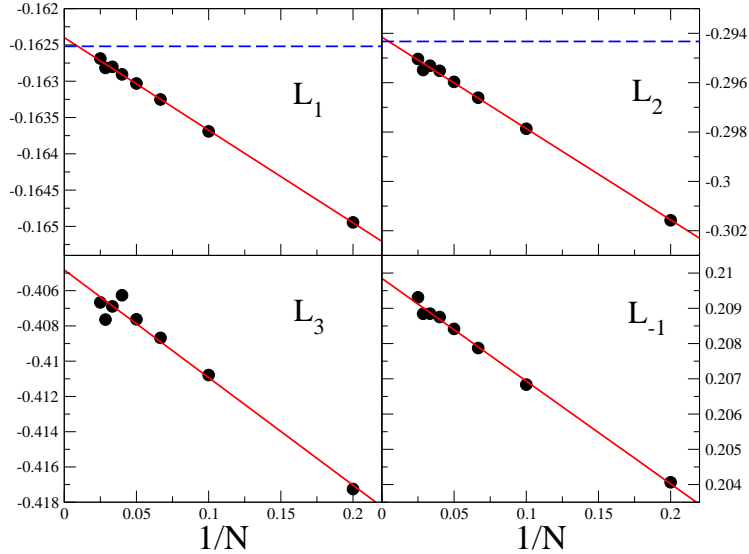}
\caption{
The numerical results for $L_{n}(N)$, $n=-1,1,2,3$, as a function of $1/N$ for $u=0.24$ and $w=0.1$.
The full lines are fits to $L_{n}(N) = L_{n} + A_n/N$. The dashed lines are the analytical values for
$L_1$ and $L_2$ computed using the replica trick. They differ from the extrapolation by about $0.1\%$.
}
\label{1suN}
\end{figure}

A direct numerical computation of the generalized Lyapunov exponents $L_n$, defined in (\ref{lns}) above,
is possible following the procedure given in \cite{crisanti}. One extracts a large number of values $\g$ at fixed $N$,
computes the average $\langle e^{n \g N} \rangle$, and plots $L_n(N)=N^{-1} \ln \langle e^{n \g N} \rangle$
as a function of $1/N$.
Usually
$L_n(N)$ turns out to be linear in $1/N$ for large $N$,
\beq
L_n(N) \sim L_n + A_n/N,
\eeq
which allows for a good extrapolation of the data to $N= \io$, see Fig.~\ref{1suN} for an example.
The values of $\g$ are extracted as follows \cite{crisanti}:
\begin{enumerate}
\item construct a random unimodular
vector $v$ (e.g., by extracting the components from a Gaussian distribution);
\item extract the numbers $q_j$ and apply the matrix $M_{q_j}$ to $v$ to obtain a vector $v' = \prod_j M_{q_j} v$;
\item compute $|v'| \simeq e^{N \g}$, which holds to leading order since the modulus of $v'$ is dominated
by the same largest eigenvalue of $\prod_j M_{q_j}$ which also dominates $\Tr \prod_j M_{q_j}$.
\end{enumerate}
This will give us a set of $\NN$ values $\g_i$ for $\g$, from which we can construct a histogram $\pi(\g)$
and extract the function $S(\g)$ at given $N$, see Fig.~\ref{Sgamma}. The coincidence of the curves for different
$N$ indicates that the asymptotic limit $N \to \io$ has been reached.
To compute $L_n$, we first have to compute the average of $\g$, $\bar\g = \frac1\NN \sum_{i=1}^\NN \g_i$,
and then use
\beq
e^{N L_n(N)} = \langle e^{n \g N} \rangle = e^{n \bar\g N} \frac1\NN \sum_{i=1}^\NN e^{n N (\g_i-\bar \g)}
\hskip10pt \Leftrightarrow \hskip10pt
L_n(N) = n \bar\g + \frac1N \ln \left[\frac1\NN \sum_{i=1}^\NN e^{n N (\g_i-\bar \g)} \right] .
\eeq
By subtracting $\bar \g$ we can compute the average as long as the quantity $N (\g_i - \bar \g) \sim \sqrt{N}$
is not too large. In practice we are limited to $N < 50$, but this is enough to get a very good linear extrapolation
to $N \to \io$.
The difference between the numerical computation and the exact analytic expressions for $L_1$ and $L_2$
is of the order of $10^{-4}$, see Fig.~\ref{1suN}.
We believe that a similar error affects the computation of $L_{-1}$ too.

\section{Fractional derivatives and inverse Laplace transforms} \label{App-fracder}

Although the notion of {\it fractional calculus} may at first appear
somewhat extravagant,
this is an old and well-studied subject in mathematics, with plenty of
applications,
as testified by the number of books on the subject (see, for instance, the
list of references
in \cite{mainardi,podlubny}). At the heart of the theory of fractional
calculus is the
definition and study of
two operators $J^{\a}:f(x)\rightarrow J_x^{\a}f(x)$ and
$D^{\a}:f(x)\rightarrow D_x^{\a}f(x)$
on a sufficiently large class of functions $\{ f(x)\}$ and for a positive
real number $\a$,
with the following properties.
\begin{enumerate}
\item When $\a=n$ is a positive integer, the operator $J_x^{\a}$ gives the
same result as $n$-fold
integration, and $D_x^{\a}$ gives the same result as the usual $n$-th
derivative
$\frac{d^n}{dx^n}$.
\item The operators of order $\a=0$ are the identity operator.
\item $J_x^{\a}$ and $D_x^{\a}$ are linear operators.
\item For any $\a$, $\b$, the semigroup property holds for $J_x^{\a}$,
namely,
$J_x^{\a}J_x^{\b}f(x)=J_x^{\a+\b}f(x)$.
\end{enumerate}

There are many inequivalent definitions of fractional derivative/integral
satisfying
these properties, which we are not going to review here. Instead, we will
recall the definition and properties of the one that turns out to be
relevant for our purposes, the so-called Weyl fractional
derivative/integral.
(Reference \cite{osler} has a short review, which also contains a clear
explanation
for why the various  definitions are inequivalent.)
The Weyl fractional integral is defined as\footnote{The subscript
``$\infty$" refers to the
extremum of integration.
Other choices of extremum would give rise to different (and inequivalent)
definitions of
fractional derivative/integral, which we would denote by $J_{x,x_0}^{\a}$,
$D_{x,x_0}^{\a}$.}
\beq \label{weyl-int}
J_{x,\infty}^{\a}f(x)=\frac{(-1)^{-\a}}{\G(\a)}\int_x^{\infty}dt f(t)(t-x)^{\a-1}.
\eeq
The natural definition of the derivative operator would be to take
$D_{x,\infty}^{\a}=J_{x,\infty}^{-\a}$
but this is divergent. A standard trick is then to define
\beq \label{weyl-der}
D_{x,\infty}^{\a}f(x)=\frac{d^{[\a ]+1}}{dx^{[\a ]+1}}\ J_{x,\infty}^{[\a
]+1-\a}f(x),
\eeq
where $[\a ]$ is the integer part of $\a$. The rationale behind this
definition is that
in this way $D_x^{\a}$ is the left inverse of $J_x^{\a}$, $i.e.$
$D_x^{\a}J_x^{\a}f(x)=f(x)$, the
same as when $\a =n$ is an integer.
With this definition it can be checked that
\beq \label{frac-exp}
D_{x,\infty}^{\a}e^{-Ax}=(-A)^{\a}e^{-Ax},
\eeq
which is exactly what we need, as explained in the
main text.

According to the analysis in the text, we need to compute
\beq
\tilde Z(x,y,\Delta t=1)=D_{\ln x,\infty}^{1/4} D_{\ln y,\infty}^{1/4} Z(x,y,\Delta t=1)\ ,
\eeq
which in the continuum limit we can write as (see Sec.~\ref{sec-hamiltonian})
\beq
\tilde Z(X,Y)=D_{-a^2 X,\infty}^{1/4} D_{-a^2 Y,\infty}^{1/4} a^2 Z(X,Y)=
\frac{a}{i} \ D_{X,\infty}^{1/4} D_{Y,\infty}^{1/4} Z(X,Y)\ .
\eeq

At leading order in $a$ we expect to have a $\d (A_1-A_2)$ in the transfer matrix.
Therefore we can take just a $\frac{1}{2}$-derivative with respect to one of the two arguments.
From the expression of $Z(X,Y)$ given in (\ref{Z-k=0}) we have at leading order ($l.o.$)
\beq
\begin{split}
\frac{a}{i}  D_{X,\infty}^{1/2} Z_{l.o.}(X,Y) &=\frac{1}{i} D_{X,\infty}^{1/2} \frac{2}{\sqrt{X+Y}}=
 -\lim_{X_0\to\infty}\frac{d}{dX}\frac{1}{\G(1/2)}\int_X^{X_0}dt \frac{2}{\sqrt{t+Y}}\frac{1}{\sqrt{t-X}}\\
 &=\frac{2}{\sqrt{\pi}}\frac{1}{X+Y}\ ,
\end{split}
\eeq
which is proportional to the Laplace transform of the delta function.
Note that the limit here has to be taken after the derivative, otherwise it would be divergent.
This is the kind of convergence problem mentioned in Sec.~\ref{sec-gluing} (the problem would
stay also if we would use two $\frac{1}{4}$-derivative instead of one  $\frac{1}{2}$-derivative).
This is not really a big problem because what we need is an operator which gives the same result
as (\ref{frac-exp}) when acting on exponential functions. Since for exponential functions the limit can be
taken both before and after the derivative, we have a freedom in choosing the order of these operations.

For the cosmological term in (\ref{Z-k=0}), interchanging limit and derivative does not give any problem,
and we get
\beq
 -\frac{a}{i} D_{X,\infty}^{1/4} D_{Y,\infty}^{1/4} \frac{\L}{(X+Y)^{3/2}}=-\frac{2 a}{\sqrt{\pi}}\frac{1}{(X+Y)^2}\ ,
\eeq
which is proportional to the Laplace transform of $a\L A_1 \d (A_1-A_2)$.

For the remaining term in (\ref{Z-k=0}) we have to choose again to take the limit after the derivative.
What we obtain is
\beq
\begin{split}
 \frac{a}{i} D_{X,\infty}^{1/4} D_{Y,\infty}^{1/4} \left( \frac{5}{6}+\frac{XY}{(X+Y)^2}\right)
 =\frac{\sqrt{\pi} (X^2-10 X Y+Y^2)}{16 (X+Y)^{5/2}}\ .
\end{split}
\eeq
(The constant term gives zero as a result.)
The inverse Laplace transform of this term presents another divergence. This may be avoided by an integration
by parts, but we prefer to take another route and use the second method presented in Sec.~\ref{sec-gluing}.

\begin{figure}[h]
\centering
\includegraphics[width=6cm]{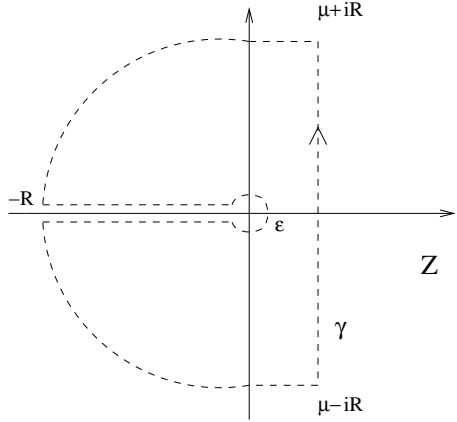}
\vspace*{13pt}
\caption{\footnotesize The Bromwich-Hankel contour used for the inverse Laplace transform.
Integrating along this contour the function $e^{Z A_1}/\sqrt{Z}$
we obtain zero because no singularities are encircled. From this, in the limit $R\to\infty$, $\ee\to 0$,
we obtain the result (\ref{cut-integral}).}
\label{contour}
\end{figure}

The method consists in performing the inverse Laplace transform on $Z(X,Y)$ directly, without use
of fractional derivatives, and then recognizing the contribution of the entropy in the Hamiltonian.
Let's see first how this works for the leading order term.
We have to compute the integral
\beq
\begin{split}
I(A_1,A_2) &\equiv \int_{-i\infty+\m}^{+i\infty+\m}\frac{dX}{2\pi i}e^{X A_1}
 \int_{-i\infty+\m}^{+i\infty+\m}\frac{dY}{2\pi i}e^{Y A_2}
 \frac{2}{\sqrt{X+Y}}=\\ &= \int_{-i\infty+\m}^{+i\infty+\m}\frac{dY}{2\pi i}e^{Y (A_2-A_1)}
  \int_{-i\infty+\m}^{+i\infty+\m}\frac{dZ}{2\pi i}e^{Z A_1}
 \frac{2}{\sqrt{Z}}=\\
 &=\d(A_2-A_1) \int_{-i\infty+\m}^{+i\infty+\m}\frac{dZ}{2\pi i}e^{Z A_1}
 \frac{2}{\sqrt{Z}} \ .
\end{split}
\eeq
Because of the square root, the integrand has a branch-cut along the negative $Z$ semi-axis.
The contour of integration for the Laplace inversion can continued to a Hankel contour
around the cut, as shown in Fig.~\ref{contour}.
We have
\beq \label{cut-integral}
\begin{split}
I(A_1,A_2) &= \d(A_2-A_1) \left( -\int_{-\infty^+}^{0^+}\frac{dZ}{2\pi i}e^{Z A_1}
 \frac{2}{i\sqrt{-Z}}- \int_{0^-}^{-\infty^-}\frac{dZ}{2\pi i}e^{Z A_1}
 \frac{2}{-i\sqrt{-Z}}\right)=\\
 &=\d(A_2-A_1) \int_{0}^{+\infty}\frac{dZ'}{\pi}e^{-Z' A_1}
 \frac{2}{\sqrt{Z'}}=\frac{2}{\sqrt{\pi}}\frac{\d(A_2-A_1)}{\sqrt{A_1}} \ ,
\end{split}
\eeq
in which we can recognize the subleading behaviour of (\ref{entropy-A}). Multiplying by $\sqrt{A_1}$
we obtain exactly the same result as from the fractional derivative.

For the cosmological term we have again a convergence problem because instead of the square root in
the denominator we have a power $3/2$ which diverges too fast at zero. This can be removed
by a formal integration by parts in the previous step, and the final result coincides
with the fractional derivative one.

The remaining terms give instead no problems.
The constant term gives a term proportional to $\d(A_1)\d(A_2)$, which is a non-propagating
and non-universal kind of term already familiar in two dimensions (see \cite{ABAB1} and references
therein), and which furthermore gives just zero when multiplied by the inverse entropy factors
$A_1^{1/4}A_2^{1/4}$, in agreement with the fractional derivative result.
The last term has no branch-cut, and is the usual term encountered in two dimensions (see,
for example, \cite{difra-calogero}). After inverse Laplace transform it gives
\beq
\hat H''_{kin}\d(A_2-A_1)=\left(-A_2\frac{\partial^2}{\partial A_2^2}-\frac{\partial}{\partial A}\right)\d(A_2-A_1)\ ,
\eeq
Dividing $\hat H''_{kin}$ by the entropy factor $1/\sqrt{A_2}$ we find the kinetic part of the auxiliary
Hamiltonian (\ref{H-aux}):
\beq
\hat H'_{kin}=-A_2^{\frac{3}{2}}\frac{\partial^2}{\partial A_2^2}
-A_2^{\frac{1}{2}}\frac{\partial}{\partial A_2}\ .
\eeq
Finally, by use of (\ref{H-inversion}) (where $\NN(A)$ has to be replaced by the subleading term
alone of (\ref{entropy-A}) because the exponential part is absorbed in the critical value of the
boundary cosmological constants), we find the kinetic term of the final Hamiltonian (\ref{H}).

We have seen that the convergence problems in the two methods are complementary.
The only problem encountered with the auxiliary Hamiltonian method is in the cosmological term,
which instead presents no problems with respect to the fractional derivative.



\newpage
\pagestyle{empty}

\backmatter


\newpage

\addcontentsline{toc}{chapter}{Acknowledgements}
\chapter*{Acknowledgements}

I am indebted to many people for my human and scientific growth during these four years in Utrecht,
of which this thesis represents a conclusion.

First of all I wish to thank my supervisor Renate Loll. For having chosen me as a PhD student,
for the help in the moments of halt, for all the feedback about research, presentations and writings,
and in particular for all the corrections that made this thesis readable.

Special thanks to Francesco Zamponi for the scientific collaboration and for the precious friendship.

I am thankful to Giovanni Amelino-Camelia for strongly supporting me in the choice to come to
Utrecht.

I acknowledge discussions with  Jan Ambj\o rn, Gerard 't Hooft,
Philippe Di Francesco, Emmanuel Guitter, Laurent Freidel, Ruben Costa-Santos, Richard Lim, and
Willem Westra (who I also thank for the Dutch translation of the summary).

My gratitude goes to all the people from the Institute for Theoretical Physics for the
stimulating scientific environment and the valuable source of knowledge they provide.

To my officemates along the years: Benham, Mathijs, Gerasimos and Artem.

To Biene, Geertje and Wilma for all the help with practical matters.

To those colleagues whose friendship has turned the work routine into fun: Daniel, Daniele, Emiliano, Fernando,
Joe, Lih-King, Pedro and Pietro.

To all the other friends I have met in the Netherlands and have made my life here memorable:
Alessandro, Alexey, Camilo, Daniel, David, Despina, Erwin, Gar-Yein, Jannis, Lu, Luisa and Aaron,
Mario, Melissa, Metka, Olga, Osmar, Sara and all the other people I met in Grote Trekdreef
(and to Claudio, who invited me there first).

To the evergreen friends in Italy (or temporarily abroad):
Simone, Smoje, Luca, Stefano, Alessandro, Francesco and Paolo.

To Yuan-Ju.

To my mother.

%


\newpage

\nocite{*}
\cleardoublepage

\fancyhf{} 
\pagestyle{fancy}
\fancyhead[LE,RO]{\thepage}
\fancyhead[RE]{\small BIBLIOGRAPHY}
\fancyhead[LO]{\small BIBLIOGRAPHY}


\addcontentsline{toc}{chapter}{Bibliography}

\bibliographystyle{unsrt}


\end{document}